\documentclass[12pt,fleqn]{article}

\usepackage{geometry} 
\usepackage{graphicx}

\usepackage{amsmath}
\usepackage{amssymb}
\usepackage{braket}

\usepackage{diagrams}
\usepackage{mathabx}
\usepackage{pdflscape}

\usepackage{outlines}

\usepackage{comment}
\usepackage{wasysym}

\makeatletter
\newsavebox{\@brx}
\newcommand{\llangle}[1][]{\savebox{\@brx}{\(\m@th{#1\langle}\)}%
  \mathopen{\copy\@brx\kern-0.5\wd\@brx\usebox{\@brx}}}
\newcommand{\rrangle}[1][]{\savebox{\@brx}{\(\m@th{#1\rangle}\)}%
  \mathclose{\copy\@brx\kern-0.5\wd\@brx\usebox{\@brx}}}
\makeatother

\makeatletter
\renewcommand\paragraph{\@startsection{paragraph}{4}{\z@}%
            {-2.5ex\@plus -1ex \@minus -.25ex}%
            {1.25ex \@plus .25ex}%
            {\normalfont\normalsize\itshape}}
\makeatother
\newtheorem{prop}{Proposition}
\newcommand{\xb}{{\boldsymbol x}}
\newcommand{\yb}{{\boldsymbol y}}

\newcommand{\ssb}{{\boldsymbol s}}  
\newcommand{\nb}{{\boldsymbol n}}  
\newcommand{\lambdab}{{\boldsymbol \lambda}}  
\newcommand{\kappab}{{\boldsymbol \kappa}}  
\newcommand{\mub}{{\boldsymbol \mu}}  
\newcommand{\anglink}{{\left \langle i \right \rangle_k^n}} 
\newcommand{\pt}{\tilde{p}}
\newcommand{\alb}{{\boldsymbol \alpha}}

\newcommand{\dkl}{\mathcal{D}_{\mathcal{K}\mathcal{L}}}

\begin{document}

\title{
Prospects for \\ Declarative Mathematical Modeling \\ of Complex Biological Systems}

\author{Eric Mjolsness
\thanks{\,Department of Computer Science, University of California Irvine CA 92697.  Email: emj@uci.edu .} 
}
\maketitle

\begin{abstract}
Declarative modeling uses symbolic expressions to represent models.
With such expressions one can formalize 
high-level mathematical computations on models
that would be difficult or impossible to perform 
directly on a lower-level 
simulation program, in a general-purpose programming language.
Examples of such computations on models
include model analysis,  
relatively general-purpose model-reduction maps, 
and the initial phases of model implementation,
all of which should preserve or approximate
the mathematical semantics of a complex biological model.
The potential advantages are particularly relevant in the case
of developmental modeling, wherein complex spatial structures exhibit dynamics at molecular,
cellular, and organogenic levels to relate genotype to multicellular phenotype.
Multiscale modeling 
can benefit from both the expressive power of declarative modeling languages
and the application of model reduction methods to link models across scale.
Based on previous work, here we define 
declarative modeling of complex biological systems
by defining the 
operator algebra semantics of an increasingly powerful series of
declarative modeling languages
including reaction-like dynamics of parameterized and extended objects;
we define semantics-preserving implementation and 
semantics-approximating model reduction transformations;
and we outline a ``meta-hierarchy'' for organizing declarative
models and the mathematical methods that can
fruitfully manipulate them.
\end{abstract}

\section{Introduction}
\label{intro}

Central to developmental biology is the genotype-to-phenotype map
required to close the evolutionary loop implied by selection on heritable variation.
However, relating genotype to phenotype in a multicellular organism is an
intrinsically multiscale and (therefore) complex modeling endeavor.
Partial automation has the potential to tame the complexity for
human scientists, provided it can address highly heterogeneous mathematical dynamics
including stochastic reaction networks, 
dynamic spatial structures at molecular and cellular scales,
and partial differential equations (PDEs) governing 
both pattern formation and
dynamic geometry within dynamic topology.
Here we outline such a mathematical modeling framework,
founded on the unifying idea of rewrite rules denoting
operators in an operator algebra.
The rewrite rules make this framework {\it declarative}:
capable of expressing mathematical ideas at a high level in a symbolically-
and computationally-manipulable form.

To this end we
propose and discuss an informal definition of declarative modeling in general, 
and provide as examples a collection of specific mathematical constructions
of processes and extended objects 
for use in declarative models of complex biological systems
and their processing by computer.
Such processing can be symbolic and/or numerical, 
including for example model reduction by coupled symbolic and machine learning methods. 
The resulting apparatus is intended for 
semi-automatic synthesis and analysis of biological models, a computational domain 
which must typically deal with substantial intrinsic complexity in the subject modeled.

The necessity for such automation is strongest for the most complex biological models,
notably those required for developmental modeling.
Examples of such dynamical spatial systems in development include
plant cell division under the influence of dynamic microtubules in the pre-prophase band;
neurite branching and somal translocation dependent on dynamic cytoskeleton
in mammalian brain development;
mitochondrial fission and fusion;
plant organogenesis in shoot (shoot apical meristem or SAM) and root (lateral root initiation);
topological changes in close-packed 2D tissues (e.g. fly wing disk) 
in response to cell division and convergent extension; 
neural tube closure;
branching morphogenesis in vascular tissues;
and many others. 
In all these cases the spatial dynamics involves nontrivial changes
in geometry and/or topology of extended biological objects;
we must be able to represent such dynamics mathematically and computationally.
Our main examples will be networks of dynamically interconnected
cells and dynamically interconnected segments of cytoskeleton within a cell.

This paper is organized as follows.
We will define declarative models in Section \ref{def_decl_mod} (informally in general but formally in particular cases)
and survey a series of declarative quantitative modeling languages of increasing expressive power
for biochemical and biological modeling as exemplified by coarse-scale models of
multicellular tissues with cell division and by cytoskeletal dynamics; also we will describe the
``operator algebra'' mathematical semantics for these languages
and the utility of structure-respecting maps among these mathematical entities.
We will
generalize the declarative modeling language/operator algebra semantics
approach to encompass extended objects in Section \ref{section_extended},
by way of 
spatially embedded
discrete graph structures (including dimension and refinement level indices) and their continuum limits;
dynamics on such structures including PDEs; 
and dynamics by which such such structures can change including graph rewrite syntactic rules
under a novel operator algebra semantics.
We will discuss progress towards a general method for nonlinear model reduction (usually across scales)
by machine learning in Section \ref{sec_mod_red}
including an application of variational calculus generalized to a higher level. 
In Section \ref{sec_Tchicoma} we will propose the elements of a larger-scale mathematically based
``meta-hierarchy'' for organizing many biological models and modeling methods,
enabled by the declarative approach to modeling and by structure-respecting maps
among declarative languages and their mathematical semantics.

Much of this paper reviews previous work, extending it 
(e.g. with the multiscale ``graded graph'' constructions of
Section~\ref{section_extended},
their dynamics of Equation~(\ref{slice_rule}),
and graph rewrite rule operator products defined by Propositions 1 and 2 in Section~\ref{graphgramprod})
and setting it in a broader context.
The aim of this paper is not to provide
a balanced summary of work in the field;
instead it is aimed mainly at 
outlining mathematically
the possibilities of particular directions for future development.

\subsection{Notation}

We will use square brackets to build sets (tuples or lists) ordered by indices, so in a suitable context
$f([x]) = f([x_k | k \in \{1, \ldots n \}]) = f(x_1, \ldots x_n)$.
Multisets are denoted $\{ ({\rm multiplicity})  ({\rm element}), \ldots \}_*$.
Indices may have primes or subsubscripts and are usually deployed as follows:
$r$ indexes rewrite rules, $i$ and $j$ index individual domain objects,
$\alpha$ and $ \beta$ index domain object types,
$p$ and $q$ index elements in either side of a rule,
$c$ indexes variables in a parameterized rule,
and
$A$ indexes a list of defined measure spaces.
Object typing will be denoted by ``$:$'', so for example
$n:{\mathbb N}$ means that $n \in {\mathbb N}$ and 
the usual set of integer arithmetic operations pertain to $n$.
Subtyping will be denoted by ``$::$''.

\section{Declarative modeling}
\label{def_decl_mod}

A distinction made in classical Artificial Intelligence (AI) by Winograd among others
is that between  {\it declarative} and {\it procedural} representations of {knowledge};
this is the AI version of a philosophical distinction between ``knowing that'' and ``knowing how'',
as it pertains to knowledge expressed in a formal language
that can be used to program intelligent systems (Winograd 1975).
Generic advantages for declarative knowledge identified by Winograd
include its greater flexibility, compactness, understandability, 
and communicability compared to procedural knowledge;
these are virtues we seek for complex biological modeling.
On the other hand declarative knowledge may be incomplete, 
as it omits for example  ``heuristic'' knowledge of domain-specific strategies for action.
\footnote{There is some history, mostly beyond the scope of this paper,
to the adaptation of the ``declarative/procedural'' distinction to both computer programming languages
and to formal modeling languages that can be executed on a computer.
The classic programming textbook by Abelson and Sussman (Abelson et al. 1996)
uses a very similar ``declarative/imperative'' distinction (knowledge of ``what is'' vs. ``how to'') 
in introducing the programming languages Scheme and Prolog, 
both with historical pedigrees in logic.
Functional programming advocated by Backus (Backus 1978) 
and exemplified in Haskell, and logic programming exemplified by Prolog,
both satisfy ``referential transparency'' 
(expressions can be substituted by equivalent expressions or values),
a technical idea from mathematical logic 
that is sometimes taken to define the ``declarative'' programming paradigm.
Even within biological modeling (Spicher et al. 2007) the 
``what'' vs. ``how'' declarative/procedural distinction has been used, 
though in a somewhat different way than we 
use it: 
in that case the relevant ``procedure''
to be omitted is fine-grain biological rather than computational process information.
We will specialize anew from the looser AI-inspired meaning: 
Declarative modeling languages 
specify the goals or criteria of a successful modeling computational simulation, 
leaving many of the procedural decisions about how exactly to pursue those goals up to other software.
This will be done by mathematically specifying the biological (or other scientific) model,
including its dynamics.
}

\begin{figure}
\begin{center}
  \includegraphics[width=0.5\columnwidth]{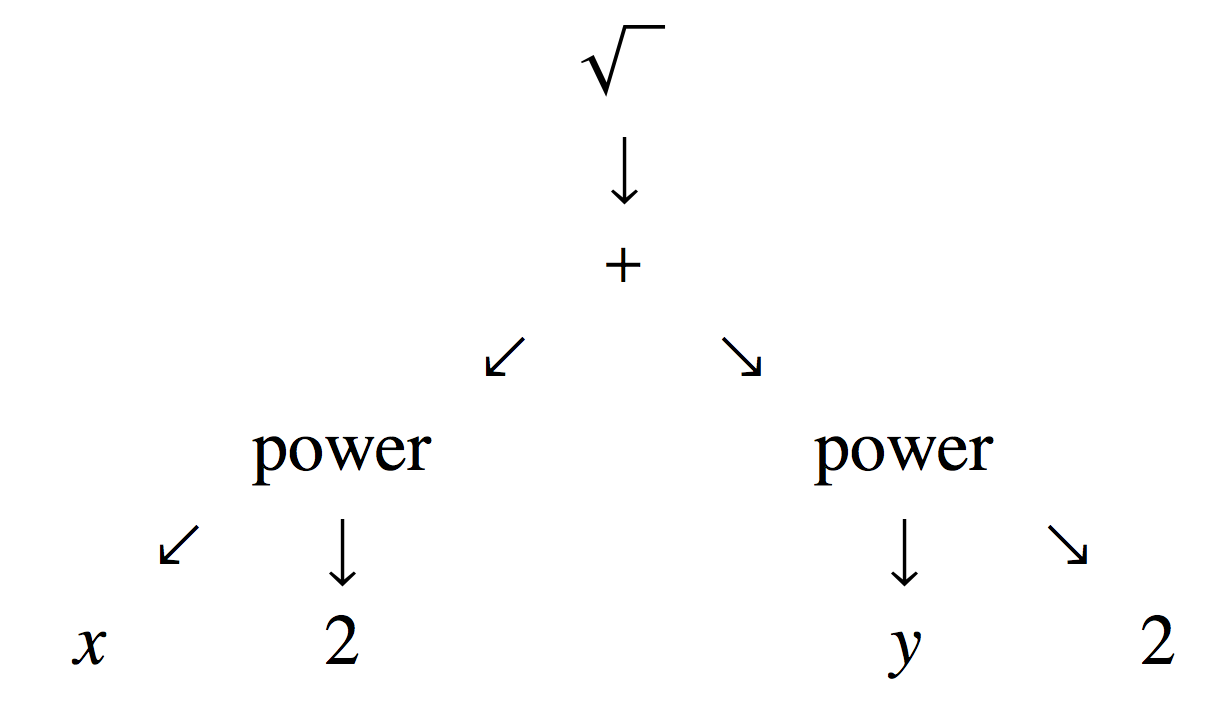}
\end{center}
\caption{ 
Schematic view of an abstract syntax tree for the algebraic expression $\sqrt{x^2+y^2}$.
It can be transformed in many ways. For example it could be numerically evaluated at
$x=5,y=12$ by successive local transformations of ASTs.
AST transformations are used in compilers for programming languages,
in computer algebra systems, and
in recent work on ``symbolic regression'' in machine learning.
 }
\label{fig:AST}       
\end{figure}

An example of a formalizable, declarative language for modeling biology is a collection of partial differential equations 
in which the variables represent local concentrations of molecular species 
and the spatial differential operators are all diffusion operators $\nabla_{\xb}\cdot(D_\alpha(\xb) \nabla_\xb$). 
Such a deterministic reaction/diffusion model can be represented in a computer by one or more abstract syntax trees (ASTs) 
including nodes for variable names, arithmetic operations, spatial differential operators such as the Laplacian, 
the first-order time derivative operator, equality constraints including boundary conditions, 
and possibly function definitions. (A simple AST is shown in Figure~\ref{fig:AST}.)
Such an AST can be used to denote a reaction-diffusion model
as a data object that can be manipulated by computer algebra.
It can also under some conditions be transformed symbolically, 
for example by separation of time scales replacing a subset of ordinary differential equations (without spatial derivatives) 
by function definitions of algebraic rate laws to be invoked in the differential equations for the remaining variables 
(e.g. (Ermentrout 2004)).
The language is ``declarative'' from a computational point of view because it doesn't
specify any algorithmic details for numerically solving the dynamical systems specified.
Implemented examples will be discussed in Section~\ref{section_extended}.

We generalize from this example. A formal language should have a ``semantics'' map $\Psi$, 
giving a mathematical meaning to some defined set of valid expressions in the language. 
For a modeling language, each valid model declaration $M$ should correspond to 
an instance $\Psi(M)$ in some space $S$ of ``dynamical systems'' interpreted broadly, 
so that such systems may be stochastic and/or infinite-dimensional. 
If some of the semantically meaningful model expressions are composed of meaningful sub-expressions, 
and their semantics can be combined in a corresponding way, then the semantics is ``compositional''; 
composition commutes with the semantic map. 
Likewise for valid transformations of model declarations, 
one would like the semantics before and after transformation to yield either the same mathematical meaning
(equivalent dynamics), 
or two meanings that are related somehow eg. by approximate equality under some conditions on 
parameters that may be partially known by proof and/or numerical experiment.

So in the context of modeling languages in general and biological modeling languages in particular,
key advantages of the declarative language style are captured by the following
informal (but perhaps formalizable) definition:

A {\it declarative modeling language} $L$ is a formal language together with
(a) compositional semantics  $\Psi:L \rightarrow S$ that maps all syntactically valid models $M$ in $L$ 
into some space $S$ of dynamical systems,  and 
(b) conditionally valid or conditionally approximately valid families of {\it transformations}
on model-denoting AST expressions in the modeling language $L$.
These AST transformations can be expressed formally in some computable meta-language, 
though the meta-language need not itself be a declarative modeling language.

Under this informal definition the utility of a declarative modeling language would depend
on its expressive power, addressed in Sections \ref{pure_rules} - \ref{deqr} and \ref{section_extended} below,
and on the range, value, and reliability of the model-to-model transformations that can be 
constructed for it, to be discussed in Sections \ref{section_extended} (implementation), 
\ref{sec_mod_red} (model reduction), and \ref{sec_Tchicoma} (wider prospects).
Multiscale modeling benefits from both the expressive power
(e.g. representing cellular and molecular processes in the same model)
and model reduction (finding key coarse-scale variables and dynamics
to approximate fine-scale ones) elements of this agenda.

Although it is plausible that ``one can't proceed from the informal to the formal by formal means'' (Perlis 1982),
so that the task of formalizing a complex biological system to create models must begin informally,
we will nevertheless try to be {\it systematic} about this task
by building the semantic map $\Psi$ up
out of ``processes'' and ``objects'' of increasing generality and structure.
These processes and objects are represented by ``expressions'' in a language, and mapped to
mathematical objects that together define a model.
So the map $\Psi$ will be concerned with expressions, processes, and objects.

\subsection{Semantic Maps}
\label{sec_semantic_map}

We show how to define a series of simple quantitative modeling languages of increasing expressive power,
accompanied by semantic maps $\Psi(M)$ to operator algebras.
To do this we need some idea concerning how to represent elementary biological processes
with syntactic expressions that can include numerical quantities. 
One such idea begins with chemical reaction notation.

\subsubsection{Pure Reaction Rules}
\label{pure_rules}

In biology, generalized versions of {\it rewrite rules} naturally specify biochemical processes
in a declarative modeling language, as well as model transformations in a metalanguage. 
The most straightforward example is chemical reaction notation.
As in (Mjolsness 2010) we could use chemical ``addition'' notation:
\begin{equation}
\Bigg(
\sum \limits_{\alpha=1}^{A_{\max }}m_{\alpha}^{\left( r\right) } A_{\alpha}
\Bigg)
\ \ \overset{k_{\left(
r\right) }}{\longrightarrow }\ \ 
\Bigg(
\sum \limits_{\beta=1}^{A_{\max }}n_{\beta}^{\left(
r\right) } A_{\beta} 
\Bigg) , %
\label{XRef-Equation-926214051}
\end{equation}
where $m_{\alpha}^{\left( r\right) } $ and $n_{\beta}^{\left( r \right) }$ are nonnegative integer-valued stoichiometries 
for molecular species $A_{\alpha}$ indexed by $\alpha$ in reaction $r$ with nonnegative reaction rate $k_{(r)}$,
and we may omit summands with $m_{\alpha}^{( r) }=0$ or $n_{\beta}^{(r)}=0$.
The left hand side (LHS) and right hand side (RHS) of the reaction arrow are 
just formal sums or equivalently
${\it multisets}$ with nonnegative
integer multiplicities of all possible reactants, defaulting to zero if a reactant is not mentioned:
\begin{equation}
\Big\{
m_{\alpha}^{\left( r\right) } A_{\alpha}
\Big\}_*
\ \ \overset{k_{\left(
r\right) }}{\longrightarrow }\ \ \
\Big\{
n_{\beta}^{\left(r\right) } A_{\beta} 
\Big\}_* . %
\label{Equation-multiset}
\end{equation}
Concisely we can summarize reaction rule $r$ as ``$ {\text{LHS}_r \rightarrow \text{RHS}_r}$''.
Either detailed syntax expresses the transformation of one multiset of symbols into another, 
with a numerical or symbolic quantitative reaction rate $k_{\left(r\right) }$.
The syntax is easily encoded in an abstract syntax tree (AST).

An example AST {\it transformation} might be a meta-rule that reverses an arrow (Yosiphon 2009)
and changes the name of the reaction rate to (for example) $k_{\left(r^{\prime}\right) }$,
for a new reaction number $r^{\prime}$,
allowing for the possibility of detailed balance to be satisfied in a collection of reactions.
Many other reaction arrow types (e.g. substrate-enzyme-product)
can then be defined
by computable transformation to 
combinations of
these elementary mass action reactions
(e.g.  (Shapiro et al. 2003; Yang et al. 2005; Shapiro et al. 2015b) 
for many examples in a declarative biological modeling context),
using either commercial (Wolfram Research 2017)
or open-source (Joyner et al. 2012) computer algebra system software.

How can we define a compositional semantics for this reaction notation?
Fortunately the operator algebra formalism of quantum field theory can be adapted to
model the case of ordinary (non-quantum) probabilities governed by the law of mass action
in a Master Equation (Doi 1976a,b; Peliti 1985; Mattis and Glasser 1998; Mjolsness 2010)
(and (Morrison and Kinney 2016) for the equilibrium case).
As in physics, each operator algebra we deal with will be generated 
by a collection of elementary operators and their commutation relations,
together with their closure under operator addition, operator multiplication,
and scalar multiplication of operators by real numbers. 
In the present case the generators are the identity operator together with,
for each molecular species or other symbol type $\alpha$ in the reaction set,
a creation operator $\hat{a}_\alpha$ and an annihilation operator $a_\alpha$
the commutation relations are
$a_\alpha \hat{a}_\beta = \hat{a}_\beta a_\alpha  - 2 \delta_{\alpha \beta} \hat{a}_\alpha a_\alpha + \delta_{\alpha \beta} I_{\alpha}$
as discussed
below (Equation~(\ref{binary_algebra_1})).
Then the
semantics $\Psi(M)$ is determined by the creation/annihilation operator monomials
\begin{equation}
{\hat{W}}_{r}
\equiv
\hat{W}_{\text{LHS}_r \rightarrow \text{RHS}_r}
\equiv
 k_{r} \left\{ \prod \limits_{\beta \in \operatorname{rhs}( r) } 
({\hat{a}}_{\beta })^{n_{\beta}^{\left(r\right) }} 
\right\} \left\{ \prod \limits_{ \alpha \in \operatorname{lhs}( r) }
(a_{\alpha })^{m_{\alpha}^{\left( r\right) }} 
\right\} \ \ %
\label{XRef-Equation-62212652}
\end{equation}
which 
specifies the non-negative flow of probabability between states
under each reaction $r$ . The states are
given by vectors $\ket{\nb} = \ket{[n_\alpha ]}$ of nonnegative integers $n_\alpha$
(distinguished from stoichiometry $n_{\beta}^{\left( r\right)}$ by its lack of a superscript), 
one for each molecular species.
The full system state is a probability distribution $p(\ket{\nb})$, on which operators act linearly.
Here $ k_{r }$ is conventional notation for a reaction rate; later we will call it $ \rho_{r }$ instead.
The usual chemical law of mass action  
is encoded in the annihilation operators $a_{\alpha}$.
Annihilation operator subscript $\alpha$ indexes the species present in the multiset on the left hand side (LHS) of Equation (\ref{Equation-multiset});
the corresponding operator is raised to the power of its multiplicity or ingoing stoichiometry $m_{\alpha}^{\left( r\right)}$ in reaction $r$,
destroying that many particles of species $\alpha$ if they exist (and contributing zero probability if they don't). 
Likewise creation operator 
${\hat{a}}_{\beta}$ has subscript $\beta$ indexing
the species present in the multiset on the right hand side (RHS) of Equation (\ref{Equation-multiset}),
and the operator is raised to the power of its outgoing stoichiometry $n_{\beta}^{\left( r\right)}$
which indicates how many particles of species $\beta$ are to be created.
The ``$\hat{W}_{\text{LHS}_r \rightarrow \text{RHS}_r}$'' notation is also used by (Behr et al. 2016).
Given Equation~(\ref{XRef-Equation-62212652}),
the actual {\it semantics} is then expressed by $\Psi(M)=W(M)$ where $W(M)$ sums over all reactions $r$ as in 
Equations~(\ref{operator_addition_eq}) and (\ref{CME}) below.

If the number of particles of a given species is any nonnegative integer, 
as we assume for molecular species in solution,
then annihilation and creation operators have infinite dimensional representations
and satisfy the commutator relations of the Heisenberg algebra
$[a_\alpha,\hat{a}_\beta] \equiv {a}_\alpha \hat{a}_\beta - \hat{a}_\beta a_\alpha = \delta_{\alpha \beta} I$, 
where $I=\prod_{\gamma} I_{\gamma}$ is the
product identity operator 
over all relevant spaces including $I_{\alpha} I_{\beta}$,
and $\delta$ is the Kronecker delta.
For each species $\alpha$ we have the matrix representation
in terms of the particle number $n_\alpha \in \{0,1,\ldots\}$ basis:
\begin{equation}
\hat{a}=\left( \begin{array}{ccccc}
 0  & 0  & 0  & 0  & \cdots  \\
 1 & 0  & 0  & 0  &   \\
 0  & 1  & 0  & 0  &   \\
 0  & 0  & 1  & 0  &   \\
 \vdots  & \ \  & \ \  & \ddots  & \ddots
\end{array}\right) = \delta _{n,m+1}\ \ \mathrm{and}\ \ a=\left(
\begin{array}{ccccc}
 0  & 1  & 0  & 0  & \cdots  \\
 0  & 0  & 2  & 0  &   \\
 0  & 0  & 0  & 3  &   \\
 0  & 0  & 0  & 0  & \ddots  \\
 \vdots  &   &   &   & \ddots
\end{array}\right) =m \delta _{n+1,m} ,
\label{integer_algebra_1}
\end{equation}
and
\begin{equation}
\left[ a, \hat{a}\right] \equiv \left( a  \hat{a}- \hat{a}a \right)
=I=\left( \begin{array}{ccccc}
 1  & 0  & 0  & 0  & \cdots  \\
 0 & 1  & 0  & 0  &   \\
 0  & 0  & 1  & 0  &   \\
 0  & 0  & 0  & 1  &   \\
 \vdots  & \ \  & \ \  & \ \  & \ddots
\end{array}\right)  ;  \hat{a}a =N \equiv \left( \begin{array}{ccccc}
 0  & 0  & 0  & 0  & \cdots  \\
 0 & 1  & 0 & 0  &   \\
 0  & 0  & 2 & 0  &   \\
 0  & 0  & 0  & 3  &   \\
 \vdots  & \ \  & \ \  & \ \  & \ddots
\end{array}\right)  .
\label{integer_algebra_2}
\end{equation}
Note: (Morrison and Kinney 2016)  suggest denoting $a_{\alpha}$ as $\check{a}_{\alpha}$, 
in which case $\hat{a}_{\alpha}$ and $\check{a}_{\alpha}$ are mnemonic of increasing
and decreasing particle number $n_{\alpha}$, respectively.

The commutator $[a_\alpha,\hat{a}_\beta]= \delta_{\alpha \beta} I$  
can be used to enforce a ``normal form'' on polynomials in which
annihilation operators precede creation operators in each monomial, 
as in Equation (\ref{XRef-Equation-62212652}).
As above we define the number operator $N_\alpha=\hat{a}_\alpha a_\alpha$,
diagonal in the ``number basis'' comprising state vectors $\ket{\nb}$.

A different
case is that in which the number of ``particles'' of any given species
must be in $\{0,1\}$, as for example if the species are 
individual binding sites occupied by a particular molecular species;
then the operators are $2 \times 2$ matrices 
$a = \bigl( \begin{smallmatrix}0 & 1\\ 0 & 0\end{smallmatrix}\bigr)$ and
$\hat{a} = \bigl( \begin{smallmatrix}0 & 0\\ 1 & 0\end{smallmatrix}\bigr)$
obeying
$[a_\alpha,\hat{a}_\beta] \equiv {a}_\alpha \hat{a}_\beta - \hat{a}_\beta a_\alpha = \delta_{\alpha \beta}(I_\alpha- 2 \hat{a}_\alpha a_\alpha)$.
Explicitly,
\begin{subequations}
\begin{align}
\hat{a} &=\left( \begin{array}{cc}
 0 & 0 \\
 1 & 0
\end{array}\right)  , a=\left( \begin{array}{cc}
 0 & 1 \\
 0 & 0
\end{array}\right)
 \; \text{implies} \;
\\
\hat{a}a &=N\equiv \left( \begin{array}{cc}
 0  & 0  \\
 0 & 1
\end{array}\right) \; , \quad 
a \hat{a} = Z \equiv I - N 
=
\left( \begin{array}{cc}
1  & 0  \\
 0 & 0
\end{array}\right)\; , \; \text{and} \\ 
[a_\alpha,\hat{a}_\beta] &= \delta_{\alpha \beta}(I_\alpha- 2 N_\alpha ) I
\quad \quad {\text{Alternative for normal form calcs:}} \\
a_\alpha \hat{a}_\beta &= \hat{a}_\beta a_\alpha  - 2 \delta_{\alpha \beta} \hat{a}_\alpha a_\alpha + \delta_{\alpha \beta} I_{\alpha} \\
& = (1- \delta_{\alpha \beta}) \hat{a}_\beta a_\alpha  + \delta_{\alpha \beta} Z_{\alpha}
\end{align}
\label{binary_algebra_1}
\end{subequations}
In this case for each particle species or object type $\alpha$ we can define the diagonal number operator
$N_\alpha=\hat{a}_\alpha a_\alpha= \bigl( \begin{smallmatrix}0 & 0\\ 0 & 1\end{smallmatrix}\bigr)$, 
the zero-checking operator
$Z_\alpha = I_\alpha- N_\alpha= \bigl( \begin{smallmatrix}1 & 0\\ 0 & 0\end{smallmatrix}\bigr)$,
and the ``erasure'' projection operator
$E_\alpha = (Z_\alpha+a_\alpha) = \bigl( \begin{smallmatrix}1 & 1\\ 0 & 0\end{smallmatrix}\bigr)$
which takes either state to the zero-particle state.
In this case also $a^2 = \hat{a}^2 = 0$.

The model semantics built on Equation (\ref{XRef-Equation-62212652})
is {\it compositional} 
over processes, hence structure-preserving, because:
\begin{itemize}
\item[(a)]{
The operators for multiple rules
indexed by $r$ in a ruleset map to an {\it operator sum}:
\begin{subequations}
\begin{align}
W &= \sum_r W_r \; , \quad \text{(rule operators sum up), where} \\ \quad \quad 
W_r & \equiv \hat{W}_r - D_r , 
\quad \text {(rules conserve probability) } \\
D_r  & \equiv  {\rm diag}({\bf 1} \cdot \hat{W}_r)   \quad \text {(total probability outflow per state) } 
\end{align}
\label{operator_addition_eq}
\end{subequations}
that specifies the combined dynamics under the chemical master equation:
\begin{equation}
\quad \dot{p}  = W \cdot p.
\label{CME}
\end{equation}
In Equation (\ref{operator_addition_eq}) 
the first statement is ruleset compositionality, the second and third ensure conservation of probability.
Equation~(\ref{CME}) is
the resulting Chemical Master Equation (CME) stochastic dynamical system
for the evolving state probability $p(\nb)$. 
Also the semantics is compositional because:} 
\item[(b)]{
The multisets on the left hand side and right hand side of a rule each map to an {\it operator product} in normal form
(incuding powers for repeated multiset elements) in Equation (\ref{XRef-Equation-62212652}).
Each product consists of commuting operators so their order is 
arbitrary.
}
\end{itemize}

Under Equation (\ref{XRef-Equation-62212652}) one may calculate that
${\rm diag}({\bf 1} \cdot \hat{W}_r)  $ equals the diagonal monomial operator 
\begin{equation}
\begin{split}
D_r &= 
\prod \limits_{\alpha \in \operatorname{lhs}( r)} {N_{\alpha}}^{m_{\alpha}}
=
\prod \limits_{\alpha \in \operatorname{lhs}( r) }
(\hat{a}_{\alpha })^{m_{\alpha}^{\left( r\right) }}
(a_{\alpha })^{m_{\alpha}^{\left( r\right) }} 
\equiv N^{(r)}
= \hat{W}_{\text{LHS}_r \rightarrow \text{LHS}_r},
\end{split}
\label{diagonal_operator}
\end{equation}
where ``$N^{(r)}$'' is a number operator for the entire left hand side of the rule, and
``$\text{LHS}_r$'' is the left hand side of reaction rule $r$;
here in the ``$\hat{W}_{\text{LHS}_r \rightarrow \text{LHS}_r}$'' notation of (Behr et al. 2016)
 the LHS appears on {\it both} sides of the arrow.
$D_r$ represents the total {\it probability outflow} from each state 
under rule $r$ and is, like $\hat{W}_r$, nonnegative in the number basis.
One may regard the linearity of Equation (\ref{operator_addition_eq})  
as a linear mapping of vector spaces
(hence as a morphism or category arrow): The source vector space is spanned
by basis vectors that correspond to ordered pairs of multisets of species symbols,
weighted by 
scalar reaction rates $k_r$,
and the target vector space 
is
a vastly larger space of possible probability-conserving operators.

The default continuous-time semantics 
$W = \Psi( M ) $
is now be defined
by Equation~(\ref{XRef-Equation-62212652}) (the specific semantic map for each rule of a model $M$)
and Equation~(\ref{operator_addition_eq}), in the context of Equation~(\ref{CME}).
Thus we have defined a ``structure respecting'' mapping $\Psi$ from pure (multiset-changing) rulesets
(each rules weighted by a nonnegative reaction rate)
to operator algebras; $\Psi$ is (at least) a linear mapping of vector spaces.

The integer-valued index $r$ we used to name the reactions is part of a meta-language
for the present theorizing, and not part of the modeling language.
One slightly confusing point is that these unordered collections of chemical reactions
are expressions in the language, but they also have a form reminiscent of
a grammar for {\it another language}
- albeit a language of multisets representing the system state,
rather than of strings or trees,
and a language that may be entirely devoid of terminal symbols
(which would represent inert products such as waste).
The CME as semantics was suggested by
(Mjolsness and Yosiphon 2006, Mjolsness 2005) and by (Cardelli 2008),
though it can be regarded as implicit in the original Doi-Peliti formalism.
There is also a projection from continuous-time (CME) semantics 
to discrete-time probabilistic semantics
in the form of a Markov chain (Mjolsness and Yosiphon 2006).

As a consequence of this semantics, two models are {\it particle-equivalent} just in
case they have the same CME solution, i.e. the same joint distribution over all collections of
particle numbers $n_{\alpha}$ observable at the same or different times $t$.
For countable collections (indexed by integer $q$) these quantities take the equal forms
\begin{equation}
\text{Pr}_{\text{\tiny CME}}([n_{\alpha(q)}(t_q)|q]) = \Big \langle \prod_q \delta (N_{\alpha(q)}(t_q)-n_{\alpha(q)} I_{\alpha(q)}) \Big \rangle_{\text{\tiny CME}}
\label{particle-equivalence}
\end{equation}
for any choice of particle numbers $\alpha(q)$ and observation times $t_q$,
where: the Kronecker delta is applied componentwise;
 the ``CME'' subscript refers to the solution of the Chemical Master Equation, Equation (\ref{CME}) above;
 and for any diagonal operator $D$ we have at a single time $t$ that
  $\langle D(t) \rangle_{\text{\tiny CME}} = {\mathbf 1} \cdot D(t) \cdot p_{\text{\tiny CME}} $; 
  unequal times require the joint distribution $\text{Pr}_{\text{\tiny CME}}$ at all the relevant times.
The $\langle \ldots \rangle$ right hand side expression isn't necessary here but will be useful in a future section.
All other observables 
$\langle f([N_{\alpha(q)}(t_q) | q ] ) \rangle_{\text{\tiny CME}} $ 
(where $f$ is applied componentwise to diagonal matrices)
follow from this linear basis, Equation~(\ref{particle-equivalence}).

From this operator algebra semantics 
and the Time-Ordered Product Expansion (TOPE) approach to
Feynman path integrals (Mjolsness and Yosiphon 2006, Mjolsness 2005)
one can derive valid exact stochastic simulation algorithms
including the Gillespie stochastic simulation algorithm (SSA)
and various generalizations,
as detailed in (Mjolsness 2013).
Such algorithms can also be accelerated exactly (Mjolsness et al. 2009),
and accelerated further by working hierarchically at multiple scales
and/or using parallel computing (Orendorff and Mjolsness 2012).
The same theory can be used to derive machine learning algorithms for the inference of
reaction rate parameters from sufficient data
(Wang et al. 2010) (cf. Golightly and Wilkinson (2011)), 
although sufficient data may be hard to obtain.
One can also develop approximate sampling algorithms by operator splitting,
justified e.g. by the Baker-Campbell-Hausdorff theorem (BCH),
and/or by moment closure methods such as those discussed in
Section~\ref{sec_mod_red} below.

A classic solvable example 
(McQuarrie 1967)
is the well-mixed chemical reaction network
$A \rightleftharpoons B$, with forward and reverse rates $k_f$ and $k_r$ respectively.
The time-evolution operator for this system is defined on the space ${\mathbb N}_A \otimes {\mathbb N}_B$ by
\[
W_{A \rightleftharpoons B} = k_f( \hat{a}_B a_A - N_B) +  k_r( \hat{a}_A a_B - N_A).
\]
This operator is a sum of four summed monomial terms: two for the forward reaction and two for the reverse.
The first term destroys a particle of type A and immediately creates a replacement particle of type B.
The second term removes the same amount of probability per unit time from the current state.
Likewise for terms three and four.
Reversibly unimolecular systems like this, with one molecule in and one molecule out of each reaction, 
can be solved analytically
by treating each molecule in the system as an independent one-particle system.
Alternatively one can use generating functions to solve most or all of the solvable small networks.

A systematic operator-algebraic solution of this example  proceeds by
(a) representing probability distributions with generating functions and mapping creation and annihilation
operators to multiplication by variables $z_{\alpha}$ and derivative operators $\partial_\alpha$ (for $\alpha \in \{A, B\}$) respectively,
obtaining a PDE;
(b) separating out the time variable by seeking 
solutions proportional to $\exp{\lambda t}$; 
(c) using conservation laws and initial conditions 
(here $n_A + n_B = \text{constant}$) 
to reduce from two integer state variables and two generating function variables to one, e.g. $\zeta = z_A/z_B$;  
(d) analytically solving the resulting differential equation 
(sometimes only possible for the steady state, $\lambda=0$); 
(e) 
impose the initial condition $g(t=0) = \prod_{\alpha} z_{\alpha}^{n_{\alpha}(t=0)}$ 
obtaining in our particular case a convolution of two binomial distributions that converge to one equilibrium binomial distribution.
For systems that are not analytically solvable, such as $A+B \rightleftharpoons C$,
there may be
an analytically tractable approximation for dynamics not only in the limit but also in the 
approach to the limit of large numbers of molecules, 
for example by approximating the most significant eigenvalues and eigenvectors of $W$
using boundary layer theory (Mjolsness and  Prasad 2013).

Associated to the continuous-time semantics $\Psi(W)$ is a 
{\it discrete-time semantics} $\Psi _{d}( W ) $ (Mjolsness and Yosiphon 2006):
\begin{equation}
\Pr (  k) =U\circ ...U\circ \Pr (  0) \equiv  U^{k}\circ \Pr (0) ,
\ \ \mathrm{for}\ \ k\in \mathbb{N} 
\label{XRef-Equation-218211639}
\end{equation}
where $U$ is related to $W$ below.
In the present treatment we will simplify matters by assuming there are no terminal states,
so the diagonal matrix $D$ can be inverted by inverting its elements.
We can define the first rule-firing update
at step $k:{\mathbb N}$ in a Markov chain as follows:
\begin{equation}
p_{k+1}= \tilde{W}\cdot {{D}}^{-1 }\cdot p_{k} 
\label{XRef-Equation-82711334}
\end{equation}
where in this simplified case $ \tilde{W}$ is just equal to $\hat{W}$.
This update is now in a form that can be iterated as a linear map
$U = \tilde{W}\cdot {{D}}^{-1 } $, and
hence can be iterated and interpreted as a stochastic algorithm
as in Equation~(\ref{XRef-Equation-218211639}).
The full treatment including terminal states, 
and the projection meta-operator
from continuous-time semantics to discrete-time semantics  $\Psi _{d}( W ) $, is given in
(Mjolsness and Yosiphon 2006).

Thus, unordered collections of pure chemical reactions provide a simple example 
of a modeling language with a compositional semantics.
But for most biological modeling, we need much more expressive power than this.

\subsubsection{Parameterized Reaction Rules}
\label{sec_param_rules}

The first modeling language escalation
beyond pure chemical reaction notation
is to particle-like objects or ``agents'' that bear numerical and/or 
discrete parameters which affect their reaction reaction rates.
For example, the size of a cell may affect its chances of undergoing cell division.
This kind of multiset rewrite rule can be generalized to (Mjolsness 2010)
\begin{equation}
{\left\{ \tau_{\alpha( p) }[ x_{p}] |p \in L_{r}\right\} }_{*}
\longrightarrow {\left\{ \tau_{\beta( q) }[ y_{q}] | q \in
R_{r}\right\}  }_{*} \quad \text{\boldmath $\mathbf{with}$}\text{\boldmath
$\ \ \ $} \rho _{r}( \left[ x_{p}\right] ,\left[ y_{q}\right] ) .
\label{XRef-Equation-924145912}
\end{equation}
Here we have 
switched from molecule-like term names $A_\alpha$
to more generic logic-like term names $\tau_{\alpha( p)}$.
The parameters $x_{p},  y_{q}$
of each term
(indexed by positions $p,q$  in their respective argument lists, 
and which may themselves be vectors ${\xb}_{p}$)
introduce a new aspect of the language,
analogous to the difference between predicate calculus and first order logic:
Each parameter can appear as a constant or as a variable, and the same variable $X_c$
can be repeated in several components of several parameter lists in a single rule.
Thus it is impossible in general to say whether two parameters in a rule are equal or not,
and thus whether two terms $ \tau _{\alpha( p) }[ x_{p}] $ in a rule are the same or not,
just from looking at the rule - that fact may depend on the values of the variables,
known only at simulation time.
The $p,q$ subindex notation is as in  (Mjolsness 2013).
The reaction rate $\rho _{r}( \left[ x_{p}\right] ,\left[ y_{q}\right] )$
now depends on parameters on one or both sides of the rewrite rule
which can be factored (automatically in a declarative environment, 
as in 
(Yosiphon 2009))
into a rate depending on the LHS parameters only and a conditional
distribution of RHS 
parameters 
given LHS parameters.

\subsubsection{Examples: Cell Division and Dynamic Cytoskeleton}
\label{subsec_examples}

Both multicellular tissue and intracellular cytoskeleton topologies change, discontinuously of course,
in ways that could be modeled with parameterized reaction rules in a flexible declarative language.
For example a stem cell of volume $V$ might divide asymmetrically yielding
a stem cell and a transit-amplifying cell in for example
mouse olfactory epithelium (Yosiphon 2009):
\begin{equation}
\begin{split}
\\\operatorname{stemcell}[ \text{${\boldsymbol x}$},V,\ldots] 
\longrightarrow &
\operatorname{TAcell}[ \text{\boldmath ${\boldmath x}$}+\Delta  \text{\boldmath
$x$},V/2,\ldots ] ,\operatorname{stemcell}[ \text{\boldmath ${\boldmath x}$}-\Delta
\text{\boldmath ${\boldmath x}$},V/2,\ldots ] 
\\  & \text{\boldmath $\mathbf{with}$}\ \ \hat{\rho }( V) 
\mathcal{N}( \Delta  \text{${ \boldsymbol x}$};c V^{1/d}) ,
\end{split}
\end{equation}
where $\hat{\rho }( V)$ is a probability per unit time or ``propensity''
for cell division depending on cell volume, and
$\mathcal{N}( \Delta  \text{\boldmath $\mathbf{x}$};c V^{1/d})$ is
a Gaussian or normal probability density function
with diagonal covariance proportional to a lengthscale set by cell volume.
It is up to the modeler to impose appropriate invariances in such a model.
In this case, Gallilean invariance is ensured by fact that the propensity
function depends on position only through $\Delta \text{\boldmath ${\boldmath x}$}$, 
a difference of cell position vectors. Rotational symmetry could be broken
by the prominent apical-basal axis in such a $d=2$ model of a pseudostratified epithelium.

Another stem cell application was to models of 
plant root growth and pattern formation regulated by the auxin growth hormone
(Yosiphon 2009, Mironova et al. 2012, Mjolsness 2013),
implemented
in a computer algebra system (Yosiphon 2009, Shapiro et al. 2013);
cf. (Julien et. al 2019) 
in this issue
for auxin-based plant shoot patterning.
These examples show that with parameters, reaction-like rewrite rules
can represent (for example) both cellular and molecular processes in the same model,
and of course their interactions,
which is a key expressiveness capability for multiscale modeling.

To further explain the 1D root growth model cited above, 
the cell division rule (one out of about a dozen in the model ``grammar'') 
was augmented to preserve 1D topology in a manner similar to the following:
\begin{equation}
\begin{array}{rl}
   & \begin{array}{l}
{\text {/* cell replication, explicitly preserving 1D structure: */ }} \\
  \operatorname{cell}[ \mathrm{curr},2,x,r,A,Y,\mathrm{prev},\mathrm{next}]
,\operatorname{cell}[ \mathrm{prev},{\mathrm{mode}}^{\prime },x^{\prime
},r^{\prime },A^{\prime },Y^{\prime },\mathrm{prevprev},\mathrm{curr}]
\\
 \ \ \ \ \operatorname{cell}[ \mathrm{next},\mathrm{mode},x^{{\prime\prime}},r^{{\prime\prime}},A^{{\prime\prime}},Y^{{\prime\prime}},\mathrm{curr},\mathrm{nextnext}]
 \\
  \longrightarrow \operatorname{cell}[ {\mathrm{new}}_{1},1,x-r\alpha , r( 1-\alpha ) ,A,Y,\mathrm{prev},{\mathrm{new}}_{2}],
 \\
 \ \ \ \  \operatorname{cell}[ {\mathrm{new}}_{2},1,x+r(1- \alpha) ,r \alpha, A,Y,{\mathrm{new}}_{1},\mathrm{next}] ,

\\
\ \ \ \  \operatorname{cell}[ \mathrm{prev},{\mathrm{mode}}^{\prime },x^{\prime
},r^{\prime },A^{\prime },Y^{\prime },\mathrm{prevprev},{\mathrm{new}}_{1}]
,
 \\
\ \ \ \  \operatorname{cell}[ \mathrm{next},\mathrm{mode},x^{{\prime\prime}},r^{{\prime\prime}},A^{{\prime\prime}},Y^{{\prime\prime}},{\mathrm{new}}_{2},\mathrm{nextnext}]
\end{array} \\
  &  \text{\boldmath $\mathbf{with}$}  \ \ \ 
\rho_Y(Y/Y_{0}) p_\alpha(\alpha) \quad {\rm  /* where } \; p_\alpha(\alpha) \; \text{\rm enforces } 
    \;  \alpha \in [ \frac{1}{2} - \Delta , \frac{1}{2}+\Delta] \quad {\rm */}
\end{array}
\label{root_subgrammar}
\end{equation}
Here $p_\alpha(\alpha)$ denotes a uniform distribution of  $\alpha$ in 
its allowed interval,
governing the relative sizes (lengths) of daughter cells compared to the parent cell.
Also $A$ (the plant growth hormone auxin) and the hypothetical molecule $Y$ are two dynamical morphogens;
``curr'', ``next'', ``prev'', etc. are unique (e.g. integer-valued) object identifiers. 
Cell positions are denoted by $x, x^{\prime}, \ldots $,
and $r, r^{\prime}, \ldots $ 
denote 1D cell radius i.e. half of cell length.
Variable ``mode'' is a discrete cell state
determining readiness for cell proliferation vs. vegetative growth;
it is determined stochastically in another rule by cell size $r$ compared to a threshold.
The discrete ``mode'' parameter could be obviated by the use of subtyping,
in which some rules such as biomechanics act on all  ``cell'' objects and other only on
``vegetative\_cell::cell'' or on ``proliferative\_cell::cell'' subtypes which under
the Liskov substitution principle of programming languages
would each be subject to the generic cell rules as well.
Random variable $\alpha \in [ \frac{1}{2} - \Delta , \frac{1}{2}+\Delta]  \subseteq  [0,1]$ 
denotes the fraction of parent cell size
inherited by one of the two daughter cells; the other gets $1-\alpha$.
Constant parameters include $Y_0$, a baseline level of $Y$, and
$\Delta$, the allowed variation of $\alpha$ away from 
$\alpha = 1/2$
which would represent spatially symmetric cell division.

There are many other cell lineage tree systems in biology in which
less committed cell types give rise to more committed cell types
during development by symmetric or and/or asymmetric cell division.
Examples include
vertebrate hematopoiesis,
and the generation of neuronal diversity in animal brain development (Holguera et al. 2018).
For example early vertebrate neural development could be modeled 
by cell division/specialization rules of the general form:
\begin{equation}
\begin{array}{rl}
\text{\small apical}\_\text{\small radial}\_\text{\small glia}({ \boldsymbol g}_1) 
	& \rightarrow \text{\small intermediate}\_\text{\small basal}\_\text{\small progenitor}({ \boldsymbol g}_2), \\
		& \quad \; \text{\small apical}\_\text{\small radial}\_\text{\small glia}({ \boldsymbol g}_3) 
			\;  \text{\bf with } \rho_{\text{aia}}( {\boldsymbol g}_1, {\boldsymbol g}_2, {\boldsymbol g}_3) \\
\text{\small intermediate}\_\text{\small basal}\_\text{\small progenitor}({ \boldsymbol g}_1) & \rightarrow   \text{\small neuron}({ \boldsymbol g}_2),  \;\; \text{\small neuron}({ \boldsymbol g}_3) 
	\;  \text{\bf with } \rho_{\text{inn}}( {\boldsymbol g}_1, {\boldsymbol g}_2, {\boldsymbol g}_3) \\ 
\text{\small apical}\_\text{\small radial}\_\text{\small glia}({ \boldsymbol g}_1)
	 & \rightarrow  \text{\small outer}\_\text{\small radial}\_\text{\small glia}({ \boldsymbol g}_2), 
	 	\; \text{\small outer}\_\text{\small radial}\_\text{\small glia}({ \boldsymbol g}_3) \\
	& \quad \quad
		 \text{\bf with } \rho_{\text{aoo}}( {\boldsymbol g}_1, {\boldsymbol g}_2, {\boldsymbol g}_3) \\
\text{\small outer}\_\text{\small radial}\_\text{\small glia}({ \boldsymbol g}_1)
	 & \rightarrow  \text{\small outer}\_\text{\small radial}\_\text{\small glia}({ \boldsymbol g}_2), 
	 	\; \text{\small neuron}({ \boldsymbol g}_3) \\
	& \quad \quad
		\;  \text{\bf with } \rho_{\text{oon}}( {\boldsymbol g}_1, {\boldsymbol g}_2, {\boldsymbol g}_3) \\
\ldots
\end{array}
\label{brain_lineage}
\end{equation}
in which the relative rates of conflicting rules 
(e.g. the first and third rules above)
are determined by
transcriptional regulation 
(gene expression level vectors ${ \boldsymbol g}_i$)
which can change upon cell division rule firings
(as formalized for example in (Mjolsness Sharp Reinitz 1991))
in accordance with some propensity functions $\rho_*({\boldsymbol g}_1, {\boldsymbol g}_2, {\boldsymbol g}_3)$,
unless of course one chooses to model 
$\rho_*({\boldsymbol g}_1, {\boldsymbol g}_2, {\boldsymbol g}_3) 
= \rho_*({\boldsymbol g}_1) \delta({\boldsymbol g}_2- {\boldsymbol g}_1)  \delta({\boldsymbol g}_3- {\boldsymbol g}_1)$
so that gene expression doesn't change upon cell division.
Further continuous-time rules would allow all cell expression vectors to evolve under
transcriptional regulation between cell division events.

The grammar of Equation~(\ref{brain_lineage}) happens to have just one object
on the LHS of each rule, so it is ``context free'' and amenable to analysis.
Generally that's not the case even for cell division grammars like
Equation~(\ref{root_subgrammar}); 
even less so for biological many-to-one transitions such as
the fusion of mitochondria or of muscle cell, or the merging
of microtubule fibers in 
cytoskeleton discussed below.

\begin{figure}
\begin{center}
  \includegraphics[width=0.9\columnwidth]{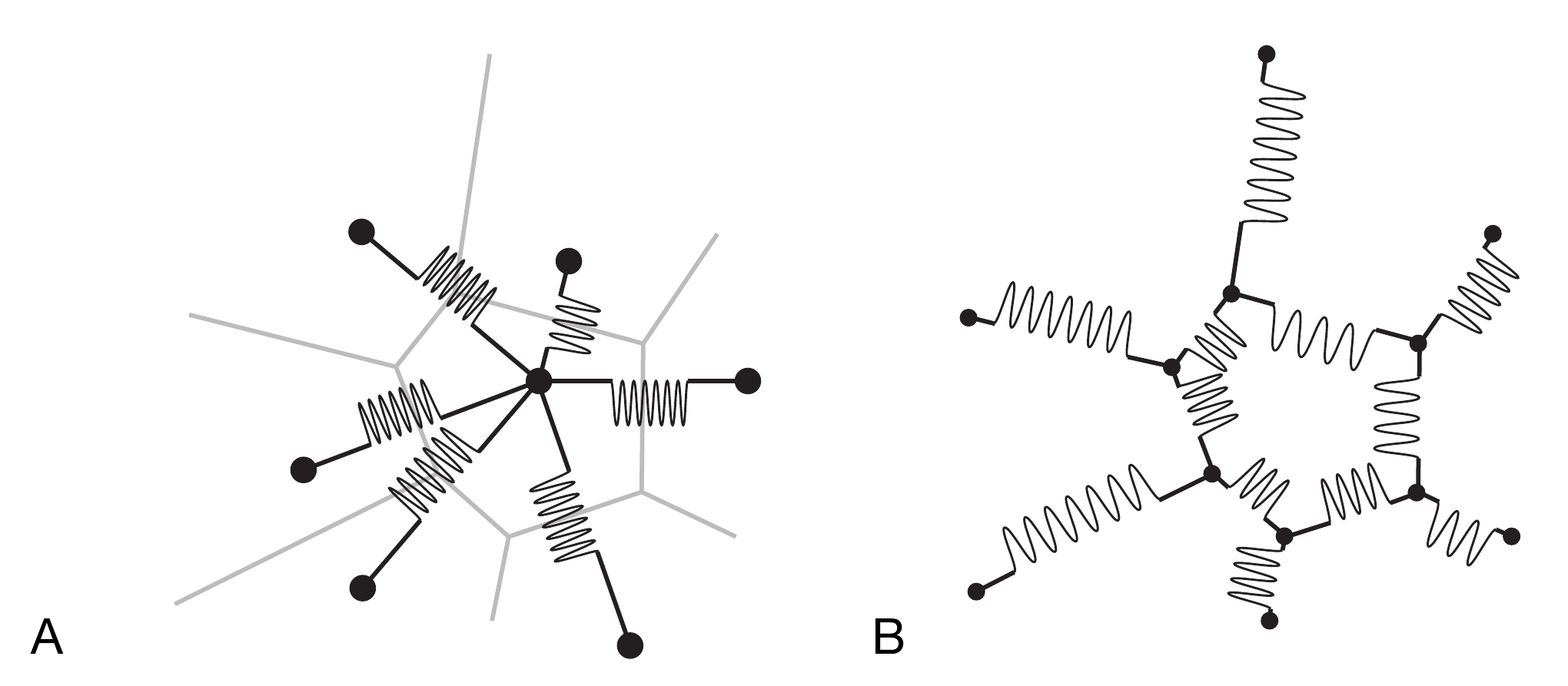}
\end{center}
\caption{ 
(A) Cell-cell spring biomechanical model.
Potential energy $V$ for each spring
can be something like e.g. a Lennard-Jones potential
that is repulsive at short range, attractive at intermediate range,
and flat (implies forceless) at long range;
another such form is a breakable spring with
$V({\vec x}_p , {\vec x}_q)= (1/2)\sum_{q \in \text{Nbrs}(p)} k_{pq} c_{pq} [(|{\vec x}_p - {\vec x}_q| - l_{pq}|^2 - {\Delta l_{pq}}^2]$,
where $c_{pq} \in\{0,1\}$ is chosen to minimize $V$ at large separations
($c_{pq} = 0$ corresponds to a broken spring with
continuity of potential energy at the breaking point).
Parameters are spring constant (strength) $k_{pq}$, resting length $l_{pq}$, and breaking stretch distance ${\Delta l_{pq}}$.
(B) Cell wall spring biomechanical model. Springs again have nonzero resting lengths, but cannot break.
Cf. more detailed polygonal tesselation models in (Wolff et al. 2019). 
Reproduced from (Shapiro et al. 2013); courtesy Bruce Shapiro. 
 }
\label{springs}       
\end{figure}

A related but more detailed approach to 2D and 3D cell division modeling was taken in 
the shoot apical meristem (SAM) dynamical patterning model of (J\"{o}nsson et al. 2006),
in which cell positions and radii were again (as in 1D) the dynamical variables,
determined by the mechanics of breakable springs in viscous media, and cell-cell interface areas
were determined by the chordal intersections of corresponding cellular regions (Figure~\ref{springs}A).
The ``Organism'' C++ code in which this model was implemented (J\"{o}nsson et al. 2018) is not fully declarative,
but is flexible enough to serve as the back end for a declarative model
by translation of input files and to output similar files.
\footnote{A simpler published plant model has been translated from declarative form 
to Organism input; see http://computableplant.ics.uci.edu/sw/CambiumOrganism/ }
It supports bidirectional coupling of regulatory networks 
such as regulated active auxin transport and/or
gene regulation network models of SAM
morphogenetic patterning
(e.g. (J\"{o}nsson et al. 2005); see also (Banwarth-Kuhn et al. 2018) in this issue)
to the biomechanics.
A similar mechanical/regulatory model with breakable springs
for the stem cell niche of mouse olfactory receptor neurons
was implemented in the Plenum prototype implementation of the Dynamical Grammars
declarative modeling language (Yosiphon 2009).
A similar cell-centered model was used recently
to model neural crest cell group migration,
with stronger springs at the rear of each cell group
representing multicellular cytoskeletal structures there.
(Shellard et al 2018).

\begin{figure}
\begin{center}
  \includegraphics[width=0.9\columnwidth]{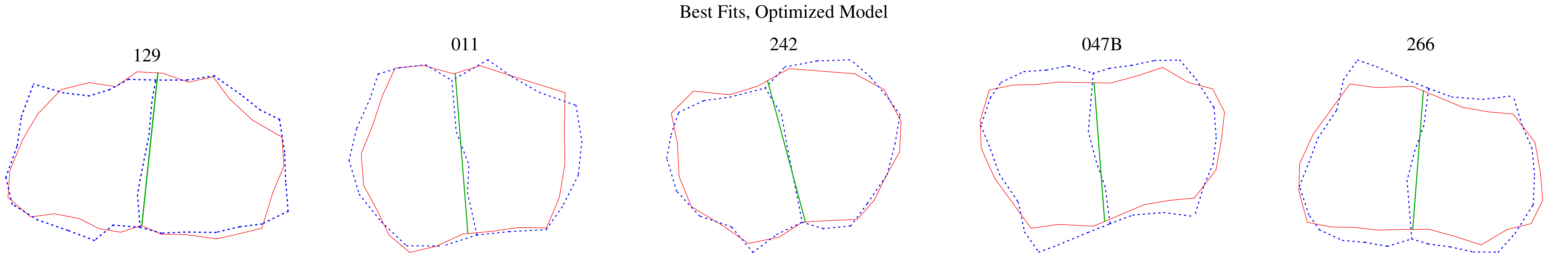}
\end{center}
\begin{center}
  \includegraphics[width=0.9\columnwidth]{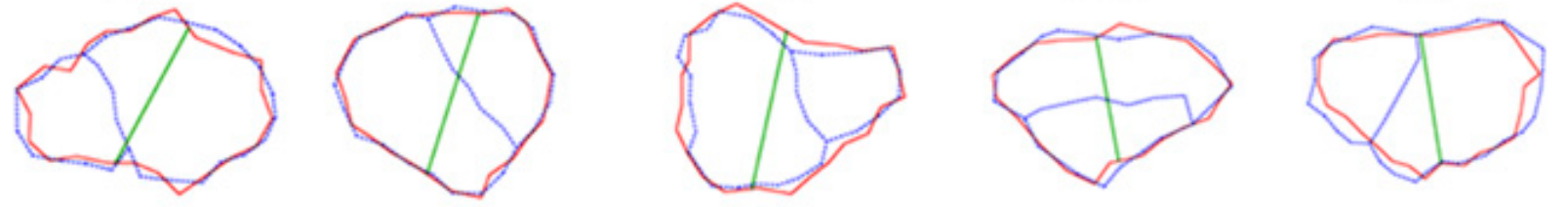}
\end{center}
\caption{
(A) Top row: Best fits within a data set of cell images, after model optimization.
(B) Bottom row: Corresponding worst fits within the same data set of cell images. 
Reprinted from (Shapiro et al. 2015a).
 }
\label{fig:variation}       
\end{figure}

\begin{figure}
\begin{center}
  \includegraphics[width=0.9\columnwidth]{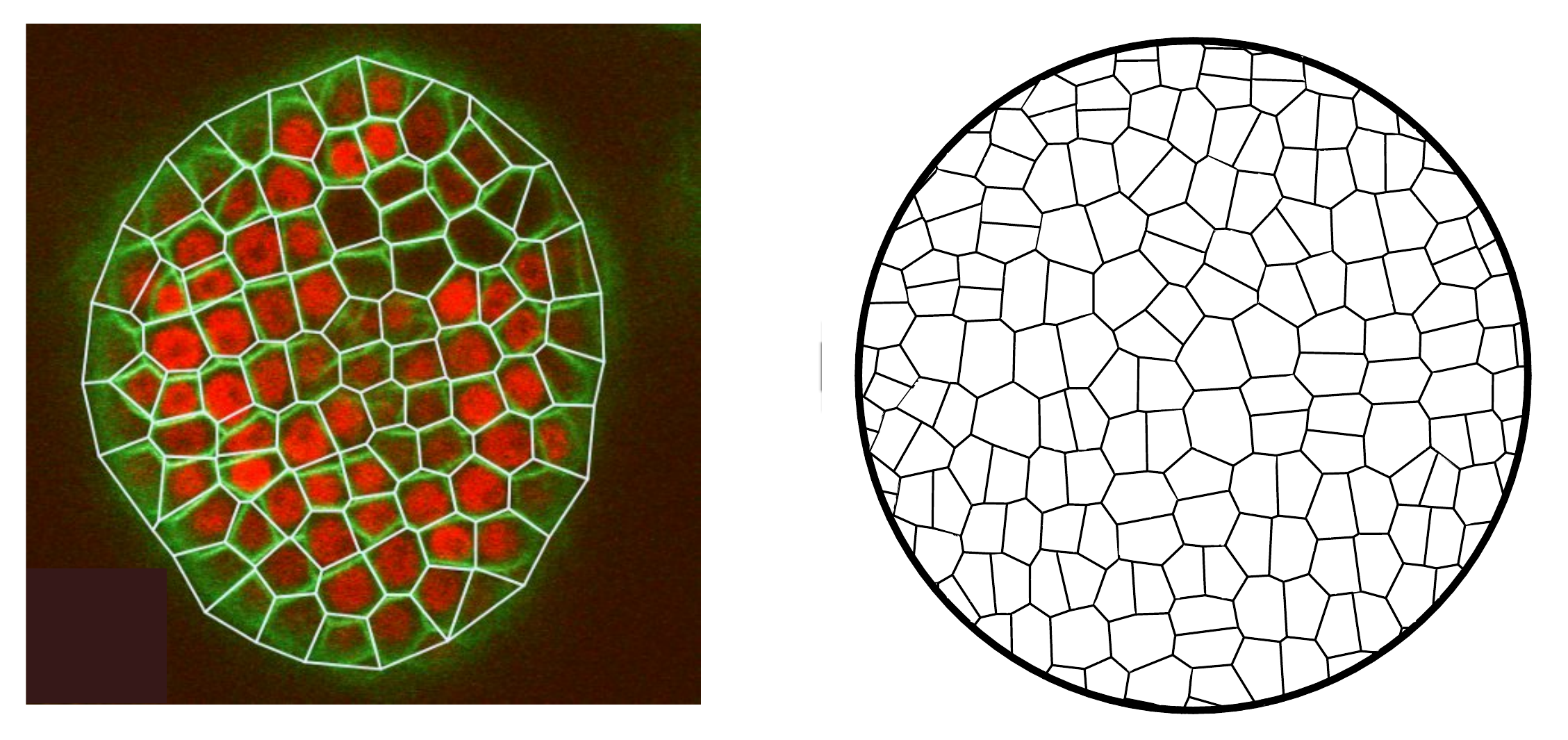}
\end{center}
\caption{ 
(A) Trimmed Voronoi diagram closely matches cell walls in 
{\it Arabidopsis thaliana} shoot apical meristem outer L1 cell layer 2D geometry
derived from confocal  laser scanning microscopy.
Green: cell wall marker. Yellow: Voronoi diagram edges. Red: nuclear marker.
Reproduced from (Shapiro et al. 2012); courtesy Bruce Shapiro and Marcus Heisler. 
(B) Growing SAM geometry produced by Cellzilla declarative model,
using optimized cell division rule as described in the text.
Reproduced from (Shapiro et al. 2013).
 }
\label{fig:tesselate}       
\end{figure}

For a two-dimensional model, one would like more than a single random variable $\alpha$ to
describe the selected geometry of a particular cell division. A variety of cell-scale phenomenological
``rules'' have been explored for this geometrical behavior in plant cells,
including rules proposed historically by Hofmeister, Errera, and Sachs.
A parameterized family of probabilistic rules based on the Boltzmann distribution
$p_{\theta}({\vec \theta}|{\vec w}) = \exp{[- \beta E({\vec \theta}|{\vec w})]}/Z(\beta)$
for phenomenological energy functions 
$E({\vec \theta}|{\vec w})=\sum_{i \in {\text{area},\text{length},\text{extension},\text{growth}}}w_i V_i({\vec \theta})$,
 encompassing modern interpretations 
of these historical rules as points ${\vec w}$ in a larger parameter space, 
was explored in (Shapiro et al. 2015a). 
Any such a Boltzmann distribution could easily be
placed in a cell division rule comparable to the foregoing 1D rule,
in place of the factor $ p_\alpha(\alpha)$.
The form chosen had several parameters learned from relevant
microscopy data for the shoot apical meristem (SAM)
(the opposite end of the plant from root apical meristem)
of the genetic model plant {\it Arabidopsis thaliana};
the optimal rule was closest to but somewhat better than 
the standard modern interpretation of Errera's rule.
Some realistic stochastic variation was captured (see Figure~\ref{fig:variation}), 
as indicated by a relatively small but nonzero 
optimal temperature parameter in the Boltzmann distribution.
Both rules were implemented in the declarative 2D
cellular tissue modeling package ``Cellzilla'' (Shapiro 2013),
and created growing convex polygonal patterns that
appear qualitatively similar to those of derived from
microscope imagery of SAM tissue
as shown in Figure~\ref{fig:tesselate}.

\begin{figure}
\begin{center}
  \includegraphics[width=0.7\columnwidth]{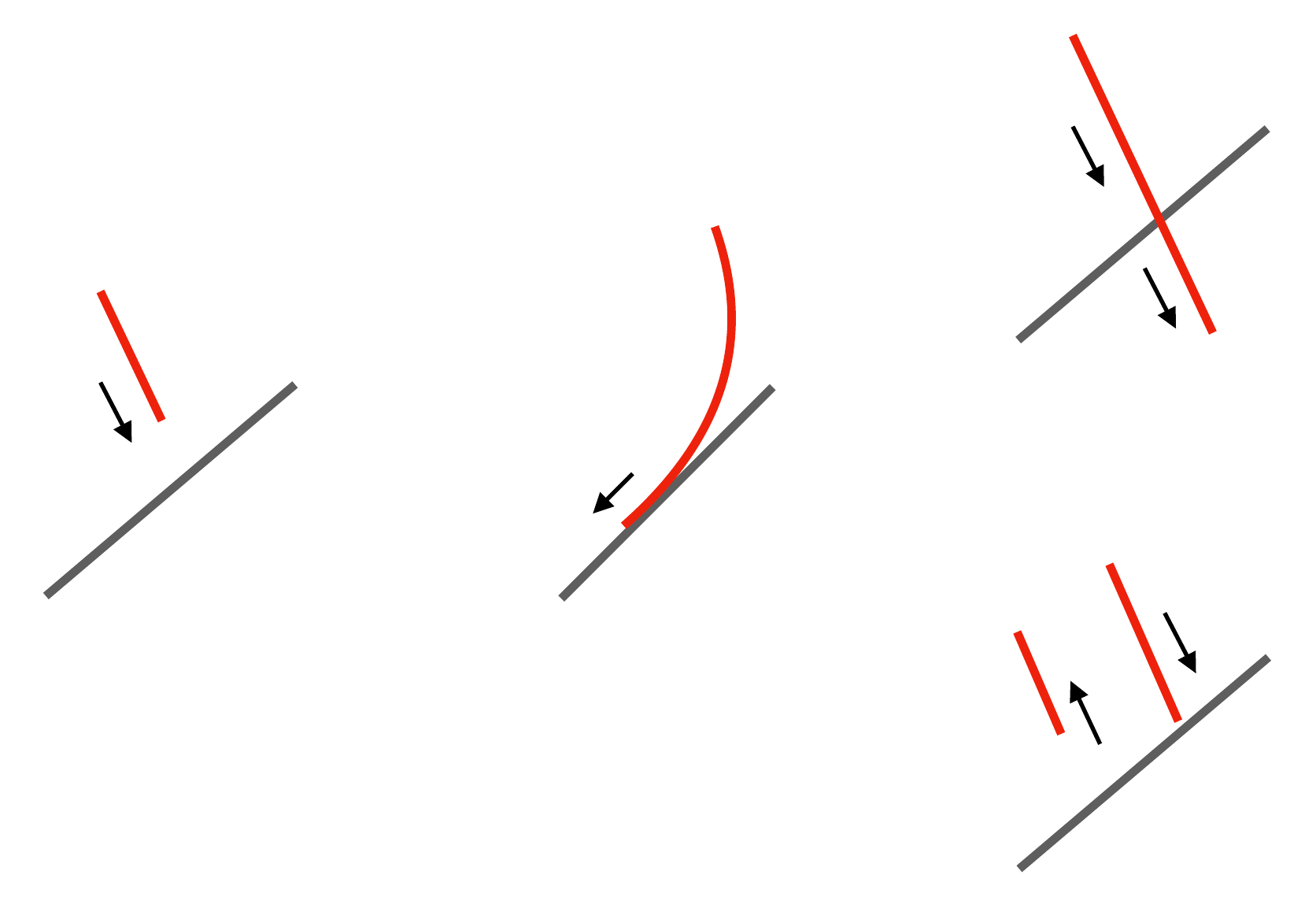}
\end{center}
\caption{
Microtubule (MT) discrete transition rules associated with the collision of
two MTs confined to a nearly two dimensional environment such 
as the cortex of a cell just inside the cell membrane.
Left: cortical MT (red) undergoing ``+ end'' growth (thin arrow) approaching the
static middle section of another cortical MT or MT fiber (gray).
Center: At low angles of incidence (Wightman and Turner 2007),
 the approaching MT may preferentially zipper or bundle into a fiber.
Right: At higher angles of incidence, the approaching MT can cross over the static one
(top right) possibly forming a junction, or (bottom right) it may undergo a state
change to catastrophic depolymerization at the + end. 
The choice may be modeled as stochastic.
Redrawn and modified from  (Chakrabortty et al., 2018).
}
\label{fig:MT_rules}       
\end{figure}

\begin{figure}
\begin{center}
  \includegraphics[width=1.0\columnwidth]{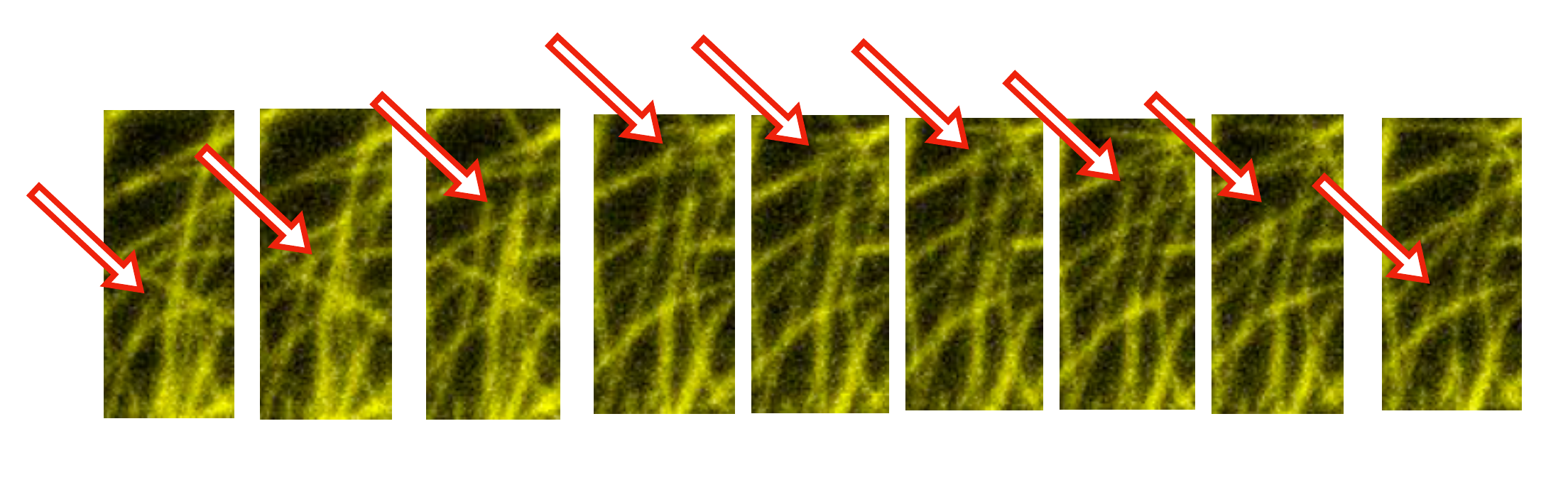}
\end{center}
\begin{center}
  \includegraphics[width=1.0\columnwidth]{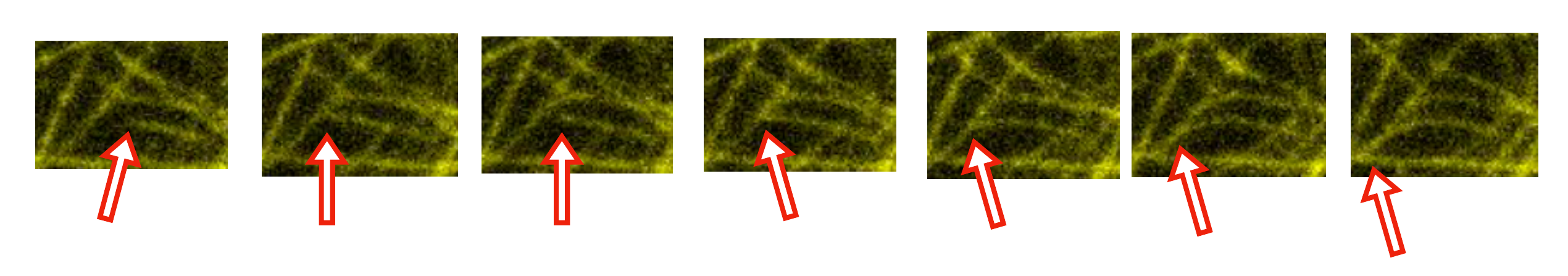}
\end{center}
\begin{center}
  \includegraphics[width=0.5\columnwidth]{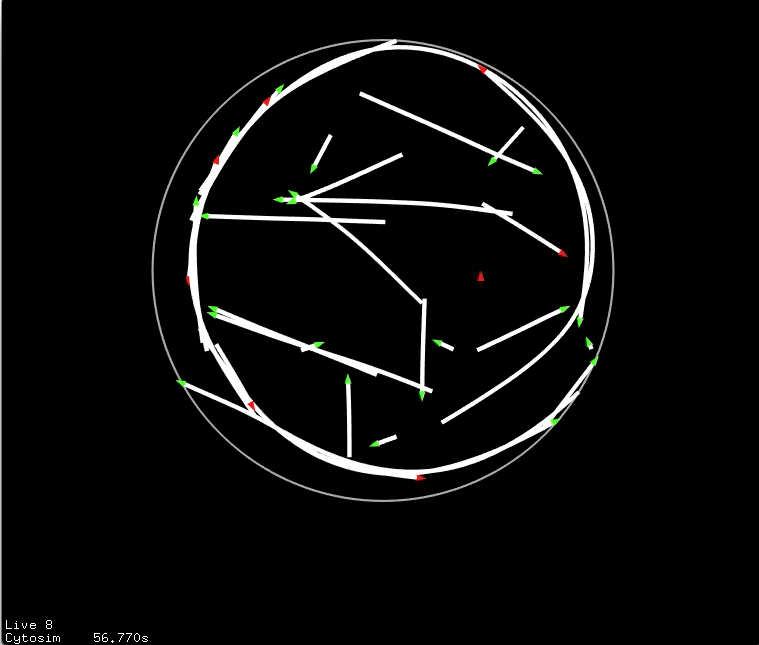}
\end{center}
\caption{
(A)-(B):
Sequences of image windows from an {\it Arabidopsis thaliana} pavement cell cortical microtubule microscopy time series.
(A) An image sequence interpretable as a collision-induced catastrophe following several crossovers.
(B) An image sequence interpretable as collision, then a short ``withdrawal'' or a large image noise event,
then regrowth followed by stable zippering.
(A-B): Thanks to Ray Wightman, Sainsbury Laboratory Cambridge University for help with
sample preparation and microscopy, using  a Leica TCS SP8 confocal laser scanning microscope in resonance scanning mode,
with {\it Arabidopsis} samples containing fluorescently labelled microtubules as described in (Wightman and Turner 2007).
(C) A larger window into a frame of a movie created in Cytosim (Nedelec et al. 2007)
exhibiting MT zippering (confluence of green arrows)
and collision-induced catastrophe (red arrows), along with stochastic MT biomechanics,
among other processes.
Thanks to Dustin Maurer, UCI for this simulation.
}
\label{fig:MT_real}       
\end{figure}

{\it Dynamic Cytoskeleton:} All of these examples of coarse-scale models of plant cell division could
and probably should be elaborated at a much finer biophysical scale.
But doing so 
requires the introduction of models of cytoskeleton in general and 
microtubule dynamics in particular, a problem of current research interest
(Vemu et al. 2018) (Chakrabortty et al., 2018). 
That is because the role microtubules play in determining the plane
of the cortical pre-prophase band, which in turn correlates well with
the subsequent septation and choice of division plane.
Here we simply observe the composable rule-like behavior of 
some of the principal processes that cortical microtubules undergo:
(1) nucleation of new MTs, often in association with old ones;
(2) treadmilling, in which tubulin subunits are added to the 
ragged growing ``+ end'' of a microtubule and removed from the ``- end'';
(3) probabilistic transitions of + end state among growth, pause, 
and catastrophic depolymerization (Shaw et al. 2003);
(4) collision of one CMT into the side of another in the 2-dimensional cell cortex,
resulting depending on collision angle in (4a) ``zippering'' or ``bundling'' into a CMT bundle
if the collision angle is small, or else (4b) an apparently stochastic choice
between (4b1) colliding + end goes into catastrophe state, or (3b2) colliding + end
crosses over and continues past the other CMT, forming a stable junction with it;
(5) katanin-induced severing of a CMT far from either end; and
(6) biomechanics of bending. Other processes may have to do
with anchoring the CMT to the cell membrane and cell wall.
Processes 4a, 4b1 and 4b2 
are illustrated in
Figure~\ref{fig:MT_rules} and  Figure~\ref{fig:MT_real}.
One way to express some of these processes ((1), (2),  part of (3), and (4a))
in a graph grammar will be discussed in Section~\ref{sec_dyn_graph}.

Even for multilevel modeling of a single-scale reaction network,
dynamic parameters allow for the possibility of aggregate objects that keep track of many
individual particles including their number. Such aggregate objects would
have their own rules as discussed in (Yosiphon 2009 Section 6.4),
possibly obtained from fine-scale reaction rules by meta-rule transformation.

\subsubsection{Parameterized Reaction Rules: Fock Space Semantics Definition}
\label{sec_param_fock}

To specify the compositional semantics for {\it parameterized} rewrite rules it is necessary
to specify the probability space in which distributions are defined.
In (Mjolsness 2010) a
sufficiently general space $S$ of dynamical systems available as targets for the semantics map $\Psi(M) \in S$ is 
specified in terms of a master equation 
governing the evolution of 
a distribution $p$ in a Fock space constructed out of elementary measure spaces for the parameters.

The semantic map $ \Psi(M) $ 
involves each species of molecule, cell type,
or other object type $\alpha$ in a way that depends on the maximum number  $n_{\alpha}^{(\max )}$ of
indistinguishable individuals possible for that species at a time.
Here 
as in Equation~(\ref{integer_algebra_1})
we will assume $n_{\alpha}$ is unbounded ($n_{\alpha}^{(\max )} = +\infty$).
This case is relevant to well-stirred chemical reaction network models in
which
a harmless simplifying assumption is that
there is no fixed upper bound to the number of indistinguishable
instances of a particular object.
However, the $n_{\alpha}^{(\max )} = 1 $ case 
(as in Equation~(\ref{binary_algebra_1}))
will be
relevant to extended object modeling below, both for spatially
localized objects and for graph-like extended objects.

In order to extend the main continuous-time model semantics $\Psi(M)$
to define the semantics of a model containing not only objects but also
dynamical {\it variables}, each generally taken to be associated with some object,
we will need to integrate over the possible values of all such variables.
Notation is as follows. As before,
$r$ indexes rewrite rules 
and $\alpha$ and $\beta$ index domain object types.
Also 
$p$ and $q$ index elements in either side of a rule,
$c$ indexes variables in a rule, and
$A$ indexes a list of defined measure spaces.
Each variable $X_c$ will have a type required by its position(s) in the argument list $[x_p]$
of one or more terms $\tau_{\alpha(p)}$, and a corresponding measure space $D_{A(c)}$
and measure $\mu_{A(c)}$. Here $A$ indexes some list of available measure spaces
including discrete measure on ${\mathbb N}$ and Lebesgue measure on ${\mathbb R}^{d:{\mathbb N}}$.

Let 
${\cal V}_{\alpha} = \bigotimes_p D_{A(\alpha, p)}$ 
be the resulting measure space for parameter lists
of terms of type $\alpha$. To summarize briefly the ``symmetric Fock space'' construction outlined in
(Mjolsness and Yosiphon 2006; Mjolsness 2010):
For each nonnegative integer $n_{\alpha}$ we define a measure space of
states that have a total of $n_{\alpha}$ ``copies'' of 
parameterized term $\tau _{\alpha}( x_{\alpha}) $:
\begin{equation}
f_{\alpha}( n_{\alpha}) =\left( \bigotimes \limits_{m=1}^{n_{\alpha}}{\cal V}_{\alpha}\right)
/\mathcal{S}( n_{\alpha}) \ \ 
\end{equation}
where $\mathcal{S}( n) $ is the permutation group on $n$ items
- in this case the elements of the $n_\alpha$ measure space factors. 
Next, any number $n_{\alpha}:{\mathbb N}$ of terms is
accommodated in a disjoint union of measure spaces $f_{\alpha}( n_{\alpha}) $,
resulting in a measure space for all terms of type $\alpha$,
and a cross product is taken over all species $\alpha$:
\begin{equation}
f_{\alpha}=\bigoplus \limits_{n_{\alpha}=0}^{\infty }f_{\alpha}( n_{\alpha}) \ \ \mathrm{and}\ \ {\cal F}=\bigotimes 
\limits_{\alpha}f_{\alpha} \;\; .
\end{equation}
This is the measurable system state space. In it, 
objects of the same type $\alpha$ are indistinguishable except by their parameters $x_p:{\cal V}_\alpha$.

The operator semantics now becomes
an integral over all the variables $X_c$:
\begin{equation}
\begin{split}
{\hat{W}}_{r}= \int \ldots \int _{[D_{c}|c]}
&
\left( \prod \limits_{k}d\mu_{A(c)} ( X_{c}) \right) \rho _{r} \Big( \left[
x_{p}( \left[ X_{c}\right] ) \right] ,\left[ y_{q}\left( \left[
X_{c}\right] \right) \right] \Big) 
\\ &\times 
\left\{ \prod \limits_{q\in \operatorname{rhs}( r) } 
{\hat{a}}_{\beta(q) }\left( y_{q}( \left[ X_{c}\right] ) \right) \right\} \left\{
\prod \limits_{p\in \operatorname{lhs}( r) }a_{\alpha( p) }\left( x_{p}(
\left[ X_{c}\right] ) \right) \right\} . %
\label{XRef-Equation-922212022}
\end{split}
\end{equation}
This expression is again in normal form, with annihilation operators preceding (to the right of)
creation operators. When other operator expressions need to be converted to normal form
one uses the same kinds of 
Heisenberg algebra
commutation relations as before except that the $\delta_{\alpha \beta}$
Kronecker delta functions are now augmented by Dirac delta functions, 
and their products,
as needed to cancel out the corresponding measure space integrals:
\begin{equation}
\begin{split}
[a_\alpha(\xb),\hat{a}_\beta(\yb)]  & = 
\begin{cases}
\delta_{\alpha \beta} \delta_{\mu_{A}}(\xb - \yb) I & \text{ for } n \in {\mathbb N} \\
 \delta_{\alpha \beta} \delta_{\mu_{A}}(\xb - \yb) (I_\alpha- 2 N_\alpha(\xb) ) I & \text{ for } n \in \{0,1\}  \\
\end{cases} \\
f(\xb) & =  \int_A \mu_A(\yb) \delta_{\mu_{A}}(\xb - \yb) f(\xb-\yb) 
\end{split}
\label{param_comm}
\end{equation}
The integrals over measure spaces act a bit like quantifiers in first order logic,
binding their respective variables. We speculate that the product of two such operator expressions
could be computed in part by using the logical unification algorithms of computational symbolic logic,
since the problem of finding the most general unifier (MGU) arises naturally when integrating over
several sets of variables during restoration of canonical form
(cf. the proof of Proposition 2 in Section~\ref{sec_dyn_graph} below)
using the commutation relations of 
Equation (\ref{param_comm}) and their delta functions.
Such full MGU computations (Martinelli et al. 1982) may incur overhead costs
which however are not large if the computation is performed 
on the relatively small AST representing the model
as part of model analysis or implementation, rather than at simulation time
when the graphs are large and there is a premium on speed.

In the Wightman axioms for quantum field theory
(QFT) the physical characteristics of
{\it locality} and {\it causality} enter through commutation relations 
similar to Equation (\ref{param_comm}) pertaining to spatiotemporal fields 
that can be built out of creation/annihilation operators (Glimm and Jaffe 1981).
What is important is that fields at points with spacelike separation
(in our case, that includes fields defined at the same time $t$, omitted in our notation,
but different places ${\vec x}$ and ${\vec y}$,
as shown in Equation~(\ref{param_comm}) above)
must commute. 
In this way different processes, or processes acting on different objects,
all happen truly in parallel.
When continuous models of space are added to the semantics our modeling languages,
e.g. differential equations as discussed in Sections~\ref{deqr} and \ref{sec_nonstandard_defs} below,
it will be important to ensure that they too respect causality.
Hyperbolic PDEs clearly do so; parabolic PDEs comprise a borderline case.

In simulation, the mapping from model time to computer time
can in principle 
be done by slicing space-time along any family of spacelike surfaces,
including but not limited to surfaces of constant model time,
while maintaining the parallelism due to 
commutation of operators at spacelike separation
as determined by any propensity functions and differential operators in the model.
This geometry provides a natural limit to the parallelism of discrete event simulations.
Thus, the operator algebra {\it of the rules} of a model,
developed further below, 
specifies the amount of computational parallelism
possible for a given model - usually high since the natural world
being modeled is intrinsically parallel.

Equations (\ref{XRef-Equation-924145912}) and (\ref{XRef-Equation-922212022}) comprise the
syntax and semantics of the basic ruleset portion of Stochastic Parameterized Grammars (SPG) language
of (Mjolsness and Yosiphon 2006; Yosiphon 2009).
A simulation algorithm is derived in (Mjolsness 2013).
{\it Equivalence} of models can again be defined as ``particle equivalence'',
Equation~(\ref{particle-equivalence}).
Other features such as submodels, object type polymorphism and graph grammars were also included.
The idea of a biological modeling language whose models take the
form of grammars goes back to L-systems (discussed below);
continuous-time versions of biomodel grammars 
to which it would be easy to add differential equation rules
goes back at least to (Mjolsness et al. 1991)
and (Prusinkiewicz et al 1993); 
in the former case there is also
in principle an optimization-based semantics for choosing
which collection of discrete-time rules to fire.

Parameterized reaction rule
notation is fundamentally more powerful than pure chemical reactions, 
because now reaction/process rates 
$\rho _{r}( \left[ x_{p}\right] ,\left[ y_{q}\right] )$
can be functions of all the parameters involved in a rule, 
{\it and} a rule firing event can change those parameters. 
It becomes possible to express sorceror's apprentice models 
which purport to accomplish an infinite amount of computing in a finite simulated time, 
though this situation can also be avoided with extra constraints on the rate functions in the language.

A related fully declarative modeling language family rides under the banner of ``L-systems'',
named after Lindenmeyer and championed by Prusinkiewicz
(e.g. (Prusinkiewicz and  Lindenmeyer 1990)).
L-systems and their generalizations have been effective in the modeling
of a great variety of developmental phenomena, particularly in plant development,
because of their declarative expressive power. But
the usual semantics of L-systems falls outside the family of
languages considered in this section and in Section~\ref{sec_rewrite_OA} below, 
for a theoretically interesting reason.
The applicable rules of L-systems are usually defined to fire
in parallel, and {\it synchronously in discretzed time}
so that one tick of a  global clock 
may see many discrete state updates performed.
This semantics seems to be incompatible with 
the {\it summation} of local, continuous-time
operators defined in Equation (\ref{operator_addition_eq}),
because ``atomic'' uninterruptible combinations of events
are represented in operator algebra by multiplication rather than by addition of operators
(and even then they are serialized),
and on the other hand continuous-time parallel processes are represented
by addition rather than multiplication of operators, in the master equation 
that describes the operation of processes in continuous time.
Despite this difference, one could seek structure-respecting mappings
between operator algebra and L-system semantics at least at the level of individual rule-firings.

A subtle point here is that the truly (to very high accuracy at least\footnote{Physical time can now be measured to one part in $10^{18}$
(McGrew et al. 2018), 
so any discretization of time must be finer than this}) 
continuous-time parallelism 
and compositionality
of the physical universe seems to be best expressed as in QFT with spacelike commutation of operators 
(as in Equation~(\ref{param_comm})) and by Equation~(\ref{operator_addition_eq})'s
summation of time-evolution operators over processes and (as shown in the next section) over space.
In terms of this elementary parallelism,
it takes further  engineering and/or computing to implement synchronous
discrete-time parallelism in terms of such continuous-time parallelism.
Modeling languages that invoke such relatively ``heavy'' discrete-time parallel
semantics at the rule-firing level include traditional L-systems,
MGS with maximal-parallel or alternative semantics (Maignan et al. 2015),
the subgrammar call feature of SPGs and Dynamical Grammars,
and the optimization-based definition of 
rule-firing choice in the
development-modeling grammars of (Mjolsness et al. 1991).
Each such semantics implicitly poses an interesting problem of efficient 
reduction to continuous-time parallel semantics,
in general or for specific biological models.

We have argued that summation of time-evolution operators corresponds 
to model compositionality in terms of processes.
To a lesser extent models are compositional in their {\it objects} as a result,
since each object participates in a limited set of processes
particularly if processes are disaggreggated by spatial position
as they will be in Section~\ref{deqr};
thus locality expressed as commutation of spacelike separated operators
helps to license a degree of decomposition by object as well.
But process compositionality is primary.

A relevant point of comparison for the semantics
of parameterized reaction network languages is the BioNetGen modeling language
(Blinov et al. 2004). This language has been applied to many problems in
signal transduction pathway modeling with discretely parameterized terms representing
multistate molecular complexes. It also represents labelled graph structures
that arise in such molecular complexes, placing it also in the class of graph rewrite rule
dynamics languages.
Another relevant graph rewrite rule modeling language is
Kappa (Danos et al. 2007))
discussed in more detail in Section~\ref{sec_dyn_graph} below.

Parameterized rewrite rule models would seem to be non-spatial, 
but particle movement through space can already be encoded 
using discrete or continuous parameters that denote spatial location. Such motion would however
have to occur in discrete steps due to discrete-time rule-firing. 
The solution to that limitation (among others) is another language escalation.

\subsubsection{Differential Equation Rules}
\label{deqr}

Another form for parameterized rewrite rules licenses locally attached 
ordinary differential equations (ODEs) for continuous parameters, 
as in (Mjolsness et al. 1991; Prusinkiewicz et al. 1993; Mjolsness 2013), 
as part of the language $L$:
\begin{equation}
\begin{split}
{\left\{ \tau_{\alpha( p) }[ x_{p}] |p \in L_{r}=R_{r} \right\} }_{*}
\longrightarrow & {\left\{ \tau_{\alpha( p) }[ x_{p}+d x_{p}] | q \in
R_{r}= L_{r} \right\}  }_{*}
\\ &
\text{\boldmath $\; \mathbf{with} \;$} 
\Bigg\{ {d x_{p}}=v_{p}( \left[ x_{k}\right] ){d t} \Bigg |  p \Bigg\}
\text{ , i.e. } \\
{\left\{ \tau_{\alpha( p) }[ x_{p}] |p \in L_{r}=R_{r} \right\} }_{*}
\longrightarrow & {\left\{ \tau_{\alpha( p) }[ x_{p}] | q \in
R_{r}= L_{r} \right\}  }_{*} 
\\ &
\text{\boldmath $\; \mathbf{with} \;$} 
\Bigg\{ \frac{d x_{p,j}}{d t}=v_{p,j}( \left[ x_{k}\right] ) \Bigg | p,j \Bigg\} ,
\label{XRef-Equation-9288735}
\end{split}
\end{equation}
where the second form uses component rather than vector notation
for all the ODEs. 
The first form is more readily generalizable 
to stochastic differential equations (SDEs)
$\{d x_{p}=v_{p}( \left[ x_{k}\right] ){d t} + w_{p}( \left[ x_{k}\right])  dB_t$ 
where $ dB_t$ is a  Brownian motion.
The ODE semantics is given
by the corresponding differential operators:
\begin{gather}
{\hat{W}}_{\mathrm{drift}}=-\int d\left\{ x\right\} \int d\left\{
y\right\}  \hat{a}( \left[ y\right] ) a( \left[ x\right] ) \left(
\sum \limits_{j}\nabla _{ y_{j}}v_{j}( \left[ y\right] ) \prod \limits_{k}\delta
( y_{k}-x_{k})  \right) %
\label{XRef-Equation-615123838}
\end{gather}
as shown for example in (Mjolsness 2013).
The SDE case is discussed in (Mjolsness and Yosiphon 2006), section 5.3.

Equations (\ref{XRef-Equation-924145912}), (\ref{XRef-Equation-9288735}) 
and (\ref{XRef-Equation-922212022}), (\ref{XRef-Equation-615123838}) comprise the
syntax and semantics respectively of the basic ruleset portion of Dynamical Grammars (DG) language
of (Mjolsness and Yosiphon 2006; Yosiphon 2009), by addition of differential equations
to SPGs. A simulation algorithm is derived 
from the Time-Ordered Product Expansion
in (Mjolsness 2013).
The generalization to operator algebra semantics for Partial Differential Equations (PDEs)
and Stochastic Partial Differential Equations (SPDEs),
as a limit of spatially discretized ODE and SDE systems, 
is outlined
in (Mjolsness 2010).

These differential equation bearing rules 
can be used to describe processes of growth and movement of individual particle-like objects,
as in ``agent-based'' modeling. 
For example, a cell may grow according to a differential equation
and divide with a probability rate
(propensity)
 that depends on its size, as in
the plant root growth model of (Mironova et al. 2012).
In general, rewrite rules can be used to describe individual processes within a model if they are augmented (as above) 
with a symbolically expressed quantitative component such as a probability distribution or a differential equation. 
Each such process has a semantic map $\Psi$ to an algebra of operators, and processes operating in parallel 
on a common pool of objects compose by operator addition (Equation (\ref{operator_addition_eq})). 
Declarative computer languages based on this and other chemical reaction arrows, 
transformed to ordinary differential equation deterministic concentration models, 
include (Shapiro et al. 2015b, 2003; Mjolsness 2013; Yosiphon 2009) among others.

The semantics as developed so far covers discrete-time transitions between parameterized objects,
stochastically interrupting continuous-time semantics given by differential equations. 
As such it is similar to ``hybrid systems''
with discrete events indicated by threshold crossings, and indeed that is one possible
implementation for the ODE portion of a dynamical grammar solution engine (Mjolsness 2013),
the hybrid SSA/ODE solver,
in which a ``warped time'' variable increases until a predefined 
randomly chosen maximum warped time when the next rule fires.
However, there are several increases in generality in the present framework.
The discrete events occur stochastically according to ODE-state-dependent
time-varying propensities, and can be specialized to behave deterministically;
the converse is not generally true.
The continuous-time ODE dynamics does not occur in a single continuous product space over parameters,
but rather in a space of intrinsically varying dimension, because discrete events
change the number and nature of parameterized objects (as in Mjolsness et al. 1991); 
we refer to this kind of dynamical system as a ``variable-structure system''.
In addition the semantics have been generalized to cover true continuous-time stochastic processes,
such as Brownian motion, specified by stochastic differential equations.
Further substantial generalizations to extended objects will be considered below.

Complex biological objects often have {\it substructure} whose dynamics
is not easily captured by a fixed list of parameters and a rate function or differential equation that depends on those parameters.
Extended objects such as molecular complexes, cytoskeletal networks, membranes, and 
tissues comprising many cells linked by extra-cellular matrix
are all cases in point. 
There can and sometimes should be several levels of substructure in a single biological model.
We now wish to extend the syntax and semantics of the foregoing class
of declarative languages to handle extended objects systematically,
by creating a compositional language and semantics for biological {\it objects}
as well as processes,
and then extending the semantics for processes accordingly.
In the case of discrete substructure this can be done with labelled graph structures
(discussed in Section~\ref{section_extended_constructive} below).
In some cases the sub-objects (such as lipid molecules in a membrane, or long polymers in cytoskeleton)
are so numerous that an approximate spatial continuum object model is justified and simpler 
than a large spatially discrete model.
Spatial continuum models with geometric objects 
can be
built out of manifolds
and their embeddings,
along with biophysical fields represented as functions defined on these geometries,
in various ways we
discuss in Section ~\ref{section_extended}.

\subsection{Refining semantic maps}
\label{sec_refining}

We have outlined how rule-like syntax can be mapped to operator algebra semantics
in several cases, though not yet for extended objects in sufficient generality
(cf. Section \ref{section_extended} below).
There is also the possibility to define several interrelated semantics maps for
one modelling language, in order to serve purposes such as analysis and computation. 
In (Mjolsness 2013, Section 3.1.2) the 
operator algebra/ master equation semantics of Equation~(\ref{operator_addition_eq}) and Equation~(\ref{CME})
for the case of stochastic chemical kinetics
was also related to a discrete-``timestep'' Markov chain semantics equivalent to the Gillespie Stochastic Simulation Algorithm,
in which the molecular state and the most recent reaction time determine the 
molecular state and physical reaction time
just following the  ``next'' reaction event on a computational time axis.
The discrete timesteps in the Markov chain map to
reaction event number $k$, not to a uniform discretization of continuous physical time $t$.
Since the distribution of {\it intervals} between reaction times depends only on molecular state,
this Markov chain can be projected further down to a
Markov chain without reaction time state information (Equation~(\ref{XRef-Equation-218211639}))
- determining what happens but not when.
But of course the additivity of time-evolution operators $W=\sum_r W_r$ 
doesn't map to additivity of Markov chains $U$ and in this sense the continuous-time
model must be primary.

Starting with such a model Markov chain which they refer to as ``stochastic semantics''
for the ``Biochemical Abstract Machine'' (Biocham) modeling language,
(Fages and Soliman 2008)
show that one can project systematically down to
yet courser semantics  for chemical reaction networks
such as a ``discrete semantics'' related to Petri nets
which forgets transition rates and hence conflates nonzero probabilities,
and a ``Boolean semantics'' which tracks only
zero vs. nonzero molecule number for each chemical species.
These coarse semantic maps are formalized as in programming language theory
by way of  a ``Galois connection'' between two lattices,
namely adjoint forward and reverse order-preserving functions.
In the case of discrete-time models there is a large literature
on programming language semantics to draw on for this purpose;
much of it uses denotational semantics based on lattice theory,
although operational semantics (Plotkin 2004) is another relevant approach.
In programming language theory, process algebras such as
the ``Calculus of Communicating Systems'' (Milnor 1980)
are designed to have a clear mathematical semantics for parallel computational processes.
In general it is useful and interesting to be able to formally map a mathematical
model semantics $\Psi$ to a computational model semantics $\Psi_C$ 
(a mapping discussed further below)
because the latter can 
be {\it implemented} in a conventional programming language; 
the resulting implementation mappings could be a formally verified
computer program implementing a mathematical model.
Formal verification can help assure not only correctness, but also computational
efficiency by making available program transformations for efficiency
that are too involved for human programmers to make at with a reasonable level of effort.

\section{Extended objects}
\label{section_extended}

Can we declaratively model non-pointlike, extended biological objects such as polymer networks, 
membranes, or entire tissues in biological development? 
To achieve constructive generality in treating such extended objects we introduce,
in Section~\ref{section_extended_constructive}, 
ideas based on discrete graphs and their possible continuum limits. 
These ideas include: graded graphs, abstract cell complexes, stratified graphs, and combinations of these ideas.
Dynamics by rewrite rules, beginning with graph rewrite rules, are developed in Section~\ref{sec_dyn_graph}.
But first we discuss the more general nonconstructive types of extended objects that
we may wish to approximate constructively.

\subsection{Nonconstructive extended objects}

Classical mathematics is not constrained by the requirement of being computationally constructive,
though it can be. This lack of constraint makes easier to establish useful mappings and equivalences
(compared for example to Intuitionist mathematics),
so it is easier to reuse what is already known in proving new theorems.
We would like to retain these advantages in a theoretical phase of work before mapping to
constructive computer simulations. 

Classical mathematical categories such as topological spaces, measure spaces, manifolds,
CW cell complexes, and stratified spaces
(the latter two composed of manifolds of heterogeneous dimension)
provide models of extended continuum objects such as biological cell membranes, cytoskeleton,
and tissues made out of many adjacent cells. 
Function spaces such as various Hilbert and Banach spaces can be used
to provide models of definable biophysical
ields (such as concentrations and biomechanical stress/strain fields)
whose values vary
over such extended objects.
Objects in such categories may (depending on the category)
have rich and useful collections of $k$-ary operators
(unary, binary, countable associative, etc. object-valued ``operators'', 
not to be confused with the probability-shifting
dynamical creation/annihilation operators of Section~\ref{sec_semantic_map})
such as category sums and products, and
even
function arrows for Cartesian Closed Categories such as compactly generated topological spaces 
(Steenrod 1967; Booth and Tillotson 1980).
Such $\oplus_C, \otimes_C$, and $\rightarrow_C$ operators for category $C$
can be targeted by compositional semantics from context-free grammar rules that generate expressions
including these operators in an AST in a modeling language; these are essentially ``type constructor'' type inference rules
in standard programming language semantics (Pierce 2002).
In principle, further invocations of category-specific
function-arrow type constructors $\rightarrow_C$ could find application
by way of variational calculus (whose ``functionals'' are functions from functions to reals) 
and even higher-order
variational calculus;
the latter has recently been applied to reaction-diffusion
models in the model reduction work outlined in Section~\ref{sec_mod_red} below
and described in detail in (Ernst et al. 2018).

Further object-generating operators may require mathematical objects in heterogeneous but related categories, 
such as 
defining new submanifolds by level sets of continuous functions using the
regular value theorem (related to the implicit function theorem),
or 
alternatively
as the image of a continuously differentiable embedding
([Hirsch 1976] Chapter 1, Theorems 3.1 and 3.2.).
Such level set functions could be biophysical fields such as concentration
of morphogen for tissue domain boundary
(as in the well-known French flag model (Wolpert 1969) for locally representable spatial information in developmental biology),
or cortical microtubules 
(as in Section~\ref{subsec_examples})
in the preprophase band 
whose placement can predict the Cortical Division Site
for plant cell division,
or they could be purely mathematical phase fields that rapidly interpolate 
between discrete values for different compartments.

With extended objects we encounter the possibility that the {\it type} of one object 
is itself another typed {\it object}. 
For example a point $x: \tilde{S}$ may be constrained
to lie on the surface of a sphere $\tilde{S}:\text{Manifold}(d=2)::\text{Manifold}$ which is itself
of type 2-dimensional manifold or more generally a Manifold. 
(As an example, a membrane-bound receptor may diffuse in the 2D membrane of a cell
which could be modeled as a manifold homeomorphic to the 2-sphere  $\tilde{S}$.)
Indeed this possibility may be taken to
define an ``extended'' object like $\tilde{S}$: It is, or it at least informs, the type
of its sub-objects. 
For consistency one might like to map biological domain objects to 
``mathematical objects'' that find their formal packaging in category theory by the same definition,
whether they are constituent objects like point $x$ or extended objects like 2D manifold $\tilde{S}$.
There are several ways to do this but 
one way is by
loose analogy with 
Homotopy Type Theory (HTT 2013): We may define
a {\it mathematical object} $x$ as an object in a category, 
if for each topological space $T$ (such as $\tilde{S}$) modeling an extended object 
(such as a topologically or even geometrically spherical model of the membrane of a protoplast)
we also define an associated category
whose category-objects are the individual {\it points} of $T$
(itself a category-object in Top or Manifold)
that might model, for example, positions of diffusible membrane-bound receptors.
Then as in HTT, 
continuous paths in a space are morphisms between its points.
An enriched category results,
based in our case on homotopy of
manifolds, CW complexes,
and stratified spaces as well as graph path homotopy. 

By such means we could nonconstructively define the mathematical semantics map $\Psi$ 
of an extended object model in terms of functors (or at least structure-representing functions)
to classical mathematical categories.
However the mathematical objects in these spaces are not generally defined in a computable way,
so we can't get all the way to  computable implementations by this route.

Our strategy to achieve computability
will be to define another fundamental kind of map in addition to semantics maps $\Psi$: 
namely, {\it Implementation} maps ${\cal I}_*$ (for various subscripts $*$ to be discussed).
Implementation maps ${\cal I}_{CM}$ go
from restricted versions of potentially nonconstructive but standard mathematical categories,
to labelled graphs. They enable the construction of implementation maps ${\cal I}$
from models to labelled graphs and their computable dynamics, by composition with $\Psi$,
which is most of what we need to run a model on a computer.
The restricted version 
of standard mathematical spaces
could for example be something like piecewise linear stratified spaces 
(Weinberger 1994),
or low-order polynomial splines to better preserve the differential topology of manifolds and stratified spaces
((Hirsch 1976) Chapter 3; (Weinberger 1994) Part II).
By abuse of terminology such implementation maps may alternatively run from 
language (with classical mathematics semantics) to language (with graph semantics),
provided the corresponding semantics diagram commutes.
In other words, implementation should commute with the semantic map, both respecting composition.
This constraint applies to ${\cal I}$ both as it acts on extended objects, and on the process models
(e.g. differential equations or Markov chains) under which they evolve.
Diagram 3 in Section~\ref{sec_slice_rewrite} will provide examples.
More implementation maps are discussed in Figure~\ref{fig:KD} of Section~\ref{sec_Tchicoma}.

Finite extended objects can be modeled by labelled graphs defined in the next section.
These are eminently computable. Infinite graphs comprise a borderline case.
One can use {\it sequences} of interrelated, finite, computable graphs
which we will formalize as ``graded graphs'' to approximate spatial continua.
(These are similar to the ``graph lineages'' together with inter-level ``prolongation maps''
defined in (Scott and Mjolsness 2019).)
In this way we can have access to continuum extended objects as declarative modeling ``semantics''
for very fine-grained biological objects, while ``implementing'' these infinite mathematical
objects in terms of explicitly finite and computable ones. Of course one's computational
resources may or may not be adequate to making the approximation needed.

A major goal for declarative modeling of continuum objects is declarative modeling
of partial differential equations (PDEs), before they have been spatially discretized into e.g.
ordinary differential equation systems. Software that already supports some PDE models
declaratively, though not in conjunction with all the other process types of 
Section~\ref{def_decl_mod} or \ref{sec_dyn_graph},
includes the ``Unified Form Language'' 
language  for specifying PDEs through their algebraic weak forms (Alnaes et al. 2014),
in the well-developed FEniCS project (Logg et al. 2012)
for finite element method (FEM) numerical solution of PDEs, 
leaving the detailed choice
of finite element solution methods to other parts of the model specification;
and  a computer algebra system (Wolfram Research 2017)
that similarly separates an algebraic model specification from the choice of solution algorithms.
Of course the choice of numerical solution methods for PDEs
is a vast area of research; here we just observe that methods for which
the mapping from continuum to discrete descriptions (e.g. meshing)
is dynamic and opportunistic,
such as Discontinuous Galerkin methods generalizing FEMs,
or Lagrangian methods including particle methods for fluid flow,
would place extra demands on the flexibility of the
graph formalism for extended objects introduced in the next section.

\subsection{Constructive extended objects via graphs}
\label{section_extended_constructive}

Discrete graphs, especially when augmented with labels,
are mathematical objects that can represent computable objects and expressions
at a high level of abstraction.

\subsubsection{Standard definitions}
An (undirected/directed) {\it graph} is a collection $V$
of nodes and a collection $E$
of edges each of which is an (unordered/ordered) pair of vertices. 
We allow self-edges (for example by defining an ``unordered pair'' to be a multiset of cardinality 2).
A {\it graph homomorphism} is a map from vertices (also called nodes) of source graph to 
vertices of target graph that takes edges (or links) to edges but not to non-edges. 
Graphs (undirected or directed) and their homomorphisms form a category. 
The category of undirected graphs can be modeled within the category of directed graphs 
by a functor representing each undirected edge by a pair of oppositely directed edges between the same two vertices.

Either category of graphs and graph homomorphisms contains resources
with which to formalize {\it labelled graphs} (specifically vertex-labelled graphs),
{\it bipartite graphs}, and {\it graph colorings} all as graph homomorphisms
(to a fully connected and self-connected graph  
${K}^+_\Lambda$ of $L=|\Lambda|$ distinct labels $\Lambda = \{ \lambda_{i \in  \{1 \ldots L\}} \}$;
to a two-node graph without self-connections;
and more generally to a  {\it clique} of $L$ distinct nodes fully interconnected but without self-connections, respectively). 
Edge-labelled graphs (as opposed to the default vertex-labelled graphs) can be 
modelled by way of a functor from graphs to bipartite graphs 
which maps vertices to red vertices and edges to blue vertices; the resulting bipartite graph can be further labelled as needed.
In a {\it multigraph}, the collection $E$ of edges is allowed to be a multiset rather than just a set;
thus a single (unordered/ordered) pair of vertices in $V$ can be connected by any nonnegative integer number of edges, not just zero or one.

As usual, a {\it tree} is an undirected graph without cycles and a Directed Acyclic Graph (DAG) is
a directed graph without directed cycles. A {\it directed rooted tree} is a DAG whose
undirected counterpart 
is a tree,
and whose (directed) edges are consistently directed away from
(or consistently towards) 
exactly
one root node.
Then Abstract Syntax Trees (ASTs)
(previously discussed as the substrate for declarative model transformations; see Figure 1)
can be represented as node- and edge-labelled directed rooted trees
with symbolic labels for which the finite number of children of any node are totally ordered
(e.g. by extra numerical edge labels), so there is a unique natural ``depth-first''
mapping of ASTs to parenthesized strings in a language.
Alternatively, ASTs can be mapped to terms in universal algebra,
facilitating abstraction.
Representing an AST as a labelled tree has the additional
advantage that one or more graph rewrite rules (Section~\ref{sec_rewrite_OA}) 
can be defined to manipulate the ASTs
so as to implement the transformations of model-denoting ASTs
foreseen in the informal definition of declarative modeling langages in Section 2.
In addition to depth-first traversal, it is also possible to totally order the nodes in a tree or DAG
in ``breadth-first'' ways that respect the directed-edge partial order,
as in task scheduling.

The automorphism group of a labelled graph is in general reduced in size from that of the corresponding unlabelled graph,
making it computationally easier to detect particular subgraphs. 
Some binary operators of type $(\text{graph},\text{graph)} \rightarrow \text{graph}$,
namely categorical disjoint sum ``$\oplus_{\text{Graph}}$'' and the 
graph cross product ``$\times$'' = categorical product ``$\otimes_{\text{Graph}}$'', 
can be defined by the usual universal diagrams applied to the Graph category. 
The  graph product ``$G \times H$'' connects a pair of product vertices $(a,b),(c,d)$
just in case $a$ connects to $c$ in $G$ {\it and} $b$ connects to $d$ in $H$.
If $G$ and $H$ happen to be two-node graphs each with a single edge between their nodes,
the resulting graph (we could call it a ``pictograph'') looks like the symbol ``$\times$'',
where $a$ and $c$ are laid out along the horizontal axis and $b$ and $d$ along the vertical.
The graph box product ``$\Box$'', essential for meshing, can also be formulated as a 
tensor product universal diagram
but not strictly within the category ((Knauer 2011) Theorem 4.3.5). 
The  graph box product ``$G \Box H$'' connects a pair of product vertices $(a,b),(c,d)$
just in case $a$ connects to $c$ in $G$ {\it and} $b$ is identical to $d$ in $H$, {\it or} vice versa.
If $G$ and $H$ happen to be two-node graphs each with a single edge between their nodes,
the resulting pictograph looks like the symbol ``$\Box$''.
The 
``$\boxtimes$'' strong graph product includes all edges licensed by either $\times$ {\it or} $\Box$
(Imrich and Klav\v{z}ar, 2000),
and again the product symbol is close to denoting its own pictograph.
Graph {\it functions} by binary operator ``$\rightarrow_{\text{Graph}}$''
that produces a graph are also possible but
require an altered definition of the category (Brown et al. 2008).

\subsubsection{Non-standard definitions}
\label{sec_nonstandard_defs}

A special case of a labelled graph is a {\it numbered graph}
with integer labels $\lambda_i^{\prime} = i$, $|\Lambda^{\prime}| \ge  |V|$,
and the labelled graph homomorphism to ${K}^+_{\Lambda^\prime}$ is one-to-one
(but not necessarily onto), 
so each vertex receives a unique number in $\Lambda^\prime = \{1 \ldots k \ge |V|\}$.
Then one way to express a labelled graph $G \rightarrow_\text{Graph} {K}^+_{\Lambda}$
is as the composition 
$G  \rightarrow_\text{Graph} {K}^+_{\Lambda^\prime} \rightarrow_\text{Graph} {K}^+_{\Lambda}$
of a graph numbering  $G  \rightarrow_\text{Graph} {K}^+_{\Lambda^\prime} $ 
followed by a relabelling determined by a mapping of sets
${\Lambda^\prime} \rightarrow_\text{Set} {\Lambda}  \mathbin{\dot{\cup}} \{\varnothing\}$,
with $\varnothing$ the value taken on ununused number labels $i$.
The possibility of unused numbers ($|\Lambda^{\prime}| \ge  |V|$) will be needed 
when consistently labelling two different graphs.

We consider graph homomorphisms to five particular integer-labelled graphs: 
\begin{equation}
\begin{split}
{\mathbb N}^+  \equiv &  ({\mathbb N}, \text{Successor}) \\
    & =  \text{ nonnegative integers } \{0, 1, \ldots\} \text{ as vertices,} \\
    & \text{ with (possibly directed) edges from each} \\ 
    & \text{ integer $i$ to its immediate successor $i+1$ and to itself} ; \\
{J}^+_D  \equiv  &
\begin{cases}
 ({\mathbb N}_D \equiv \{0,... D \ge 0\}, \ge) & \text{ directed graphs;} \\
    \quad \quad = \text{ integers } \{0,... D \} \text{ with } (i , j) \text{ edge  iff } i \ge j ; \\
    {K}^+_{{\mathbb N}_D}  = {K}^+_{\{0,...D\}} \text{ (fully connected w. self-edges)} & \text{ undirected graphs} \\
\end{cases} \\
{{\mathbb N}}_D^{+ \text{op}} \equiv & 
\begin{cases}
 \text{ integers } \{0,... D \} \text{ with } (i , j) \text{ edge  iff } i = j+1 \text{ or } i=j  \;\;  \text{ directed graphs} ; \\
 \text{ integers } \{0,... D \} \text{ with } (i , j) \text{ edge  iff } |i - j| \le 1 \quad \text{ undirected graphs} \\
\end{cases} \\
C_D  \equiv & {\mathbb N}^+ \Box {J}^+_{D} \\
{\tilde{C}}_D  \equiv & {\mathbb N}^+ \Box {{\mathbb N}}_D^{+ \text{op}} \\
\end{split}
\end{equation}
(In this notation the `+' exponents ${G}^+$ refer to the addition of self-edges.)
Graph homomorphisms to these graphs will provide the core of the next few definitions.
Following ((Ehrig et al. 2006) Chapter 2), such homomorphisms themselves form useful graph-related categories.
These definitions can all be used in either the undirected or directed graph contexts.
The two contexts are
related by the standard functor in which an undirected edge corresponds to two oppositely directed edges,
defining a mapping from undirected graphs to directed graphs.

A {\it graded graph} is a homomorphism from a graph $G$ to ${\mathbb N}^+$. 
It labels vertices by a level number.
Additional assumptions may allow an undirected graded graph to ``approach'' 
a continuum topology such as a manifold or other metric space, in the limit of large level number. 
The metric can be derived via a limit of a diffusion process on graphs.
For example, $D$-dimensional rectangular meshes approach ${\mathbb R}^D$  this way 
(e.g. (Mjolsness and Cunha 2012)).
A conflicting definition for the term ``graded graph'', disallowing connections within a level,
was given in (Fomin 1994) but we need both $\Delta l = 0$ and $\Delta l = +1$ edges.
In the category of undirected graphs, $\Delta l = \pm 1$ are indistinguishable
except by consulting the vertex labels  $l$.
Of course, $({\mathbb{N}^+}, id_{{\mathbb{N}^+}})$ is itself a graded graph.

A {\it stratified graph} of dimension $D$ is a homomorphism from a graph $G$ to ${J}_D^+$. 
It labels vertices by a ``dimension'' of the stratum of the stratum they are in.
The reason for reversing the edges in the directed graph case, compared to the graded graph definition,
is that the level-changing edges then correspond to standard boundary relationships from higher to lower dimensional strata.
In the undirected graph case, such a homomorphism
is equivalent to a graph whose vertices are labelled by dimension number, since it imposes no constraints on edges. 

{\it Abstract cell complex}:
A special case of a stratified graph of dimension $D$ is a graph homomorphism from $G$ to ${\mathbb N}_D^{+ \text{op}}$, 
the directed graph of integers $\{0, \ldots D\}$ with self-connections and immediate predecessor connections $i\rightarrow i-1$ within the set.
Since ${\mathbb N}_D^{+ \text{op}}$ maps to 
${J}^+_{D}$ by graph homomorphism, this is (equivalent to) a special case
of stratified graphs.

A {\it graded stratified graph} of dimension $D$ is a graph homomorphism from $G$ to  $C_D = {\mathbb N}^+ \Box {J}^+_{D}$.  
Two nodes in $C_D$ are connected, permitting corresponding edge connections in $G$,
iff either their level numbers are equal
(and, in the directed case, the source node dimension is $\ge$ the target node dimension), 
or if their level numbers differ by one and their dimensions are equal. 
Because $G$  projects to both ${\mathbb N}^+$ and ${J}^+_{D}$, 
a graded stratified graph maps straightforwardly (by functors) to both a graded graph and to a stratified graph. 
However, it enforces an additional consistency constraint on level number and dimension: 
they cannot both change along one edge
(and, in the directed case, the source node dimension must be $\ge$ the target node dimension).

A {\it graded abstract cell complex} of dimension $D$ is likewise 
a graph homomorphism from $G$ to $\tilde{C}_D = {\mathbb N}^+ \Box {\mathbb N}_D^{+ \text{op}}$.

Our aim with these definitions is to discretely and computably model continuum
stratified spaces (including cell complexes) in the limit of sufficiently large level numbers. 
Stratified spaces are topological spaces decomposable into manifolds in more general ways than CW cell complexes are.
In addition, we would like each manifold to be a differentiable manifold with a metric related to a Laplacian,
since that will be the case in modeling spatially extended biological or physical objects.

Thus we arrive at the constructive {\it slice categories}  $S(H): G \rightarrow H$ with morphisms
$h: G \rightarrow H$ that makes a commutative diagram 
$\varphi_G = h \circ \varphi_{G^{\prime}}$:
\begin{diagram}
G &  &\rTo^{h}  && G^{\prime} \\
&\rdTo^{\varphi_G} & &\ldTo^{\varphi_{G^{\prime}}} &&&& (\text{Diagram }1) \\ 
& & H & \\
\end{diagram}
where  $H = {\mathbb N}^+, \;\; {J}_D^+,  \;\; {\mathbb N}_D^{+ \text{op}},  \;\; C_D, \;\; {\rm or } \;\; \tilde{C}_D$. 
Here it is essential that  
each possible target graph $H$
includes all self-connections.
In particular a
homomorphism of graded graphs
is a graph homomorphism
with $H={\mathbb{N}}^+$,
so graded graphs
together with level-preserving graph homomorphisms form a slice category.

For a discrete approximation to a stratified space 
by a stratified graph, we can identify the ``graph strata'', 
and their interconnections by boundary relationships, as connected components of constant dimension.
So, by eliminating links between nodes of different dimension, and then finding the connected components that remain, 
we identify the {\it strata} in a stratified graph (or in a stratified labelled graph). 
In this commutative diagram:
\begin{diagram}
G &  &\rTo^{h}  && G_S \\
&\rdTo^{\chi_G} & &\ldTo^{\chi_{G_S}} &&&& (\text{Diagram }2) \\ 
& &{J}_D^+ & \\
\end{diagram}
each inverse image $(\chi_{G_S}^{-1})(d)$ 
must be a 
fully disconnected graph, with each vertex then corresponding
(by $h^{-1})$ to a $d$-dimensional stratum 
(maximal connected $d$-dimensional component)
in $G$.
The directed graph associated to $G_S$ 
(directed by dimension number in the case of undirected graphs) is a DAG,
due to disconnection within each dimension. The graph homomorphism $h$
becomes a homomorphism of stratified graphs.

{\bf Observation 1} The resulting $G_S$ in Diagram 2 is the graph of strata, a minimal structure for modeling complex geometries. 
{\it It is therefore a natural graph on which to specify rewriting rules for major structure-changing processes} 
such as biological cell division, mitochondrial 
fission/fusion, 
neurite or cytoskeletal branching and other essential processes of biological development. 

The geometry of cytoskeleton, in particular, is better captured by stratified graphs and stratified spaces than by cell complexes,
because 1D and 0D cytoskeleton is often embedded directly into 3D cytosol rather than into 2D membrane structures.
(This violates the CW complex assumption that cells are $d$-dimensional balls, whose boundaries map 
continuously into finite unions of lower dimensional balls,
both because dimension 2 is skipped between 3 and 1, and because 3D cytosol is not generally homeomorphic to a ball
if it can be punctured by a 1D+0D cytoskeleton with 1D loops and/or multiple anchor points in the biological cell's surface.)

In  many developmental biology systems the spatial dynamics involves nontrivial changes
in geometry and/or topology of extended biological objects.
By using rewrite rules for the graph of strata, 
together ODE-bearing rules for the parametric embeddings of individual strata into 3D space,
we now have in principle a way to represent such dynamics mathematically and computationally.

{\bf Observation 2} In a graded stratified graph
(and therefore also in a graded abstract cell complex)
a useful special case occurs if $G_S$ restricted by level number
stabilizes beyond a constant number of levels, so the description in terms of strata has a continuum limit $G^*_S$.
In that case, {\it the limiting  $G_S$ is also the natural locus for verification of compatible boundary conditions between PDEs} 
of different dimensionality governing the evolution of biophysical fields defined on continuum-limit strata.
Here ``compatibility'' includes local conservation laws.

The edges remaining in $G_S$, connecting strata of unequal dimensionality,
model in-contact relations such as ``boundary'' and/or ``inside''.
Further constraints are needed to disentangle these alternatives.
Similar ideas to $G^*_S$ may be involved in the ``persistent homology''
approach to unsupervised learning of data structure (Bendich et al. 2007).

Another useful special case of a stratified graph 
occurs if $G_S$ obeys the constraint that 
(a) only adjacent dimensions connect.
In that case one has the discrete graph structure
which we have defined as an abstract cell complex (ACC),
although there are several related claimants for that phrase and we make
no claim for the superiority of this one.
This condition results from using the Hasse diagram for the
dimension-labelled strata, 
with other possible boundary relationships among cells
being obtained by multiple steps along $\Delta d= -1$ ACC edges.
In most geometric applications one may in addition observe that 
(b) two strata of dimensionality labels
$d-1$ and $d+1$ are always mutually adjacent to either zero or two strata of dimension $d$ (Lane 2015),
though in common with the ``abstract complexes'' of MGS (Giavitto and Michel 2001, Giavitto and Spicher 2008)
we do not make this part of the definition.
ACCs have been used for declarative developmental modeling at least of
plant development (Lane 2015) and neural tube development (Spicher 2007),
the latter using MGS.

Given an extended object $G$ constructed as outlined here, it is generally also necessary
to define some dynamics that run ``in'' or ``on'' such an object:
diffusing or otherwise moving particles with position $x \in G$ described by a spatial probability distribution $p(x,t)$,
or other dynamical fields $f(x)$ at a given moment of time $t$. 
To this end,
recall that the inverse image of a node in $G_S$ identifies the corresponding stratum in $G$.
It is natural to use the graph Laplacian on each such stratum to define:
(a) geometric distances, using the Green's function of the Laplacian operator;
(b) regularizers for regression of functions from sparsely provided data (cf. Poggio and Girosi (1990));
(c) a Sobolev space $H^2$ of functions in the infinite-graph limit, for biophysical fields; 
(d) the definition of {\it functional integrals} in the infinite-graph limit, using a kernel 
$k_{m,\lambda}(x,y) = (m-\lambda \nabla^2) \delta_{\rm Dirac}(x-y)$ to form a statistical mechanics partition function such as
the Gaussian
functional
integral (e.g. Mandl and Shaw (2010)):
\begin{equation}
\begin{split}
Z[m,\lambda,J] & = \int D[f] \exp{\left[- {1 \over 2} \int \int dx dy f(x) k_{m,\lambda}(x,y) f(y)+ \int dx J(x)f(x)\right]} \\
  &= \int D[f] \exp{\left[- {1 \over 2} \int dx  [ m f(x)^2  + \lambda  (\nabla f(x))^2]+ \int dx J(x)f(x) \right]} ; 
\end{split}
\end{equation}
and/or (e) Graph Convolutional Networks, a generalization of deep convolutional neural networks
for machine learning from rectangular grids to general graphs (Hammond et al. 2011, Kipf and Welling 2016).
But in order to recover the expected continuum properties from the graded graph
of meshes associated with a stratum, it may be necessary to reweight the edges of the graph
(and hence its Laplacian) according to the local scale of its embedding into for example
the three dimensions of flat biological space.
Laplacians and their heat kernels on Riemannian manifolds are sufficient to
recover the local geometric structure by way of triangulated
coordinate systems (Jones et al. 2008). 
Graph heat kernels converge to the manifold ones in
cases where a sequence of graphs ``approximate'' the manifold
(Coifman and Lafon 2006; cf. Mjolsness and  Cunha 2012).

\paragraph{Relation to PDEs:} 
If some of the strata in $G_S$ are host to partial differential equations 
(respecting possibly dynamical boundary conditions at adjacent lower and higher dimensional strata)
then $G_S$ will be too minimal for solution algorithms like finite elements or finite volumes, 
and those strata may have to be meshed. In that case the strata of $G_S$ 
may need to be subdivided sufficiently finely into patches that can each host 
a local coordinate system, compatible with its neighboring patches of different dimension and 
(under one possible strategy) separated by 
extra artificial boundary strata patches from its neighbors of the same dimension. 
Compatible finer meshes (computed eg. by programs such as Tetgen 
((Si and Gartner 2005); illustrated for plant tissue by (Mjolsness and  Cunha 2012))
could be aligned with the local coordinate systems.
As a simpler example, a cell complex graph whose cells embed into 3D as polyhedra
can be subdivided into cuboids whose main diagonals each stretch from a vertex
of the starting polyhedron to its centroid; 
but this decomposition may or may not have good numerical properties
such as condition number of a meshed mechanical stiffness matrix.
A problem for developmental biology, as for fluid simulation,
is to maintain mesh quality while the system simulated undergoes large deformations.
For example very close to corner-like boundaries it may become hard to avoid mesh cells with
extreme angles and poor numerics, 
but conforming meshes that fence off such boundaries at short distance from them
(Rand and Walkington 2009, Murphy et al. 2001, Engwirda 2016)  together with analytical PDE dimension reduction 
to such surfaces may provide an alternative path forwards.
Meshing for PDEs is a vast field of applied mathematical research to which we cannot do justice;
the point here is that stratified graded graphs 
(and perhaps other graph slice categories)
provide a way to formalize 
many of the problems and capabilities that have to be 
represented explicitly for declarative modeling to be applied.

Abstract cell complex graphs are also the key structure for Finite Element Methods (Hughes 2000),
and for Discrete Exterior Calculus (DEC) discretizations of PDEs 
(Desbrun et al. 2005) and the related Finite Element Exterior Calculus (FEEC) (Arnold et al. 2010).
DEC and FEEC allow for the separate discrete representation of $k$-forms
and the full exploitation of 
the generalized Stokes' theorem,
including symplectic PDE integrators and the Helmholtz/Hodge decomposition 
of function spaces for possible PDE solutions.
DEC can be combined with subdivision (de Goes et al. 2016)
in the manner of a graded stratified graph.
(Giavitto and Spicher 2008) show how to encode oriented abstract cell complexes
and discretized differential operators 
similar to DEC in the MGS declarative modeling language.
On the other hand, analysis on more general stratified spaces is also possible
including differential operators for corners with singularities allowed on
the approach to lower dimensional strata (Schulze and Tarkhanov 2003).

By these various means one would like to generalize from ODE-bearing rules
to PDE-bearing rules which would be of two basic types:
(1) PDEs for the evolution of biophysical fields such as diffusion within a manifold 
(possibly including a hyperbolic term for finite speed {\it causal} information propagation,
as in the hyperbolic Telegrapher's Equation 
derivable for stochastic diffusion (Kac 1974)
which has as a limit the parabolic
Heat Equation), and
(2) PDEs for the evolution of the embeddings of strata into higher strata,
such as cytoskeleton mechanics (1D into 2D or 3D) or the biomechanics
of 2D membranes embedded into ordinary three dimensional Euclidean space.
This could be accomplished by way of level sets and local stress fields, for example.

Such dynamic biophysical fields can also influence the (discontinuous and usually much slower)
topology-changing dynamics by which the number and connectivity
of a model's geometric strata change - such as in plant or animal cell division.
We study such dynamic graph structure next.

\subsection{Dynamics of Graphs}
\label{sec_dyn_graph}

There are at least two ways to mathematically
define the semantics of graph rewrite rules
for use in biological modeling of extended objects.
The operator algebra approach
pursued in this paper, extending the form of rewrite rule semantics given in
Section~\ref{def_decl_mod}, has been used as the theory behind some molecular complex modeling
and tissue-level developmental modeling methods
(Johnson et al 2015, Mjolsness 2013). 
It takes advantage of 
the algebra
of operators to blend with continuous-time process models by scalar multiplication of
rates and operator addition of parallel processes, thereby also
gaining compatibility with quantitative methods of statistical mechanics and field theory in physics,
and with machine learning by continuous optimization.
The essential step is to express natural graph-changing operations,
including a collection of graph rewrite rules, in terms of an operator algebra
generated by the operators for the individual rules.
A second approach, the graph homomorphism pushout diagram approach championed in (Ehrig et al. 2006),
has been used in molecular complex modeling (Danos et al. 2012)
by providing a mathematical semantics for the ``Kappa'' modeling language (Danos et al. 2007).
It takes advantage of the category theory formulation of graphs, discussed above.
We will first develop operator algebra semantics for graph grammars
in continuous time in detail.
Then we will briefly compare it with the pushout semantics approach
which leverages category theory in a way similar to our
discussion above.

\subsubsection{Graph rewrite rule operators}
\label{sec_rewrite_OA}

By generating unique (e.g. integer-valued) ``ObjectID'' parameters for each new object
created in a parameterized grammar rule, it is possible to implement
graph grammar rules by parameterized grammar rules (Section~\ref{sec_param_rules}) alone,
just using repeated ObjectID values to represent graph links.
Since parameterized grammar rules are mapped by semantic map $\Psi(M)$ to an operator algebra,
the composition of two maps ${\cal I} \circ  \Psi$ 
defines an operator algebra semantics
for  graph grammars.
This route was detailed in (Mjolsness 2005, Mjolsness and Yosiphon 2006),
and implemented declaratively in (Yosiphon 2009).
But it is also possible (Mjolsness 2010) to define graph grammar rewrite rule semantics directly
as a continuous-time dynamical system,
using operator algebra as in the previous semantics definitions
of Section~\ref{def_decl_mod}.

Suppose we have two labelled graphs 
$G _1: G^{\text{pure}}_1 \rightarrow_\text{Graph} K^+_{\Lambda_1}$ and
$G_2: G^{\text{pure}}_2  \rightarrow_\text{Graph} K^+_{\Lambda_2}$.
We decompose them each into a numbered graph 
$G^{\text{num}}_i : G^{\text{pure}}_i  \rightarrow_\text{Graph} K^+_{\{1,...k_i \ge |V_i|\}} $
and a relabelling
$ K^+_{\{1,...k_i  \ge |V_i|\}} \rightarrow_\text{Graph} K^+_{\Lambda_i}$
determined by a mapping of sets
$\{1,...k_i \} \rightarrow_\text{Set} {\Lambda_i}$,
determined in turn by an ordered listing of labels
$\lambdab_i$, possibly augmented by the nullset symbol $\varnothing$. 
The whole decomposition can be denoted
$G^{\text{num}}_i\llangle \lambdab_i \rrangle$.
We are interested in graph rewrite rules 
$G^{\text{num}} \llangle {\lambdab} \rrangle \rightarrow G^{\prime \text{ num}} \llangle {\lambdab^{\prime}} \rrangle $
that respect a single consistent numbering of vertices of the two numbered graphs before their relabellings.
In that case vertices in $G_1$ and $G_2$ that share a vertex number are regarded as
``the same'' vertex $v$, before and after rewriting
(similar to the shared graph ``$K$'' in the double pushout approach discussed in 
Section~\ref{sec_pushout} below), 
so that any graph edges contacting $v$ and not mentioned
in the rewrite rule are preserved.

\vspace{-10pt}
\paragraph{Graph rewrite rule examples}

Here 
(Equation (\ref{triangle_rule}) below)
is an example pertaining to refinement of triangular meshes in 2D.
This is one of four rewrite rules that suffice to implement a standard
triangular mesh refinement scheme. 
(Similar examples were studied in (Maignan et al. 2015).) 
Three of those rules including this one deal with
partially refined triangle edges, an intermediate state produced by the
previous refinement of nearby triangles. It can also be interpreted as an
(unlabelled) graded graph rewrite rule since it preserves the 
graded graph constraints on level numbers $l$, if they are satisfied initially.
(The other rewrite rules are similar but deal with 
the cases of zero, two, or three partially refined triangle sides.)
The labelled graph rewrite rule is:
\begin{equation}
\begin{split}
\left( 
\begin{diagram}[size=1em]
&& & & 1 & \\
&& &\ldLine & &\rdLine(4,4) &&&&  \\ 
& & 4 & \\
&\ldLine & &&&  \\ 
\; 2 &  \rLine(7,0)  &&&&&&& 3 \;\;  &  \\
\end{diagram}
\right) 
\llangle
l_1,l_2,l_3,l_4
\rrangle
\\
 \longrightarrow  & 
\left( 
\begin{diagram}[size=1em]
&& & & 1 & \\
&& &\ldLine & &\rdLine &&&&  \\ 
& & 4 & \rLine(2,0) &&&6 \\
&\ldLine & &\rdLine &&\ldLine && \rdLine \\ 
\; 2 &  \rLine(2,0)  &&&5& \rLine(2,0) &&& 3 \;\;  &  \\
\end{diagram}
\right) 
\begin{split}
\llangle
l_1,l_2, & l_3,l_4,  \max(l_1,l_2,l_3,l_4),  \\
& \max(l_1,l_2,l_3,l_4)  
\rrangle    
\end{split}
\end{split}
\label{triangle_rule}
\end{equation}
Note that in this example there is a shared numbering of nodes of the two graphs,
and node numbers 1-4 occur in both graphs. 
This is equivalent to identifying
the shared subgraph $K$ in the double pushout semantics of Diagram 4 in 
Section~\ref{sec_pushout} below.
Any extra edges that contact nodes 1-4 in a subgraph of the pool graph,
e.g. parts of other nearby triangles,
will remain after this rule fires.

Of the other three grammar rules, 
two are analogs of (\ref{triangle_rule}) for two or three hanging side nodes like node 4 above;
only one rule increases the maximal level number:
\begin{equation}
\begin{split}
\left( 
\begin{diagram}[size=1em]
&& & & 1 & \\
&& &\ldLine(4,4) & &\rdLine(4,4) &&&&  \\ 
& &  & \\
& & &&&  \\ 
\; 2 &  \rLine(7,0)  &&&&&&& 3 \;\;  &  \\
\end{diagram}
\right) 
\llangle
l_1,l_2,l_3
\rrangle
\\
 \longrightarrow  & 
\left( 
\begin{diagram}[size=1em]
&& & & 1 & \\
&& &\ldLine & &\rdLine &&&&  \\ 
& & 4 & \rLine(2,0) &&&6 \\
&\ldLine & &\rdLine &&\ldLine && \rdLine \\ 
\; 2 &  \rLine(2,0)  &&&5& \rLine(2,0) &&& 3 \;\;  &  \\
\end{diagram}
\right) 
\begin{split}
\llangle
& l_1,l_2,  l_3,  \max(l_1,l_2,l_3)+1,  \\
& \max(l_1,l_2,l_3)+1, \max(l_1,l_2,l_3)+1  
\rrangle
\end{split}
\end{split}
\label{triangle_rule_level_up}
\end{equation}
In the resulting four-rule grammar, the order in which triangles
are refined does not interfere with the potential refinement of
other triangles. If for example a global maximum-level constraint $l \le L$ 
were imposed via the firing rate of each rule,
then the refinement process would converge by many alternative paths to the 
same fully refined terminal triangular mesh.

A more complicated rewrite rule can maintain not only $l=$ level number but also
$d=$ dimension in a cell complex representation, during this kind of triangle refinement.

To implement a 2D version of the  the polyhedron $\rightarrow$ cuboid
mesh refinement scheme mentioned in
Section~\ref{sec_nonstandard_defs}, 
one could 
alternatively 
start with four triangle $\rightarrow$  quadrilateral refinement rules;
for example instead of rule (\ref{triangle_rule})
one would have:
\begin{equation}
\begin{split}
\left( 
\begin{diagram}[size=1em]
&& & & 1 & \\
&& &\ldLine & &\rdLine(4,4) &&&&  \\ 
& & 4 & \\
&\ldLine & &&&  \\ 
\; 2 &  \rLine(7,0)  &&&&&&& 3 \;\;  &  \\
\end{diagram}
\right) 
\llangle
l_1,l_2,l_3,l_4
\rrangle
   \\
 \longrightarrow  & 
\left( 
\begin{diagram}[size=1em]
&& & & 1 & \\
&& &\ldLine(2,3) & &\rdLine(2,3)  &&&&  \\ 
&&&&& \\
& & 4 & & &&6 \\
& \ldLine(2,4) &  & \rdLine &&\ldLine&&\rdLine(2,4)  \\
& &  & &7&& \\
& &  & &\dLine && \\
\; 2 &  \rLine(2,0)  &&&5& \rLine(2,0) &&& 3 \;\;  &  \\
\end{diagram}
\right) 
\begin{split}
\llangle
& l_1,l_2,  l_3,  l_4, \max(l_1,l_2,l_3,l_4),  \\
& \max(l_1,l_2,l_3,l_4), \max(l_1,l_2,l_3,l_4)  
\rrangle
\end{split}
\end{split}
\label{another_triangle_rule}
\end{equation}
However, one is then committed to designing a larger grammar that can refine both triangles and quadrilaterals
with varying numbers of hanging side nodes like node number 4 above,
possibly requiring six more quadrilateral refinement rules since there are two ways to place two such side nodes.
Abstraction via sub-grammars may then become important.
Further notational innovations could also reduce the number of rules needed.

In addition to the forgoing geometric example, 
we consider briefly how one might express some of the  dynamics
of visible plant cortical microtubule bundles previously described,
in particular growth at a growing end (whether + or - ends of invididual
MTs in the bundle); retraction at a retracting end,
and bundling following front-to-side collision,
in terms of graph grammar rules.
Let a microtubule be an extended object represented as a chain of
super-particles 
(each much larger than a tubulin dimer, representing
a roughly straight cylindrical segment of one MT 
of length approximately on a lengthscale $L$ that is
several times an MT diameter,
{\it or} of a parallel and/or antiparallel bundle of a few such cylindrical segments).
Continuous parameters of such a fiber segment super-particle will include
its center-of-mass position, and a unit vector pointing towards the 
growing end and away from the retracting end of an end segment. 
(Interior segments will have lengthwise unit vectors too, but their
sign shouldn't matter).
Discrete parameters will include
a four-valued categorical label 
$s \in \{\operatorname{internal}, \operatorname{grow\_end}, \operatorname{retract\_end}, \operatorname{junct} \}$ 
(or $s \in \{\Circle, \CIRCLE, \blacksquare, \blacktriangle \}$ in diagrams)
for status as interior segment,
growth-capable end segment, retraction-capable end, or junction segment
respectively.

A diagrammatic presentation of an MT graph grammar,
with subscripts for the rule-local arbitrary but consistent numbering of
vertices in left- and right-hand side graphs of each rule,
is here:
\begin{equation}
\begin{split}
&  
\left( 
\begin{diagram}[size=1em]
\; 
\CIRCLE_1 & & &  \\
\end{diagram}
\right) 
\llangle
(\text{${ \boldsymbol x}$}_1,\text{${ \boldsymbol u}$}_1)
\rrangle
 \longrightarrow  
\left( 
\begin{diagram}[size=1em]
\; 
\Circle_1 & \rTo(2,0) & & 
\CIRCLE_2 & \\
\end{diagram}
\right) 
\llangle
(\text{${ \boldsymbol x}$}_1,\text{${ \boldsymbol u}$}_1),(\text{${ \boldsymbol x}$}_2,\text{${ \boldsymbol u}$}_2)
\rrangle
\\  &  \quad \quad \text{\boldmath $\mathbf{with}$} \ \ 
	\hat{\rho }_{\text{grow}}( [\text{\small tubulin}])
	\mathcal{N}( \text{${ \boldsymbol x}$}_1 -  \text{${ \boldsymbol x}$}_2 ; L \text{${ \boldsymbol u}$}_1, \sigma ) 
	  \mathcal{N}( \text{${\boldsymbol u}$}_2; \text{${ \boldsymbol u}$}_1 /(|\text{${ \boldsymbol u}$}_1 |+\epsilon), \epsilon) , \\
&  
\left( 
\begin{diagram}[size=1em]
\; 
\blacksquare_1 & \rTo(2,0) & 
\Circle_2 &  \\
\end{diagram}
\right) 
\llangle
(\text{${ \boldsymbol x}$}_1,\text{${ \boldsymbol u}$}_1),(\text{${ \boldsymbol x}$}_2,\text{${ \boldsymbol u}$}_2)
\rrangle
  \longrightarrow  
\left( 
\begin{diagram}[size=1em]
 &  
 \blacksquare_2 & \\
\end{diagram}
\right) 
\llangle
(\text{${ \boldsymbol x}$}_2,\text{${ \boldsymbol u}$}_2)
\rrangle
\\  &  \quad \quad \text{\boldmath $\mathbf{with}$} \ \ 
	\hat{\rho }_{\text{retract}} \\
&  
\left( 
\begin{diagram}[size=1em]
\; \Circle_1& \rTo(2,0) & & 
\Circle_2 & \rTo(2,0) & 
\Circle_3 & \\
& & &&&  \\ 
& 
\CIRCLE_4  & & \\
\end{diagram}
\right) 
\llangle
(\text{${ \boldsymbol x}$}_1,\text{${ \boldsymbol u}$}_1),(\text{${ \boldsymbol x}$}_2,\text{${ \boldsymbol u}$}_2),
  (\text{${ \boldsymbol x}$}_3,\text{${ \boldsymbol u}$}_3),(\text{${ \boldsymbol x}$}_4,\text{${ \boldsymbol u}$}_4)
\rrangle   \\
& \quad \longrightarrow   
\left( 
\begin{diagram}[size=1em]
\; \Circle_1 & \rTo(2,0) & & 
\blacktriangle_2 & \rTo(2,0) & 
\Circle_3 & \\
& & \ruTo &&&  \\ 
& 
\; \Circle_4 & & \\
\end{diagram}
\right) 
\llangle
(\text{${ \boldsymbol x}$}_1,\text{${ \boldsymbol u}$}_1),(\text{${ \boldsymbol x}$}_2,\text{${ \boldsymbol u}$}_2),
  (\text{${ \boldsymbol x}$}_3,\text{${ \boldsymbol u}$}_3),(\text{${ \boldsymbol x}$}_4,\text{${ \boldsymbol u}$}_4)
\rrangle
\\  & \quad \quad \text{\boldmath $\mathbf{with}$} \ \ 
	 \hat{\rho }_{\text{bundle}} (| \text{$ {\boldsymbol u}_2$} \cdot \text{${\boldsymbol u}_4$} |/ |\cos \theta_{\text{crit}}|) 
			\exp(-| \text{${\boldsymbol x}$}_2 - \text{${\boldsymbol x}$}_4 |^2/2 L^2) \\
 &  
\left( 
\begin{diagram}[size=1em]
\; \blacksquare_1 & \rTo(2,0) & 
\CIRCLE_2 &  \\
\end{diagram}
\right) 
\llangle
(\text{${ \boldsymbol x}$}_1,\text{${ \boldsymbol u}$}_1),(\text{${ \boldsymbol x}$}_2,\text{${ \boldsymbol u}$}_2)
\rrangle
    \longleftrightarrow  
\varnothing 
     \quad \quad  \text{\boldmath $\mathbf{with}$} \ \ 
	\Big( \hat{\rho }_{\text{retract}} ,
\\ & \quad \quad \quad \quad 
	\hat{\rho }_{\text{nucleate}}( [\text{\small tubulin}] )  
		\mathcal{N}(  \text{${ \boldsymbol x}$}; \mathbf{0} , \sigma_{\text{broad}} ) 
		\delta_{\text{Dirac}}(| \text{${ \boldsymbol u}$}_1 | -1 )
		 \delta_{\text{Dirac}}( \text{${ \boldsymbol u}$}_1 - \text{${ \boldsymbol u}$}_2 ) \Big) \\
 &  
\left( 
\begin{diagram}[size=1em]
\; \CIRCLE_1 &  \\
\end{diagram}
\right) 
\llangle
(\text{${ \boldsymbol x}$}_1,\text{${ \boldsymbol u}$}_1)
\rrangle
    \longleftrightarrow  
\left( 
\begin{diagram}[size=1em]
\; \blacksquare_1  &  \\
\end{diagram}
\right)  
\llangle
(\text{${ \boldsymbol x}$}_1,\text{${ \boldsymbol u}$}_1)
\rrangle
\\  &  
\quad \quad \text{\boldmath $\mathbf{with}$} \ \ 
	( \hat{\rho }_{\text{retract} \leftarrow \text{growth}} ,
	\hat{\rho }_{\text{growth} \leftarrow \text{retract}} )   \\
\end{split}
\label{bundling_rule_2}
\end{equation}

A corresponding textual presentation of the dynamic graph grammar (DGG) is here:

\begin{equation}
\begin{split}
\operatorname{grow\_end}[ \text{curr}, \text{${\boldsymbol x}, {\boldsymbol u}$}, S_{\text{in}},\varnothing] 
\longrightarrow &
\operatorname{internal}[ \text{curr},  \text{${\boldsymbol x}, {\boldsymbol u}$}, S_{\text{in}}, \{ \text{new}\}] , 
\\ &
\operatorname{grow\_end}[ \text{new}, \text{\boldmath ${\boldmath x}$}+\Delta  \text{\boldmath $x$},\text{${\boldsymbol u}$}^{\prime},\{ \text{curr}\},  \varnothing ] ,
\\  & \text{\boldmath $\mathbf{with}$} \ \ 
	\hat{\rho }_{\text{grow}}( [\text{tubulin}])
	\mathcal{N}( \Delta  \text{${ \boldsymbol x}$}; L \text{${ \boldsymbol u}$}, \sigma ) \\
	& \quad \times \mathcal{N}( \text{${\boldsymbol u}$}^{\prime}; \text{${ \boldsymbol u}$} /(|\text{${ \boldsymbol u}$} |+\epsilon), \epsilon) , \\
\operatorname{ retract\_end}[ \text{prev}, \text{${\boldsymbol x}_{\text{p}}, {\boldsymbol u}_{\text{p}}$}, S_{\text{in}}, \{ \text{curr}\} ] ,
\\ 
\operatorname{internal}[ \text{curr},  \text{${\boldsymbol x}, {\boldsymbol u}$}, \{ \text{prev}\}, S_{\text{out}}] 
\longrightarrow &
\operatorname{ retract\_end}[ \text{curr}, \text{\boldmath ${\boldmath x}$},\text{${\boldsymbol u}$}^{\prime}, \varnothing , S_{\text{out}}],
\\  & \text{\boldmath $\mathbf{with}$} \ \ 
	\hat{\rho }_{\text{retract}} \\
\operatorname{internal}[ \text{prev}, \text{${\boldsymbol x}_{\text{p}}, {\boldsymbol u}_{\text{p}}$}, S_{\text{in}}, \{ \text{curr}\} ] ,
\\ 
\operatorname{internal}[ \text{curr},  \text{${\boldsymbol x}, {\boldsymbol u}$}, \{ \text{prev}\}, \{ \text{next}\}] ,
\\ 
\operatorname{internal}[ \text{next},  \text{${\boldsymbol x}_{\text{n}}, {\boldsymbol u}_{\text{n}}$}, \{ \text{curr}\}, S_{\text{out}}] ,
\\ 
\operatorname{grow\_end}[ \text{sport}, \text{${\boldsymbol x}, {\boldsymbol u}_{\text{s}}$}, S_{\text{in}}^{\prime},\varnothing] 
\longrightarrow &
\operatorname{internal}[ \text{prev}, \text{${\boldsymbol x}_{\text{p}}, {\boldsymbol u}_{\text{p}}$}, S_{\text{in}}, \{ \text{curr}\} ] ,
\\ &
\operatorname{junct}[ \text{curr},  \text{${\boldsymbol x}, {\boldsymbol u}$}, \{ \text{prev}, \text{sport}\}, \{ \text{next}\}] ,
\\ &
\operatorname{internal}[ \text{next},  \text{${\boldsymbol x}_{\text{n}}, {\boldsymbol u}_{\text{n}}$}, \{ \text{curr}\}, S_{\text{out}}] ,
\\ &
\operatorname{internal}[ \text{sport}, \text{${\boldsymbol y}, {\boldsymbol u}_{\text{s}}$}, S_{\text{in}}^{\prime},  \{ \text{curr}\}] 
\\  & \text{\boldmath $\mathbf{with}$} \ \ 
	 \hat{\rho }_{\text{bundle}} (| \text{$ {\boldsymbol u}$} \cdot \text{${\boldsymbol u}_{\text{s}}$} |/ |\cos \theta_{\text{crit}}|) 
	 		\\ & \quad \times \exp(-| \text{${\boldsymbol x}$} - \text{${\boldsymbol y}$} |^2/2 L^2) \\
\operatorname{ retract\_end}[ \text{prev}, \text{${\boldsymbol x}_{\text{p}}, {\boldsymbol u}{\text{p}}$}, S_{\text{in}}, \{ \text{curr}\} ] , \\ 
\operatorname{grow\_end}[ \text{curr}, \text{${\boldsymbol x}, {\boldsymbol u}$}, S_{\text{in}},\varnothing]  
\longleftrightarrow & \varnothing
\;  \\  & 
	\text{\boldmath $\mathbf{with}$} \ 
	( \hat{\rho }_{\text{retract}} ,
	\hat{\rho }_{\text{nucleate}}( [\text{tubulin}] ) )  \\
		& \quad \mathcal{N}(  \text{${ \boldsymbol x}$}; \mathbf{0} , \sigma_{\text{broad}} ) 
		\delta_{\text{Dirac}}(| \text{${ \boldsymbol u}$}|-1 )  \\
\operatorname{grow\_end}[ \text{curr}, \text{${\boldsymbol x}, {\boldsymbol u}$}, S,\varnothing]  \\ 
\longleftrightarrow 
& \operatorname{ retract\_end}[ \text{curr}, \text{${\boldsymbol x}_{\text{p}}, {\boldsymbol u}$}, \varnothing, S ]  
\;  \\  & 
	\text{\boldmath $\mathbf{with}$} \ 
	(\hat{\rho }_{\text{growth} \rightarrow \text{retract}},
	 \hat{\rho }_{\text{retract} \rightarrow \text{growth}}  )   \\
\end{split}
\end{equation}

Here the notation
``$\operatorname{grow\_end}[ \text{curr}, \text{${\boldsymbol x}, {\boldsymbol u}$}, S_{\text{in}},\varnothing] $''
is equivalent to \\
``$\CIRCLE[ \text{curr}, \text{${\boldsymbol x}, {\boldsymbol u}$}, S_{\text{in}},\varnothing] $''
or to
``$\operatorname{segment}[ \CIRCLE, \text{curr}, \text{${\boldsymbol x}, {\boldsymbol u}$}, S_{\text{in}},\varnothing] $'', etc..
The parameter $ [\text{tubulin}]$ is the concentration of tubulin dimer, here taken to be a constant
but in general dynamic.
Parameter $\epsilon$ ($0<\epsilon \ll 1$) represents a small amount of stochastic ``wobble'' in the direction vector ${\boldsymbol u}$ 
per increment $\approx L$ in MT length; approximately unit  vectors ${\boldsymbol u}$  are renormalized before use in the first rule,
with protection against division by zero.
Spatial step standard deviation parameter $\sigma$ could be on the order of $L \epsilon$.
Notation $\mathcal{N}(x; \mu, \sigma)$ denotes a vector Gaussian or Normal distribution 
for vector $x$ with vector mean $\mu$ and standard deviation $\sigma$ with covariance matrix proportional to the identity.
The last two rules use bidirectional arrows as
shorthand notation for a pair of ordinary unidirectional rules.

A key feature of this dynamical graph grammar is that a large number of growth and/or retraction steps 
(essentially, MT treadmilling in bundles)
are expected to occur in between any of the events that change the number or nature of MTs:
collision-induced junction formation, or MT birth or death in the fourth (bidirectional) rule.
These first two MT length-altering but not MT number-altering rules
form an solvable subsystem - analytically solvable as a recursion equation
in the case of flat 2D geometry
and noise $\epsilon \rightarrow 0$, otherwise a form of random walk which is also tractable
analytically and easy to sample numerically.
So the system decomposition $W_{\text{MT}} = W_{\text{treadmilling}} + W_{\text{birth/death/bundling}}$
can be approached numerically by the (operator-algebraic) TOPE
method mentioned in Section \ref{deqr}.
In the zero noise limit, one could approximate the treadmilling graph grammar rules
by differential equation rules in a different graph grammar that
for example omits the interior nodes entirely but adds real-valued length 
parameters functions to one or both end nodes.
This method would properly consider the effect of time-varying treadmilling velocities
and other time-dependent propensities.
If in addition the treadmilling velocities are constant then the straight line MT trajectories can 
be projected analytically to their intersections in space and time, 
which would result in an event-driven
simulation algorithm similar to that of (Tindemans et al. 2014).

\paragraph{Graph rewrite rule theory}  

{\it In general} now, suppose $G $ and $G^{\prime} $ are numbered graphs 
sharing a common numbering of their vertices,
with index-ordered 
adjacency matrices  $[g_{p q} | p, q]$ and $[g^{\prime}_{p^{\prime} q^{\prime}} | p^{\prime}, q^{\prime}]$
whose elements take values in $\{0,1\}$.
Then an unambiguous graph rewrite rule can be expressed as:
\begin{equation}
G \llangle {\lambdab} \rrangle \rightarrow G^{\prime} \llangle {\lambdab^{\prime}} \rrangle .
\label{DGG_syntax}
\end{equation}
where the double angle brackets denote label substitutions:
$\sigma(1) \mapsto \lambda_1, \sigma(2) \mapsto \lambda_2, \dots $
and
$\sigma^{\prime}(1) \mapsto \lambda^{\prime }_1, \sigma^{\prime }(2) \mapsto \lambda^{\prime }_2, \dots $
where $\sigma: {\mathbb N} \rightarrow {\mathbb N} $
and $\sigma^{\prime}: {\mathbb N} \rightarrow {\mathbb N} $
are strictly monotonically increasing mappings of initial segments of integers into the shared index space.

Given the foregoing graph rewrite rule syntax
for graphs $G$ and $G^{\prime}$ with adjacency matrices $g$ and $g^{\prime}$,
we now define an operator algebra semantics
sufficient to bring all such graph rewrite rules into the general operator algebra/master equation 
formalism of previous sections. This is 
of course 
necessary to 
incorporate
all the capabilities
of the previously discussed languages
by summing up time evolution operators of the corresponding old and new kinds.

We give graph grammar rule 
operator
semantics
for the case of directed graphs
in terms of binary state vectors for
node and edge existence; 
in this case creation and annihilation operators all have dimension $2 \times 2$.
As before, the notation is that indices may have primes or subsubscripts and are usually deployed as follows:
$r$ indexes rewrite rules, $i$ and $j$ index individual domain objects (now nodes or vertices in a graph),
$\alpha$ and $ \beta$ are generic indices or multiple indices,
and $p$ and $q$ index elements in either side of a rule.
Assuming there is at least one label
in the label space 
that
can be used to indicate node allocation from available memory, 
and provided the global state is initialized to have zero probability of active edges for unused nodes, 
then the off-diagonal portion of the operator algebra semantics is
slightly modified from (Mjolsness 2010):
\begin{equation}
\begin{split}
{\hat{W}}_{r}
\;  \propto
\; \rho _{r}({\lambdab} , {\lambdab^{\prime }}) &
\sum \limits_{{\langle i_{1},... i_{k}\rangle}_{\neq}}
\Bigg[  \prod \limits_{p \in \operatorname{lhs}(r) \setminus  \operatorname{rhs}(r)} 
 \Big( \prod_i E_{i_{p} \; i}  E_{i \; i_{p}} \Big) 
\Bigg]  \\ & \times
\Bigg [ \prod \limits_{p^{\prime}, q^{\prime} \in \operatorname{rhs}( r) } {\left( {\hat{a}}_{i_{p^{\prime}}
i_{q^{\prime}}}\right) }^{{g^{\prime }}_{p^{\prime} \; q^{\prime} }} \Bigg] 
\Bigg[  \prod \limits_{p^{\prime} \in
\operatorname{rhs}( r) }  
{\hat{a}}_{i_{p^{\prime}} {\lambda^{\prime }}_{p^{\prime}}} \Bigg] \\ & \times
\Bigg[ \prod \limits_{p, q \in \operatorname{lhs}( r) } {\left( a_{i_{p}
i_{q}}\right) }^{g_{p \; q}} \Bigg] 
\Bigg[  \prod \limits_{p \in \operatorname{lhs}(r) }  
a_{i_{p} \lambda_{p}} \Bigg] . %
\end{split}
\label{XRef-Equation-818134830}
\end{equation}
The sum over indices
$\sum_{\langle i_{1},... i_{k}\rangle_{\neq}}$
means that none of the indices are allowed to be equal to any of the others, 
in the sum.
As in Equation~(\ref{XRef-Equation-922212022}) there could also be an integration over 
all the possible values of some rule variables, in this case a subset of the incoming and outgoing
labels ${\lambdab}$ and $ {\lambdab^{\prime }}$.
\begin{equation}
{\hat{W}}_{r} = \int d X    {\hat{W}}_{r}({\lambdab}(X) , {\lambdab^{\prime }}(X)) 
\label{lambda_integral}
\end{equation}
where $ {\hat{W}}_{r}({\lambdab}, {\lambdab^{\prime }})$ is given by Equation~(\ref{XRef-Equation-818134830}).
However, we will not have occasion to use this extra flexibility.

Here in Equation~(\ref{XRef-Equation-818134830})
we note that the set $\Lambda$ of available labels $\lambda$
can be augmented with an extra unique formal label $\varnothing$
to make the full label set $\Lambda^+ = \Lambda  \mathbin{\dot{\cup}} \{ \varnothing \}$.
Then we can enforce a unary ``winner-take-all'' (WTA) or ``one-hot''
encoding by coupling each $\Lambda$ label addition or removal with
an opposite change of state for the $\varnothing$ label:
$ a_{i , \, \lambda}  = {\hat b}_{i , \,\varnothing}  b_{i , \,\lambda} $, 
$\hat{a}_{i , \, \lambda}  =   \hat{b}_{i , \,\lambda} b_{i , \, \varnothing} $.
After using $b_{\alpha}^2 = 0 = \hat{b}_{\alpha}^2$,
and since $b$ operators with different indices commute, the form of
Equation~(\ref{XRef-Equation-818134830}) is preserved by this mapping.
Under it, inductively every vertex $i$ will have 
exactly one label $\lambda \in \Lambda^+$ present, $n_{i \lambda}=1$;
this constrains the subspace that the dynamical system will move through.
A vertex $i$ is regarded as ``allocated'' or ``active'' if  any of its label states 
$(i, \lambda \in \Lambda)$ are present ($n_{ i \lambda}=1$ for some $\lambda \in \Lambda$); otherwise not;
in other words, if it has $n_i = \sum_{\lambda \in \Lambda} n_{i \lambda}=1$  
rather than zero.  
Since $\sum_{\lambda \in \Lambda} n_{i \lambda} + n_{i, \varnothing} =1$ is an invariant of dynamics using this WTA encoding,
then for a (partly or completely) unlabelled graph
 the presence of the extra ``$\varnothing$'' label will indicate a vertex is inactive.

The main four factors in lines 2 and 3 of 
Equation~(\ref{XRef-Equation-818134830})
are as in (Mjolsness 2010) and act in an
analogous way to the previous rule semantics definitions:
first (reading operator products right-to-left) 
all the vertex labels, hence all the vertices, of the incoming (LHS) graph are annihilated in an arbitrary order,
``then'' (instantaneously) all the edges of the incoming graph are annihilated in an arbitrary order,
then all the vertices and vertex labels of the outgoing (RHS) graph are created,
and then all the edges of the incoming graph are created.
However, these ordered operations all happen with zero time delay in the model, and
with the same binding of the indices $i_*$.
We note that the RHS vertex indices $i_{p^{\prime}}$ get assigned uniquely since $(\hat{a}_i)^2 = {\mathbf 0}$.

In addition to the main four factors,
we now consider allocation and deallocation of integer-valued graph vertex indices $i$ from a single central index list.
The leading (rightmost) operator
could be preceded by a rightmost memory-checking ``pre-factor'' such as
$ X_r = \big[  \prod \limits_{p \in \operatorname{rhs}(r) \setminus  \operatorname{lhs}(r)} Z_{i_{p} \lambda_{p}} \big ]$
to ensure that only currently unused memory gets allocated for new use,
but since $\hat{a}_{i \lambda} Z_{i \lambda}  = \hat{a}_{i \lambda} $ and $\hat{a}_{i \lambda} $ is present in the second line and unobstructed by $a_{i \lambda} $ in the third line
whenever it is present in this pre-factor,
and since we are in sector ${\cal S}_\text{graph}$ with at most one label per node activated,
such a pre-factor is not needed.

The indices $i_p$ should be each averaged rather than summed over,
since it doesn't matter {\it which} unused node is brought into service;
hence the proportionality rather than equality in (\ref{XRef-Equation-818134830}).
If only a finite range of index values $i_p \in \{0, \ldots I_{\rm overflow}-1\}$ are possible and they are all used up,
then only rewrite rules that don't create new nodes will be able to fire.
One is naturally interested in the infinite storage space limit $I_{\rm overflow} \equiv N_{\rm globalmax} \rightarrow +\infty$
for which this problem doesn't occur;
in that case one could argue that the
$E$ post-factors (in right to left order) become unimportant 
since for any finite state of the graph rewriting system,
the chance of accidentally reusing a node approaches zero.
Here however we will conservatively retain the $E$ post-factors on line 1. 
These trailing (leftmost) ``garbage collection''  erasure operators 
$ \Big( \prod_i E_{i_{p} \; i}  E_{i \; i_{p}} \Big)$
erase any edges $(i_p,  i)$ or $(i,  i_p)$ 
dangling from 
or associated with
nodes $i_p$ that have just
been deleted (factor $a_{i_{p} \lambda}$ on the third line), making those nodes suitable for reuse in future rule firings.
An erasure factor $E_{i_{p} \lambda_p}$ is not needed since $E_{\alpha}  a_\alpha = a_\alpha$, and $a_{i_{p} \lambda_p}$ is present on line 3
and unobstructed by $\hat{a}_{i_{p} \lambda_p}$ (for $p \in \text{rhs}(r)$) on line 2 whenever it would be present on line 1.

These erasure operators maintain the invariance of the statements that 
(a) every vertex has either 0 or 1 labels present (1 defining an ``active'' vertex), and
(b) edges that are present must connect two active vertices.
In operator-state language the predicate $S_{\text{graph}}$  is:
\begin{subequations}
\begin{align}
\forall_i \quad &\Big(\sum_{\lambda \in \Lambda} N_{i \lambda} \Big) \ket{\nb} = n_i \ket{\nb}, \quad n_i \in \{0,1\} \\
\forall_{i,j}  \quad &N_{i j} \ket{\nb} =  n_{i j}\ket{\nb}, \quad n_{i j} \leq n_i n_j
\end{align}
\label{graph_sector}
\end{subequations}
(Here the label space $\Lambda$ could be designed as a product space 
$\Lambda = \otimes_A \Lambda_A$, so no intrinsic expressivity is lost under condition (a)).
We assume these invariant statements are true of the initial condition; 
for example they are true in 
the ``vacuum'' state in which all $n_*=0$,
and of course in any other 
state reachable from the vacuum by one or more grammar rule firings.
These conditions $S_{\text{graph}}$ define a linear subspace ``sector'' ${\cal S_{\text{graph}}}$ of the Fock space ${\cal{F}}$.
Starting from the null space $\ket{\nb = {\mathbf 0}}$, this sector is preserved by the operators $\hat{W}_r$.
The exact form of the $X,E$ pre- and post- factors (though we don't use $X$ here), 
and the invariant predicate ${\cal S_{\text{graph}}}$ they maintain,
can be formulated in several different ways that differ just by the hidden indices $i_p$
that end up representing particular labelled graphs.

We may calculate $D_r   \equiv  {\rm diag}({\bf 1} \cdot \hat{W}_r) $ 
for Equation~(\ref{XRef-Equation-818134830})
using ${\bf 1} \cdot E = {\bf 1}$,
where $G_r$ is the labelled graph on the LHS of rule $r$:
\begin{equation}
D_r = N^{(G_r)} \equiv
\sum \limits_{{\langle i_{1},... i_{k}\rangle}_{\neq}}
 \Bigg[ \prod_{p \in {\text{lhs}_r}}
 N_{i_{p} \; {\lambda}}  \Bigg] 
 \Bigg[ \prod \limits_{p, q \in \operatorname{lhs}( r) } {\left( N_{i_{p}
i_{q}}\right) }^{g_{p \; q}} \Bigg] .
\label{graph-number_op}
\end{equation}

Equations (\ref{DGG_syntax}) and (\ref{XRef-Equation-818134830}),
elaborated here from (Mjolsness 2010) which omitted 
$E$ factors,
provide the built-in syntax and semantics for 
the basic ruleset portion of a proposed Dynamical Graph Grammar (DGG) generalization of 
Stochastic Parameterized Grammars (\ref{XRef-Equation-924145912}) and (\ref{XRef-Equation-922212022})
or 
Dynamical Grammars (\ref{XRef-Equation-924145912}),(\ref{XRef-Equation-9288735}) 
and (\ref{XRef-Equation-922212022}),(\ref{XRef-Equation-615123838}),
all as outlined in (Mjolsness 2010, Mjolsness and Yosiphon 2006, Mjolsness 2005).
The latter two references also map graph grammars to operator algebras by way of 
unique ObjectID node labels - similar to indices $i_p$ in that all that matters about them
for the graph grammar is not their numerical values, but just whether two such numbers are equal or not.

Equivalence of models, previously defined by ``particle equivalence'' under master equation semantics in
Equation~(\ref{particle-equivalence}), must be modified for this graph grammar semantics
to account for the fact that the index numbering $i_p$ of graph vertices is unimportant; 
only their labels and edges matter.
Any permutation of indices should yield an equivalent state, as should
any subpermutation which also alters the choice of which indices are active,
holding fixed their number.
So we would like to define ``equivalent'' models in an index subpermutation-invariant way.
Subpermutation invariance (with respect to indexing) is achieved by using the index-invariant number operators 
$N^{G(q)}$ of Equation~(\ref{graph-number_op})
for each labelled graph $G_q$ in any collection indexed by $q$ (usually taken to be finite or at least countable),
and then seeking equality of all joint probabilities:
\begin{equation}
\text{Pr}_{\text{\tiny graph}}([n_{(G_q)}(t_q)|q]) =  \Big \langle \prod_q \delta (N^{(G_q)}(t_q)  -n_{(G_q)}(t_q) I_{G(q)}) \Big \rangle_{\text{\tiny graph}},
\label{graph-equivalence}
\end{equation}
just as in Equation~(\ref{particle-equivalence}), but with graph number operators now defined
in the index-subpermutation-invariant manner of Equation~(\ref{graph-number_op}).
As before, other graph grammar observables $\langle f([N^{(G_q)}(t_q) | q ] ) \rangle $ can be defined from these,
where any function $f$ acts on the components of a diagonal matrix (which is consistent with any polynomial expansion or power series $f$ may have).
The resulting equivalence relation on labelled graph grammar models could be called {\it graph-equivalence}.
As a further elaboration of the graph case, graph-equivalence could be restricted
to some sector ${\cal S}$ of the Fock space, preserved under time evolution of a model,
for example ${\cal S_{\text{graph}}}$ defined above.
By a natural extension, two operators are graph-equivalent if they can be added to a third operator in common,
yielding two graph-equivalent model time-evolution operators.

Closely related concepts have been proposed 
recently by (Behr et al. 2016)
and nicely re-expressed, 
{\it without} an explicit analog of our grounding 
and hence implementation map ${\cal I}$
in terms of elementary creation/annihilation operators $\hat{a},a$,
but instead more abstractly.
As in (Mjolsness 2010), they
suggest that Fock spaces and master equations built
from a collection of rule-level operators 
generalizing the Heisenberg algebra creation/annihilation operators 
could express graph rewrite rules.
They introduce operator representations similar to our number basis
by way of many graph-counting ``observable'' operators 
$\hat{W}_{\text{LHS}_r \rightarrow \text{LHS}_r }$ %
($ = N^{(\text{LHS}_r)}= D_r$ in our notation - see Corollary 2 below).
They also derive differential equation dynamics for graph moments in terms of other, 
generally higher order moments, which raises the classic moment closure problem of statistical mechanics.
(Behr et al. 2016) also derive their operator construction from 
the ``double pushout'' category-theoretic approach to defining graph grammar semantics,
which we will discuss briefly in Section~\ref{sec_pushout} below.

By comparison, in this work we express $\hat{W}_r$ in terms of products of elementary 
creation and annihilation operators and thus provide an explicit implementation in
terms of Boolean and/or integer-valued random variables.
This implementation map 
(a) enables the integration of graph rewrite semantics with non-graph modeling language semantics defined in previous sections; 
(b) enables the mechanical computation of fundamental commutation relations for graph rewrite rules as in Propositions 1 and 2 and the Corollaries below,
thus in principle permitting the derivation and
analysis of operator splitting simulation methods underwritten
by e.g. the Baker-Campbell-Hausdorff (BCH) theorem;
and (c) supports the derivation of model reduction based moment closure methods
such as those of Section~\ref{sec_mod_red} below.
In addition we include labels for the graph vertices and show how
other categories of spatially extended objects can be 
mapped to and hence implemented in terms of
such labelled graphs.

As is the case for Equation (\ref{XRef-Equation-62212652}),  Equation (\ref{XRef-Equation-818134830})
can be taken as a normal form for rewrite rule dynamics, but now applied to graphs.
We will show next that
(somewhat akin to the ``Concurrency theorems'' of the double pushout approach 
discussed in Section~\ref{sec_pushout} below,
but more general)
the product and commutator
of two such forms can be rewritten as a (possibly large) integer-weighted
sum of expressions having the same form, or a form of equivalent meaning
with extra factors of 
$E$ that don't affect the active node set;
however, some of the weights may be negative.

\subsubsection{Product of Graph Grammar Rules}
\label{graphgramprod}

We approach the mulitplication of grammar rule operators in two steps.
First we consider the simpler 
form omitting cleanup post-factors:
\begin{equation}
\begin{split}
{\hat{W}}_{r} \; \propto
\; \rho _{r}({\lambdab} , {\lambdab^{\prime }}) &
\sum \limits_{\langle i_{1},... i_{k}\rangle_{\neq} } 
\Bigg [ \prod \limits_{p^{\prime}, q^{\prime} \in \operatorname{rhs}( r) } {\left( {\hat{a}}_{i_{p^{\prime}}
i_{q^{\prime}}}\right) }^{{g^{\prime }}_{p^{\prime} \; q^{\prime} }} \Bigg] 
\Bigg[  \prod \limits_{p^{\prime} \in
\operatorname{rhs}( r) }  
({\hat{a}}_{i_{p^{\prime}} {\lambda^{\prime }}_{p^{\prime}}})^{h^{\prime}_{p^{\prime}}} \Bigg] \\ & \times
\Bigg[ \prod \limits_{p, q \in \operatorname{lhs}( r) } {\left( a_{i_{p}
i_{q}}\right) }^{g_{p \; q}} \Bigg] 
\Bigg[  \prod \limits_{p \in \operatorname{lhs}(r) }  
(a_{i_{p} \lambda_{p}})^{h_{p}} \Bigg] . %
\end{split}
\label{weaker_form}
\end{equation}
where $h_{i_p} \in \{0,1\}$ is an indicator function for inclusion of vertex $i_p$ independent of its edges.
Again, the sum over indices
$\sum_{\langle i_{1},... i_{k}\rangle_{\neq}}$
means that none of the indices are allowed to be equal to any of the others, 
in the sum.
This form can leave and/or delete undeleted hanging edges, owing to the lack of erasure post-factor.
If all $h_{i_p}=1$ this is the form used in (Mjolsness 2010).
The advantages of this form are that it is (a) subpermutation-invariant with respect to indexing, 
like Equation~(\ref{XRef-Equation-818134830}), and 
(b) already in normal form (monomial in $\hat{a}_*$ times monomial in $a_*$)
and therefore the product of two such expressions takes the same general form, 
by repeatedly using  the basic commutation relations
of Section~\ref{pure_rules} or 
Equation~(\ref{binary_algebra_1}d):

\begin{prop}
The product of two operators taking the form of
Equation (\ref{weaker_form}) 
can be rewritten as an
signed-integer-weighted
sum of expressions taking the same form.
The product and the sum are 
equal, and graph-equivalent, and each is subpermutation-invariant with respect to indexing.
\end{prop}

{\it Proof:}
First we remark that the form of Equation~(\ref{weaker_form}), in which all combinations
of index values are summed over, is related by an invertible linear bijection with integer coefficients
to sums of monomial basis operators
(again with creation/annihilation exponents forced to be 0 or 1 since higher exponents result in zero operators)
in which none of the indices $\left\{ i_{1},... i_{k}\right\}$ are allowed to be equal. 
In the forward linear map, an unrestricted sum over indices maps to a sum of
index-equality-restricted sums over unequal indices;
these sums over unequal indices each have weight one and are in 1-1 correspondence
with the partitions of $k$ indices into groups 
that are constrained to be equal within a group
and unequal to other groups, resulting in sums of sums each taking the form
$\sum_{\langle j_{1},... j_{k^{\prime} \leq k}\rangle_{\neq}} \ldots$ .
Since there is only one partition of $k$ into $k$ different blocks,
by induction on $k$
this linear map can be inverted by successive substitutions
resulting again in integer coefficients.

Thus it suffices to prove the proposition for the related special form:
\begin{equation}
\begin{split}
{\hat{W}}_{r} \; \propto
\; \rho _{r}({\lambdab} , {\lambdab^{\prime }}) &
\sum \limits_{\left\{ i_{1},... i_{k}\right\}
} 
\Bigg [ \prod \limits_{p^{\prime}, q^{\prime} \in \operatorname{rhs}( r) } {\left( {\hat{a}}_{i_{p^{\prime}}
i_{q^{\prime}}}\right) }^{{g^{\prime }}_{p^{\prime} \; q^{\prime} }} \Bigg] 
\Bigg[  \prod \limits_{p^{\prime} \in
\operatorname{rhs}( r) }  
({\hat{a}}_{i_{p^{\prime}} {\lambda^{\prime }}_{p^{\prime}}})^{h^{\prime}_{p^{\prime}}} \Bigg] \\ & \times
\Bigg[ \prod \limits_{p, q \in \operatorname{lhs}( r) } {\left( a_{i_{p}
i_{q}}\right) }^{g_{p \; q}} \Bigg] 
\Bigg[  \prod \limits_{p \in \operatorname{lhs}(r) }  
(a_{i_{p} \lambda_{p}})^{h_{p}} \Bigg] . %
\end{split}
\label{weaker_form_tangled}
\end{equation}

The product of two expressions of the form of Equation~(\ref{weaker_form})
initially takes the general form
\begin{equation}
\begin{split}
{\hat{W}}_{r_2} {\hat{W}}_{r_1} \; & \propto \; 
\big( \rho _{r_1}({\lambdab_1} , {\lambdab_1^{\prime }}) \rho _{r_2}({\lambdab_2} , {\lambdab_2^{\prime }}) \big) 
\sum \limits_{\left\{ i_{1},... i_{k_1}\right\}} 
\sum \limits_{\left\{ j_{1},... j_{k_2}\right\}} 
 \\  & 
\Bigg [ \prod \limits_{p^{\prime}, q^{\prime} \in \operatorname{rhs}( r_2) } {\left( {\hat{a}}_{i_{p^{\prime}}
i_{q^{\prime}}}\right) }^{{g^{\prime }}_{2, p^{\prime} \; q^{\prime} }} \Bigg] 
\Bigg[  \prod \limits_{p^{\prime} \in
\operatorname{rhs}( r_2) }  
({\hat{a}}_{i_{p^{\prime}} {\lambda^{\prime }}_{2, p^{\prime}}})^{h^{\prime}_{2, {p^{\prime}}}} \Bigg] 
\\ & \times
\Bigg[ \prod \limits_{p, q \in \operatorname{lhs}( r_2) } {\left( a_{i_{p}
i_{q}}\right) }^{g_{2, p \; q}} \Bigg] 
\Bigg[  \prod \limits_{p \in \operatorname{lhs}(r_2) } 
(a_{i_{p} {\lambda}_{2, p}})^{h_{2, {p}}} \Bigg] 
\\  & \times%
\Bigg [ \prod \limits_{p^{\prime}, q^{\prime} \in \operatorname{rhs}( r_1) } {\left( {\hat{a}}_{j_{p^{\prime}}
j_{q^{\prime}}}\right) }^{{g^{\prime }}_{1, p^{\prime} \; q^{\prime} }} \Bigg] 
\Bigg[  \prod \limits_{p^{\prime} \in
\operatorname{rhs}( r_1) }  
({\hat{a}}_{j_{p^{\prime}} {\lambda^{\prime }}_{1, p^{\prime}}})^{h_{1, {p^{\prime}}}} \Bigg] 
\\ & \times
\Bigg[ \prod \limits_{p, q \in \operatorname{lhs}( r_1) } {\left( a_{j_{p}
j_{q}}\right) }^{g_{1, p \; q}} \Bigg] 
\Bigg[  \prod \limits_{p \in \operatorname{lhs}(r_1) } 
(a_{j_{p} {\lambda}_{1, p}})^{h_{1, {p}}} \Bigg] ,
\label{super_naive_product}
\end{split}
\end{equation}
with each $g,h \in \{0,1\}$.
Recall that all $a,\hat{a}$ commutators are either zero, when operator types or indices don't match,
or they are diagonal and a linear combination of the identity and a normal form $N =  \hat{a} a $ matrix,
multiplied by a delta function that eliminates one or more indices from the sum over indices.

We repeatedly commute factors of $\hat{a}$ on line 4 to the left of factors of $a$ on line 3 
until normal form - all creation operators to the left of all annihilation operators - is restored.
Each out-of-order product $a_{\alpha} \hat{a}_{\beta}$ is replaced in turn using
Equation~(\ref{binary_algebra_1}d), 
which has 
several summands;
distributing multiplication over addition at each step,
each summand has the property of reducing the total (finite)
number of out-of-order pairs by at least one pair; 
convergence to termination at a  
finite sum of operator terms generated by symbolic commutation, 
each in normal order, 
and each subpermutation invariant due to the sum over its remaining indices, is thus ensured by induction.
Upon termination each elementary operator $\hat{a}_{\alpha}$ or $a_{\alpha}$ will
appear linearly, to the power zero or one, in each summand,
since $a_{\alpha}^2={\mathbf 0}=\hat{a}_{\alpha}^2$ for all indices ${\alpha}$;
this fact eliminates summands that aren't multilinear 
(since all creation operators commute with each other 
and hence can be collected by subscript within the normal form,
and likewise for annihilation operators.)
Thus the final operator expression is multilinear in these elementary operators, 
up to and including the final normal form.
From that fact one can pick out appropriate 0/1-valued matrices 
$g_{1;2}$ and $g^{\prime}_{1;2}$ on the edge labels,
and 0/1-valued vectors $h_{1;2}$ and $h^{\prime}_{1;2}$ on the vertices.
Likewise with labelled vertices.
Operator equality implies graph-equivalence.
Subpermutation invariance of each normal order operator term,
and therefore of both the starting product ${\hat{W}}_{r_2} {\hat{W}}_{r_1} $ 
and the ending sum of such terms,
follows from 
the outer sum over all indices $i_*, j_*$ that remain in each operator term,
after some indices have been eliminated by Kronecker delta factors.
QED.

\vspace{10pt}

Alternatively 
as in the graph rule semantics of Equation~(\ref{XRef-Equation-818134830})
we may wish to eliminate hanging edges as part of the mechanics
of the grammar operator algebra. So we consider the more general form:
\begin{equation}
\begin{split}
{\hat{W}}_{r} \; \propto
\; \rho _{r}({\lambdab} , {\lambdab^{\prime }}) &
\sum \limits_{\langle i_{1},... i_{k}\rangle_{\neq}}
\Bigg[ 
 \Big(  \prod \limits_{p \in B_r} \; \prod_{ i \neq i_q | \forall q \in \bar{B}_{r p}} E_{i_{p} \; i}  \Big) 
  \Big(  \prod \limits_{p \in C_r} \; \prod_{ i  \neq i_q | \forall q \in \bar{C}_{r p} } E_{i \; i_{p}} \Big) 
\Bigg]  \\ & \times
\Bigg [ \prod \limits_{p^{\prime}, q^{\prime} \in \operatorname{rhs}( r) } {\left( {\hat{a}}_{i_{p^{\prime}}
i_{q^{\prime}}}\right) }^{{g^{\prime }}_{p^{\prime} \; q^{\prime} }} \Bigg] 
\Bigg[  \prod \limits_{p^{\prime} \in
\operatorname{rhs}( r) }  
({\hat{a}}_{i_{p^{\prime}} {\lambda^{\prime }}_{p^{\prime}}})^{h_{p^{\prime}}} \Bigg] \\ & \times
\Bigg[ \prod \limits_{p, q \in \operatorname{lhs}( r) } {\left( a_{i_{p}
i_{q}}\right) }^{g_{p \; q}} \Bigg] 
\Bigg[  \prod \limits_{p \in \operatorname{lhs}(r) }  
(a_{i_{p} \lambda_{p}})^{h_{p}} \Bigg] . %
\end{split}
\label{general_form}
\end{equation}
where sets $B_r$, $C_r$, $\bar{B}_r$, and $\bar{C}_r$  are finite (assuming $\text{lhs}(r)$ and $\text{rhs}(r)$ are both finite);
this form encompasses Equation~(\ref{XRef-Equation-818134830})
in which case $B_r = C_r = \text{lhs}(r) \setminus \text{rhs}(r)$ and $\bar{B}_{r p}= \bar{C}_{r p} = \varnothing $ for all $p$.
Owing to the sum over indices $\{i_1, \ldots i_k\}$ is is again subpermutation-invariant.

\begin{prop}
The product of two operators taking the form of
Equation (\ref{general_form}) 
can be rewritten as an
signed-integer-weighted
sum of expressions taking the same form.
The product and the sum are 
equal, and graph-equivalent, and each is subpermutation-invariant with respect to indexing.
\end{prop}

Note: Less formally, the product of two graph rewrite rule operators
is an integer-weighted sum of other graph rewrite rule operators
in a slightly expanded operator algebra.
This proposition establishes 
 {\it an operator algebra of natural graph-rewriting operations},
including all operations in any given collection of graph rewrite rules,
as the operator algebra
generated by the probability inflow operators ${\hat{W}_r}$ for the individual rules.
The probability outflow operators $D_r$ are also encompassed,
by a variant of Equation~(\ref{diagonal_operator}) shown in Corollary 2 below.

{\it Proof}:

As in the proof of Proposition 1,
we first convert the sum over unequal indices into an integer-weighted
sum of sums over arbitrary indices, of the following form:
\begin{equation}
\begin{split}
{\hat{W}}_{r} \; \propto
\; \rho _{r}({\lambdab} , {\lambdab^{\prime }}) &
\sum \limits_{\left\{ i_{1},... i_{k}\right\}
} 
\Bigg[ 
 \Big(  \prod \limits_{p \in B_r} \; \prod_{ i \neq i_q | \forall q \in \bar{B}_{r p}} E_{i_{p} \; i}  \Big) 
  \Big(  \prod \limits_{p \in C_r} \; \prod_{ i  \neq i_q | \forall q \in \bar{C}_{r p} } E_{i \; i_{p}} \Big) 
\Bigg]  \\ & \times
\Bigg [ \prod \limits_{p^{\prime}, q^{\prime} \in \operatorname{rhs}( r) } {\left( {\hat{a}}_{i_{p^{\prime}}
i_{q^{\prime}}}\right) }^{{g^{\prime }}_{p^{\prime} \; q^{\prime} }} \Bigg] 
\Bigg[  \prod \limits_{p^{\prime} \in
\operatorname{rhs}( r) }  
({\hat{a}}_{i_{p^{\prime}} {\lambda^{\prime }}_{p^{\prime}}})^{h_{p^{\prime}}} \Bigg] \\ & \times
\Bigg[ \prod \limits_{p, q \in \operatorname{lhs}( r) } {\left( a_{i_{p}
i_{q}}\right) }^{g_{p \; q}} \Bigg] 
\Bigg[  \prod \limits_{p \in \operatorname{lhs}(r) }  
(a_{i_{p} \lambda_{p}})^{h_{p}} \Bigg] . %
\end{split}
\label{general_form_tangled}
\end{equation}
By integer-weighted linearity we need only prove the Proposition for this form, rather than Equation~(\ref{general_form})

Now the product of two expressions of the form of Equation~(\ref{general_form_tangled})
initially takes the general form:

\begin{equation}
\begin{split}
{\hat{W}}_{r_2} {\hat{W}}_{r_1} \; & \propto \; 
\big( \rho _{r_1}({\lambdab_1} , {\lambdab_1^{\prime }}) \rho _{r_2}({\lambdab_2} , {\lambdab_2^{\prime }}) \big) 
\\ & \times
\sum \limits_{\left\{ i_{1},... i_{k_1}\right\}} 
\sum \limits_{\left\{ j_{1},... j_{k_2}\right\}} 
\Bigg[   
 \Big( \prod \limits_{p \in B^{\prime}_r} \; \prod_{ i   \neq i_q | \forall q \in \bar{B}^{\prime}_{r p}} E_{i_{p} \; i}  \Big)
  \Big( \prod \limits_{p \in C^{\prime}_r} \; \prod_{ i  \neq i_q | \forall q \in \bar{C}^{\prime}_{r p} } E_{i \; i_{p}} \Big) 
\Bigg]  \\ 
& \times
\Bigg [ \prod \limits_{p^{\prime}, q^{\prime} \in \operatorname{rhs}( r_2) } {\left( {\hat{a}}_{i_{p^{\prime}}
i_{q^{\prime}}}\right) }^{{g^{\prime }}_{2, p^{\prime} \; q^{\prime} }} \Bigg] 
\Bigg[  \prod \limits_{p^{\prime} \in
\operatorname{rhs}( r_2) }  
({\hat{a}}_{i_{p^{\prime}} {\lambda^{\prime }}_{2, p^{\prime}}})^{h^{\prime}_{2, {p^{\prime}}}} \Bigg] 
\\ & \times
\Bigg[ \prod \limits_{p, q \in \operatorname{lhs}( r_2) } {\left( a_{i_{p}
i_{q}}\right) }^{g_{2, p \; q}} \Bigg] 
\Bigg[  \prod \limits_{p \in \operatorname{lhs}(r_2) } 
(a_{i_{p} {\lambda}_{2, p}})^{h_{2, {p}}} \Bigg] 
\\  & \times%
\Bigg[   
 \Big( \prod \limits_{p \in B_r} \; \prod_{ i  \neq i_q | \forall q \in \bar{B}_{r p}} E_{i_{p} \; i}  \Big) 
  \Big( \prod \limits_{p \in C_r} \; \prod_{ i  \neq i_q | \forall q \in \bar{C}_{r p} } E_{i \; i_{p}} \Big) 
\Bigg]  \\ 
& \times
\Bigg [ \prod \limits_{p^{\prime}, q^{\prime} \in \operatorname{rhs}( r_1) } {\left( {\hat{a}}_{j_{p^{\prime}}
j_{q^{\prime}}}\right) }^{{g^{\prime }}_{1, p^{\prime} \; q^{\prime} }} \Bigg] 
\Bigg[  \prod \limits_{p^{\prime} \in
\operatorname{rhs}( r_1) }  
({\hat{a}}_{j_{p^{\prime}} {\lambda^{\prime }}_{1, p^{\prime}}})^{h_{1, {p^{\prime}}}} \Bigg] 
\\ & \times
\Bigg[ \prod \limits_{p, q \in \operatorname{lhs}( r_1) } {\left( a_{j_{p}
j_{q}}\right) }^{g_{1, p \; q}} \Bigg] 
\Bigg[  \prod \limits_{p \in \operatorname{lhs}(r_1) } 
(a_{j_{p} {\lambda}_{1, p}})^{h_{1, {p}}} \Bigg] .
\label{naive_product}
\end{split}
\end{equation}

This expression should equal a sum of expressions that 
take the same form as
Equation (\ref{general_form}),
although possibly with negative signs and 
altered
factors of $E$ 
as stated,
for various choices of graph rule matrices
$g_{1;2 v}$ and $g^{\prime}_{1;2 v}$ ($v$ indexing the summands that result)
and likewise for $h$.
The leading factors of $\rho$ multiply properly to give a new leading $\rho$
for each summand, possibly to be multiplied by integers arising from commutation relations.
The $\sum \limits_{\left\{ i_{1},... i_{k_1}\right\}}  \sum \limits_{\left\{ j_{1},... j_{k_2}\right\}} $
will become a new $\sum \limits_{\left\{ i_{1},... i_{k_{1;2} \leq k_1+k_2} \right\}} $
after some nonnegative number of index collisions involving commutators proportional
to $\delta_{i_{p_1} i_{p_2}}$ and $\delta_{j_{q_1} j_{q_2}}$ for various $p$ and $q$ subindices 
reduce the number of indices summed over by demanding index equality.

The proof work occurs in two steps: 
(1) commuting the post-factors of $E$ in line 5 to join those in line 2, 
and 
(2) commuting factors of $\hat{a}$ from line 6 past (to the left of) factors of $a$ in line 4
to restore normal form,
more specifically for (2a) the edges and (2b) the node or vertex labels.
We will discuss these two steps in reverse order,
since Step 2 is simpler.
Indeed, in the absence of 
$E$ factors, the proof of the simplified form of Proposition 2 reduces
to Proposition 1. 

Recall
from Equation~(\ref{binary_algebra_1})
 that all $a,\hat{a}$ commutators are either zero, when operator types or indices don't match,
or they are diagonal and a linear combination of the identity and a normal form $N =  \hat{a} a $ matrix,
multiplied by a delta function that eliminates one or more indices from the sum.

{\it Step 2.}
We repeatedly commute factors of $\hat{a}$ to the left of factors of $a$,
as in the proof of Proposition 1, until normal form is reached.
This determines the 0/1-valued entries of $g, g^{\prime}, h,$ and $h^{\prime}$ for
each summand.

{\it Step 1.}
The products of $E$ factors on line 5 of Equation~(\ref{general_form})
can commute freely to the left and upwards to line 1
except for possible interference with 
\[
\prod \limits_{p^{\prime}, q^{\prime} \in \operatorname{rhs}( r_2) } {\left( {\hat{a}}_{i_{p^{\prime}}
i_{q^{\prime}}}\right) }^{{g^{\prime }}_{2, p^{\prime} \; q^{\prime} }} 
\]
on line 3.
Those $E$ factors that get to line 1 either augment the sets $B^{\prime}$ and/or $C^{\prime}$
(possibly having lost some index values $i$ to Kronecker delta functions with $i_*$, as recorded in sets $\bar{B}^{\prime}$ and/or $\bar{C}^{\prime}$)
if they are new, 
or owing to $(E_{\alpha})^2= E_{\alpha}$ they make no change to line 1.
To determine when $E$ does not commute past line 3,
we express $E_{\alpha}=I_{\alpha}-N_{\alpha}+a_{\alpha} = I_{\alpha} - \hat{a}_{\alpha} a_{\alpha} + a_{\alpha}$
and use the commutation relations
Equation~(\ref{binary_algebra_1}d),
in which the commutation is the first summand and the second and third summands
comprise a correction multiplied by a symbolic Kronecker delta function that enforces the equality of
(in the case of edges) not just one but two index quantities $i_*$ and/or $i$.
So at least one of the sub-index quantities $i_*$ will be removed by the Kronecker delta,
along with its index summation,
ensuring eventual termination in the normal form process as in Step 1 or the proof of Proposition 1.
Thus, the obstructed terms arising from line 5
can be fully absorbed into the (annihilation-first) normal form factors
that obstructed them,
using the creation/annihilation commutators
Equation~(\ref{binary_algebra_1}d) to reduce all factors to normal form,
thus joining in the normal form reduction process of Step 2.
The result is a sum of operator expressions that are
signed integer multiples of expressions in the form of
Equation~({\ref{general_form}). 
Sub-permutation invariance and graph-equivalence are established as in the proof of Proposition 1.
QED.

\vspace{10pt}

Propositions 1 and 2 are probably not the tightest or best 
formulations possible of the respective graph rewrite rule operator algebras,
since they admit a wider class of operators than seems necessary 
(except in for purposes of the proof) 
and since they don't maintain as much control over the signs
of the integer weights as seems possible in particular cases.
But they do show that graph rewrite rule semantics is embedded in
an operator algebra in the manner specified, and in a way that
could be computed automatically.

\vspace{10pt}

{\bf Corollary 1}
The commutator $[\hat{W}_{r_1}, \hat{W}_{r_2}]$ of two operators 
under the semantics of Proposition 2 (taking the form of Equation~(\ref{general_form}))
can also be rewritten as an
integer-weighted
sum of expressions taking the same form.
The product and the sum are 
equal, and graph-equivalent, and each is subpermutation-invariant with respect to indexing.
Likewise, the commutator  $[\hat{W}_{r_1}, \hat{W}_{r_2}]$ of two operators 
under the semantics of Proposition 1 (taking the form of Equation~(\ref{XRef-Equation-818134830}))
can also be rewritten as an
integer-weighted
sum of expressions taking the same form (\ref{XRef-Equation-818134830}).
Compared to the product $\hat{W}_{r_1} \hat{W}_{r_2}$,
however, many summands may cancel
in a commutator.


{\bf Corollary 2}
The product and the commutator of
two full graph rewrite rule operators $W_{r_1}, W_{r_2}$ 
(including their negative diagonal terms $-\hat{D}_{r_1}, -\hat{D}_{r_2}$)
under the semantics of Proposition 2 (taking the form of Equation~(\ref{general_form}))
can also be rewritten as an
integer-weighted
sum of expressions taking the same form.
Likewise the product and the commutator of
two full graph rewrite rule operators $W_{r_1}, W_{r_2}$ 
under the semantics of Proposition 1 (taking the form of Equation~(\ref{XRef-Equation-818134830}))
can also be rewritten as an
integer-weighted
sum of expressions taking the same form. 
In either case, the product (or commutator) and the sum are 
equal, and graph-equivalent, and each is subpermutation-invariant with respect to indexing.

{\it Proof:} The diagonal terms $D_r$ are equal to $\hat{W}_{r^{\prime}}$ for a new rule $r^{\prime}$, 
not included in the model,
in which the LHS of $r$ is both the LHS and the RHS of $r^{\prime}$.
The reason is that the LHS $\setminus$ RHS postfactor and any
RHS $\setminus$ LHS prefactor of $\hat{W}_{r^{\prime}}$ are both empty,
so Equation~(\ref{graph-number_op}) for $D_r$ also equals $\hat{W}_{r^{\prime}}$
from Equation~(\ref{XRef-Equation-818134830}).
Thus $D_r = \hat{W}_{LHS_r \rightarrow LHS_r}$, as in Equation~(\ref{diagonal_operator}) for particle semantics, but now for graphs.
Proposition 2 then applies to $\hat{W}_{r}$ and $D_r$ alike, for all rules $r$ in the model.

{\bf Observation 3}
In this sense (Propositions 1 and 2 and Corollaries 1 and 2),
there is a higher-level algebra, generated by any collection of graph rewrite rule operators,
together with the identity operator $I =N^{(\varnothing)}= \hat{W}_{\varnothing \rightarrow \varnothing}$
that can be ``implemented'' by (mapped compositionally by operator algebra homomorphism to)
a sufficiently large indexed collection of binary state variables with their own 
lower-level state-changing operator algebra.

{\bf Observation 4}
{\it Semantics alternatives.}
An alternative semantics to
Equation (\ref{XRef-Equation-818134830}) 
could include factors of $\prod_{LHS} (Z_{i_p i_q})^{\bar{g}_{p q}}$ where $\bar{g}$
is a 0/1-valued matrix representing a second type of graph edge that
identifies {\it prohibited} connections on the LHS, and likewise 
$\prod_{RHS} (Z_{i_p^{\prime} i_q^{\prime}})^{\bar{g}^{\prime}_{p^{\prime} q^{\prime} }}$ for the RHS.
The normal form could put these new $Z$ product factors just to the right of (acting just before)
the corresponding factors for $g$ and $g^{\prime}$ in Equation (\ref{XRef-Equation-818134830}).
If corresponding entries of $g$ and 
$\bar{g}$ 
both take the value 1,
that inconsistency has no effect since
their product has a factor $a Z = a (I-\hat{a}a) = a - (a \hat{a})a= a - (I-\hat{a}a)a = a - a + \hat{a}a^2 = 0$.

Instead of the creation and annihilation operators for Boolean edge variables
we could use creation and annihilation operators for ${\mathbb N}$-valued numbers 
of identical particles in the definition (\ref{XRef-Equation-818134830}) of $\hat{W}_r$.
But (1) the handling of memory allocation
and deallocation by $E$ 
factors might have to be revised, and 
(2) graph grammar rules could have unintended semantics in terms of multigraphs:
graphs with nonnegative integer edge weights.
On the other hand, multigraphs and multigraph grammar rules can
also be useful, if that is the intended semantics.

\subsubsection{Slice Rewrite Rule Operators}
\label{sec_slice_rewrite}

The slice categories of Diagram 1, with 
$H = {\mathbb N}^+, \; {J}_D^+,  \;  {\mathbb N}_D^{+ \text{op}}, \; C_D, \; {\rm or }  \; \tilde{C}_D$
for graded graph, stratified graph, abstract cell complex, graded stratified graph, and graded abstract cell complex respectively,
are all variants of the category of labelled graphs $\varphi: G \rightarrow H$ 
whose labels $\kappa$ are nodes of $H$, with extra constraints added on
the integer-valued labels. 
We can encode these constraints in each case with a predicate
$P_H({\varphi})$,
and  enforce them with a real-valued ``gating'' indicator function $\Theta(P_H({\varphi}))$
which takes the value 1 if the predicate is satisfied and zero otherwise.
If an ordering on the nodes of $G$ is established, as we have assumed, then
these objects become 
$P_H({\kappab})$ and an indicator function $\Theta(P_H({\kappab}))$.
In the foregoing graph rewrite rule semantics, such an ordering is established
by the arbitrary indexing scheme of $p$ and $q$. So we may generalize
the graph rewrite rule to cover these slice categories as well by mapping
$\rho _{{\rm slice} \; r}({\kappab} , {\kappab^{\prime }})$
to a corresponding 
$\rho _{{\rm graph} \; r}(({\kappab, \lambdab}) , ({\kappab^{\prime }, \lambdab^{\prime }}))$:
\begin{equation}
\rho _{{\rm graph} \; r}(({\kappab, \lambdab}) , ({\kappab^{\prime }, \lambdab^{\prime }})) = 
\Theta(P_H({\kappab}))  \times 
\Theta(P_H({\kappab^{\prime }}))  \times 
\rho _{{\rm slice} \; H,\; r}({\lambdab} , {\lambdab^{\prime }}) .
\label{slice_rule}
\end{equation}
The first indicator function in Equation (\ref{slice_rule}) could be omitted if the initial condition gives nonzero
probability only to valid $H$-slice graphs and all rules in the grammar are valid
$H$-rewrite rules, maintaining the validity conditions $P_H$ using the second indicator function in  in Equation (\ref{slice_rule}).
The remaining operator products in Equation (\ref{XRef-Equation-818134830}) can remain the same,
yielding the operator algebra semantics of these (and potentially other) slice category rewrite rules,
complete with provision for extra domain-model specific labels $\lambda$.

In this way we could {\it implement} special graph grammar syntax for slice graph rewrite rules,
and thereby achieve summable operator algebra semantics for modeling languages with
rewrite rules at the level of graded graphs, stratified graphs, abstract cell complexes,
graded stratified graphs, and/or graded abstract cell complexes 
that could implement selected continuum limits such as mesh-approximable
stratified spaces and CW complexes (as suggested in Diagram 3 below).

For example in the case of undirected graded graphs, we could let directed edges represent $\Delta l = +1$
edges and undirected edges represent $\Delta l = 0$. 
The triangle 2d mesh refinement example
would then become:
\begin{equation}
\left( 
\begin{diagram}[size=1em]
&& & & 1 & \\
&& &\ldTo & &\rdLine(4,4) &&&&  \\ 
& & 4 & \\
&\ruTo & &&&  \\ 
\; 2 &  \rLine(7,0)  &&&&&&& 3 \;\;  &  \\
\end{diagram}
\right) 
\longrightarrow  
\left( 
\begin{diagram}[size=1em]
&& & & 1 & \\
&& &\ldTo & &\rdTo &&&&  \\ 
& & 4 & \rLine(2,0) &&&6 \\
&\ruTo & &\rdLine &&\ldLine && \luTo \\ 
\; 2 &  \rTo(2,0)  &&&5& \lTo(2,0) &&& 3 \;\;  &  \\
\end{diagram}
\right) 
\label{slice_triangle_rule}
\end{equation}

Directed graded graphs could be represented too,
with just one bit of edge label information to record whether there
is a change of level number along a directed edge or not.
In both cases the integer-valued level number edge labels are removed, to be
restored automatically by an implementation map ${\cal I}$ from slice graph grammar
rule syntax to ordinary graph grammar rule syntax.
(This implementation map could even be expressed
in the form of a 
declarative model transformation
meta-grammar rule, mapping rules
like (\ref{slice_triangle_rule}) to a slight variant of
rules like (\ref{triangle_rule}) with an AST for the labels.)
Similarly,
for 
stratified graphs one could label edges with $\Delta d$
and allow the interpretation process to restore $d$.
For abstract cell complexes one instead needs only to record one extra bit
of edge information regarding $d$: $\Delta d \in \{0, -1\}$.
As in the case of rule (\ref{triangle_rule}), 
with these extra edge labels rule (\ref{slice_triangle_rule})
could be made more elaborate by retaining all relevant
strata and their connections
at smaller level numbers rather than just the graded graph ``frontier''
comprising the deepest substrata in each stratum.

{\bf Observation 5}
In all the foregoing slice graph category cases the implementation mapping on rewrite rules
should match the implementation mapping on their semantics
as proposed above, so that slice implementation
at the model level
commutes with semantics:
\begin{diagram}
\text{Slice Rules} &  &\rTo^{\Psi}  && \text{Gated Op. Alg.} && \\
\dTo{{\cal I}_m} &  &  & & \dTo{{\cal I}_m} & \quad \text{    (Diagram 3 )}\\
\text{GraphGram Rules} &  &\rTo^{\Psi}  && \text{Operator Algebra} \\
\end{diagram}
where, again, the slice rewrite rules can pertain to
graded graphs, stratified graphs, abstract cell complexes,
graded stratified graphs, or graded abstract cell complexes; the latter two
could be used to support continuum limits such as mesh-approximable
stratified spaces and CW complexes.

\subsubsection{Pushout semantics}
\label{sec_pushout}

A very different approach to defining a similar idea of graph rewrite rule semantics
is provided by the ``double pushout'' category-theoretic construction
(Ehrig et al. 2006), 
using the category of graphs and graph transformations.

A pure (unlabelled) graph grammar rule $G \rightarrow_K G^{\prime}$ 
can replace graph $G$ with graph $G^{\prime}$,
holding common subgraph $K$ constant, anywhere that $G$ (and its subgraph $K$) occurs
as a subgraph inside of some ``host''  graph $C$  within the current pool of one or more graphs
(indeed $C$ can be taken to be the entire pool of graphs in the current state since that's just a big, possibly disconnected graph).
The result will be an altered version $C^{\prime}$ of the host or pool graph.
Using graph homomorphism arrows,
the double pushout diagram for the firing of a graph grammar rule
relates all these graphs as follows:
\begin{diagram}
G & \lTo & K & \rTo & G^{\prime}  \\
 \dTo && \dTo & &  \dTo &&&& (\text{Diagram }4) \\
C  &\NEpbk  \lTo &  D &\rTo \NWpbk  & C^{\prime}  \\
\end{diagram}
The diagram is to be universal at $C$ and $C^{\prime}$
(so any rival occupant for either position is essentially just a homomorphic image of the universal occupant),
hence is double pushout.
If such a $C^{\prime}$ exists, then it is a candidate outcome for the effect of a rule firing
of the given rule on the given pool.

What is nice about this diagram is that, as in Section~\ref{section_extended_constructive} above,
its objects and arrows can be reinterpreted in other graph-related categories 
including all the foregoing slice categories
such as labelled graphs, graded graphs, stratified graphs, abstract cell complexes, 
and their various combinations in Section~\ref{sec_nonstandard_defs} above; 
and any other typed attributed graphs.
We also imagine that it should be possible to {\it implement}
a slice graph category rewrite rule under the double pushout semantics
in terms of ordinary (double pushout) labelled graph rewrite rule firings,
by use of suitable labels.
A possible sticking point is verifying implementation of the
universalities of the pushout construction in the slice category.
In this paper we only studied the analogous question 
for operator algebra semantics (Section~\ref{sec_slice_rewrite} above).

There is also a closely related single pushout diagram version of the semantics,
and a collection of ``independence'' conditions
for two successive rule firings to have an order-independent result
(Ehrig et al. 2006).
The work by (Behr et al. 2016) discussed above combines in one paper 
and connects together both double-pushout and 
master equation semantics, using a restricted subset of the operator algebra 
implied by Propositions 1 or 2. 
Their work provides evidence in favor of some version
of the following idea:

{\bf Conjecture 1} Since the operator algebra and pushout diagram semantics are alternative
ways of defining the ``same'' operation, they must be related.
So we conjecture there is a map $\Psi_2$ making Diagram 5 commute:
\begin{diagram}
& & \text{GraphGram Rules} & \\
&\ldTo^{\Psi_1} & &\rdTo^{\Psi} &&&& (\text{Diagram } 5 ) \\ 
\text{Pushout} &  &\rTo^{[\Psi_2]}  && \text{Operator Algebra} \\
\end{diagram}
The conjectured semantics reduction mapping $\Psi_2$ can be defined as 
tranferring probability 
to the subspace of states of the Fock space 
corresponding to a $C^{\prime}$ post-firing pool graph,
from the subspace of states of the Fock space 
corresponding to a $C$ pre-firing pool graph.
The operator algebra of Propositions 1 and 2 and its corollaries,
possibly different in detail from that of (Behr et al. 2016),
is in turn mapped by an implementation map ${\cal I}$ 
to a computational substrate
including stochastic simulation algorithms.

\subsubsection{Efficient implementations}

We just saw that slice graph grammar rules can be implemented (efficiently) in terms
of ordinary labelled graph grammar rules.

The efficient implementation of graph grammars rules themselves can also be considered.
We have mentioned that they can be and have been implemented in terms of parameterized grammars with parameters
devoted to recording integer-valued ObjectIDs. 
That implies that worst-case performance for parameterized grammars can be
as bad as finding small unlabelled subgraphs in large unlabelled graphs, though finding subgraphs in
practice is a lot easier than finding them in worst case, and labels help substantially.
So one option is just to deploy algorithms that match small symbolic expressions,
or use computer algebra systems that have done the same. 
But another option is available specifically for declarative modeling:
to find 
the rules with
the most commonly occurring rule firings in a model,
and 
to use meta-grammars or a meta-language (discussed in Section~\ref{def_decl_mod})
to transform those rules
into submodels comprising rules taking only special forms
that can be compiled into special-case efficient simulation code.
Examples of special-form rules amenable to special-case simulation code
include parameterless rules, terms with parameter sets that take only a few values,
rules that consist only of differential equations, 
context-free grammar rules such as in Equation~(\ref{brain_lineage}),
1D chain preserving rules such as Equation~(\ref{root_subgrammar}),
and many other possibilities.
Then, use strategies like operator splitting to simulate quickly most of the time,
slowing down only for occasional higher-cost operations 
like cell division in a tissue model
or bundling/zippering in a microtubule network model.

\subsection{Meta-Hierarchy via Graphs}
\label{meta-hierarchies}

If we seek models in discrete mathematics for the idea of a ``hierarchy'' such as a hierarchy
of biological systems and subsystems, 
or a hierarchy of modeling methods,
the simplest possibility is a tree: a directed graph whose undirected counterpart has no cycles.
This graph could be labelled with the names of the subsystems, methods, or other concepts
in the hierarchy. Such a restrictive definition could be appropriate for a compositional hierarchy,
or for a strict classification aimed at reconstructing clades,
but not in general for a hierarchy of specializations or subsets in which a node may have several parent nodes.
For that case a less restrictive possibility is to model a hierarchy as
a labelled DAG (directed acyclic graph), which has no cycles as a directed graph. 
Thus a labelled DAG is a natural model for the idea of a hierarchy that is more general than a tree.

However, as the foregoing examples show,
a hierarchy may be composed of items related in several different ways
(composition, specialization, mutually exclusive specialization, and so on.)
This fact suggests a further generalization.
If the edge labels in a labelled DAG 
take values in a further DAG of possible relationships, 
themselves related
only
by specialization, the resulting compound structure
can be called a
{\it meta-hierarchy} since it encodes a hierarchy of interrelated hierarchies.
This kind of structure has precedent in eg. the more general
``typed attributed graphs'' of (Ehrig et al. 2006).

In Section \ref{sec_Tchicoma} we will consider a meta-hierarchy whose vertices index into (are labelled by) formal languages
for modeling aspects of biology.

\section{Model Reduction}
\label{sec_mod_red}

Model reduction can be a reasonable strategy to deal with biological complexity.
Instead of picking out the most important variables and processes to include in a model
{\it a priori}, one can include some reasonable representation of many variables and processes
(although the model still won't be complete) 
with reasonable initial parameter values 
in a fine-scale model, and then computationally seek a smaller, coarser scale model that behaves
in approximately the same way,
on some some set of ``observables'' or ``quantities of interest'',
in some relevant region of parameter space.

In addition to eliminating conditionally unnecessary state variables for simplicity's sake,
model reduction has the potential to: 
(a) enable scaling up to very large models through increased computational efficiency in simulation;
(b) mathematically connect predictive models across scales of description
for both causal authenticity and greater accuracy at each scale;
(c) enable the study of the great diversity of possible emergent phenomena,
as a function of the parameters, structures, and initial conditions of fine-scale models.
Repeated model reduction can result in a {\it hierarchical stack of inter-related models},
with which to systematically maximize these advantages.

How can we use machine learning to 
perform
the computational search for reduced models?
Given enough data from a pure (parameterless) stochastic chemical reaction network,
and the correct structure of the network, it is possible to learn the reaction rates
(Wang et al. 2010; Golightly and Wilkinson 2011). 
In Section~\ref{sec_dyn_boltz} below we summarize how to
learn not just reaction rates but an {\it effective reduced-state space model}
in the form of a time-varying Boltzmann distribution
in at least some examples by following very general principles,
for the parameterless case (Johnson et al. 2015)
and for the case in which parameters include spatial positions (Ernst et al. 2018).

The general theme of using machine learning to create computationally
efficient reduced models is rapidly advancing.
In computational chemistry for example (Smith et al. 2017)
develop a neural network for learning from, interpolating, and much more efficiently
applying the energy and therefore force information computed in Density Functional Theory fine-scale calculations.
Likewise, other work (e.g. (Burkardt et al. 2006)) addresses difficult problems in fluid flow.

As for the general semantics ($\Psi$) and implementation (${\cal I}$) families of structure-respecting mappings,
we will denote model reduction mappings by a mapping family symbol ${\cal R}$.

\subsection{Learning Boltzmann Distribution Dynamics}
\label{sec_dyn_boltz}

In the parameterless rewrite rule (e.g. a chemical reaction as in Equation (\ref{Equation-multiset}))
case we learn a coarse-scale approximation $\tilde{p}$ of $p$ as 
a time-varying version of a Boltzmann distribution
or Markov Random Field (MRF):
\begin{equation}
\pt({\ssb}, t; [ \mu_\alpha | \alpha] ) = \frac{1}{\mathcal{Z}(\mu(t))} \exp [ - \sum_\alpha \mu_\alpha(t) V_\alpha(s_i \in C_\alpha) ] .
\label{eq:gccd}
\end{equation}
Here each potential function $V_\alpha$ 
is a function of
a subset or clique $C_\alpha$ of the
components of ${\ssb}$, creating a bipartite ``factor graph'' of variable-nodes indexed by $i$
and factor-nodes indexed by $\alpha$ (Lauritzen 1995; Frey 2003). 
If the factor graph is not connected then its connected
components all factorize into independent probability distributions whose product is $\pt$.

To get the dynamics of ${\vec \mu}$
one can minimize the KL divergence between $\tilde{p}$ and $p$,
where $\tilde{p}$ evolves under a differential equation 
\begin{equation}
\frac{d \mu_\alpha}{ dt}  = F_\alpha( [\mu_\beta|\beta] ) = F_\alpha({\mub}) 
\label{eq:gccd_dyn}
\end{equation}
whose right hand sides $[ F_\alpha | \alpha ]$ can be taken to be a learned
combination of a large number of hand-designed basis functions
(Johnson et al. 2015; Johnson 2012),
optimizing a KL divergence between distributions $p$ and $\pt$.
This is the ``Graph-Constrained Correlation Dynamics'' (GCCD) model reduction method.
It was used to achieve a substantial reduction in number of variables
for modeling a molecular complex in synapses.

The goal of training for model reduction
may be summarized as minimal degradation over time
of the approximation on a set of observables:
\begin{diagram}[size=2em]
\pt(t) &&  \rTo^{\Delta t} & & \pt(t+\Delta t)  \\
& \rdTo & & &   \uTo & \rdTo &\\
\uTo<R && \simeq O(t) & \rDashto & \HonV &   &\approx O(t+\Delta t) & \; & \quad \quad (\text{Diagram } 6) \\
& \ruTo &  & & \uLine<R   & \ruTo &  \\
p(t)  &   &    \rTo^{\Delta t}  & & p(t+\Delta t) \\
\end{diagram}
Here $R$ is a restriction operator mapping fine-scale to coarse-scale states,
and $\Delta t$ indicates the passage of time under model dynamics.
This diagram can be stacked horizontally, for teacher-forcing model training,
or vertically, for application across more than two scales.
This definition of model reduction is discussed more extensively
in (Johnson et al. 2015).

This model reduction method can be extended to the case of continuous spatial parameters 
(Equations (\ref{XRef-Equation-924145912}) and (\ref{XRef-Equation-9288735})) as follows
(Ernst et al. 2018):
Instead of a discrete state vector $\ssb$ for all the numbers of all the possible 
(usually molecular) species, 
we have a representation comprising the total number $n:{\mathbb N}$ of objects (e.g. molecules) present,
indexed by $i \in \{1, \ldots n\}$, together with an $n$-dimensional vector $\alb$ of discrete species types $\alpha_i$ 
and an $n$-dimensional vector $\xb$ of continuous spatial parameter vectors $x_i$ such as $d=3$ dimensions of space
(though orientation could contribute ${d \choose 2}=3$ more components). 
Then a pure state vector (probability 1 concentrated on one state)
can be denoted by the ``ket'' basis vector $\ket{n,\alb,\xb,t}$, analogous to a single vector $\ket{\nb}$ in GCCD above,
and the full state of the system is a probability mixture of the basis states.
Similar to GCCD, construct a coarse-scale mixed state based on an energy function and a
Boltzmann distribution that sums over 1-particle contributions 
$\nu_1 (\alpha_{i_1}, x_{i_1}, t)$ to the energy, summed over one particle index $i_1$,
plus 2-particle contributions 
$\nu_2 ( (\alpha_{i_1 },\alpha_{ i_2}), (x_{i_1 }, x_{ i_2 }), t)$ 
that obey permutation invariance,
summed over two indices $i_1 < i_2$ , 
and so on up to $k$-particle contributions of order $k=K$:
\begin{equation}
\begin{split}
\ket{ [ \nu_k | k \in \{1, \dots K\}], t } &= 
\sum_{n=0}^\infty \sum_\alb \int d\xb 
\pt (n,\alb,\xb,t) 
\ket{n, \alb, \xb,t} ,
\\ &
 \text{  with mixture probabilities }   \pt (n,\alb,\xb,t) :\\
 \pt = \braket{n,\alb,\xb, t }{  [\nu_k | k ] , t } 
&= \frac{1}{
\mathcal{Z} \left [ [\nu_k | k]  \right ]
}
\exp [ - \sum_{k=1}^K \sum_\anglink \nu_k (\alb_\anglink, \xb_\anglink, t) ]
\label{eq:reducedKet}
\end{split}
\end{equation}
where $\anglink = \{ i_1 < i_2 < \dots < i_k : i \in [1,n] \}$ denotes ordered subsets of 
$k$ indexes
each in $\{ 1,  \dots, n \}$.
Here the partition functional
$\mathcal{Z} \left [ [\nu_k | k]  \right ]$ has two nested square brackets, the outer brackets indicating that
$\mathcal{Z}[\ldots]$ depends on its arguments as a functional depends on functions rather
than as a function depends on numbers, and the inner brackets
indicating that the $\nu_k$ functions indexed by $k$ should all be included
in the argument list.

The goal as in Diagram 6 is to find $\tilde{p}$ that approximates the solution to the master equation
$\dot{p}(n,\alb,\xb,t) = {W} \cdot p(n,\alb,\xb,t) $,
where $W$ sums over all processes that affect the state such as chemical reactions and diffusion
(studied in Ernst et al. (2018))
but eventually also active transport, crowding effects and so on.

To define the time evolution of the reduced model $\tilde{p}(t)$, 
introduce a set of functionals $[ {\cal F}_k | k \in \{1 \ldots K\}] $
(ODEs, PDEs, or even other forms for ${\cal F}$ could be tried)
to create a governing dynamical system
for the interaction functions $[ \nu_k  | k]$:
\begin{equation}
\frac{d}{dt} \nu_k (\alb_\anglink, \xb_\anglink, t) 
= {\cal F}_k \left [ [ \nu_{k^{\prime}} (\alb, \xb,t) |  k^{\prime}] \right ] .
\label{eq:ade}
\end{equation}
The criterion for choosing the functions ${\cal F}$ is to minimize an 
{\it action}
that integrates a dissimilarity measure
(KL-divergence)
between $p$ and $\tilde{p}$:
\begin{equation}
\begin{split}
S &= \int_0^\infty dt \; \dkl(p || \pt) , \quad \text{ where } \quad \\
\dkl (p || \pt) &= \sum_{n=0}^\infty \sum_\alb \int d\xb \; p(n,\alb,\xb,t) \ln \frac{p(n,\alb,\xb,t)}{\pt (n,\alb,\xb,t)} .
\end{split}
\end{equation}
In principle this minimization is a higher-order kind of variational calculus,
in which one optimizes a higher order functional $S[\![{\cal F}]\!]$ of a functional $\cal F$
whose arguments are functions $\nu_k(\alb,\xb)$ of continuous space $\xb$;
in practice this particular problem is well approximated by reduction
to a tractable algorithmic problem in terms of real-valued parameters
by PDE-constrained optimization and a spatial mesh.

Assuming this objective has been minimized, (Ernst et al. 2018)
use the chain rule of calculus to
show:
\begin{prop}
Given a reaction network and a 
fixed collection of $K$  
interaction functions $\{\nu_k\}_{k=1}^K$, the linearity of the CME in reaction operators 
$\dot{p} = \sum_r {W}^{(r)} p$ extends to the functionals ${\cal F}_k = \sum_r {\cal F}_k^{(r)}$.
\end{prop}

In this very limited technical sense, there is a vector space homomorphism 
(which is a particular kind of structure-respecting map ${\cal R}$)
from the spatial operator algebra semantics
to the Dynamic Boltzmann reduced model semantics.

\subsection{Speculation: Expression Dynamics for Model Search}

One more significant connection between modeling languages and graph grammars is the possible
use of graph grammars for structural model learning. Reaction and rewrite rule model classes share
with artificial neural networks, Markov Random Fields, and many other machine-learnable model classes
the property that model architecture is determined by a weighted graph (e.g. the bipartite graph
of reactions and reactants) whose structure can be sculpted by setting nonzero weights (e.g. reaction rates) to zero
or vice versa, with some expectation of continuity in the neighborhood of zero weight in an optimization formulation of training.
This operation would correspond to the deletion or insertion of weak rules.
However, other ``structural'' moves in graph space are bolder and may have higher potential optimization gain.
For example if a rule is {\it duplicated} while maintaining the sum of the rates, then the model behavior and objective
function are unchanged but at least one of the daughter rules may be freed up to drift and adopt new functions.
Analogous ``duplicate and drift'' mechanisms are believed to enable neofunctionalization
resulting in selective advantage in genetic evolution (Moore and Purugganan 2003; Thompson et al. 2016).
Such an operation is easy to encode with a graph grammar meta-rule.

Duplication with arrow-reversal of one daughter rule, on the other hand, would require for continuity
that the unreversed rule retain nearly its full strength and the reversed one enter the ruleset at very small strength.
But as discussed in Section~\ref{sec_semantic_map} such a 
reversed rule could subsequantly evolve in strength to establish detailed balance. 
Other meta-rules could implement the mutation, crossover and three-parent rule-generation 
operations of genetic algorithms and differential evolution. Semantic word embeddings
as used in current machine learning methods for natural language processing
may provide a way to learn the inter-substitutability of individual symbols in rewrite rules
based on similarity of vectors evolved under previous successful substitutions,
as has been done recently in the context of symbolic regression (Arabshahi et al. 2018).
This kind of mechanism also promotes the evolution of evolvability. 
So in addition to the generic metarules we have listed,
modeling-language specific or domain-specific rules could be optimized.
All of these mechanisms for evolutionary structural optimization of models
are local in the model AST and so can be encoded and studied using graph grammars.

\section{A Meta-Hierarchy for Declarative Modeling}
\label{sec_Tchicoma}

We have developed the ideas of declarative modeling 
including various formal modeling languages, together with
structure-respecting maps (some of them category morphisms or functors) 
between formal languages and related mathematical objects
for semantics, implementation, and model reduction.
We have also defined a ``meta-hierarchy'' as a DAG 
whose edges are labelled by a DAG of types of relationships,
the relationship types forming a specialization hierarchy.

\begin{figure}
\begin{center}
  \includegraphics[width=1.02\columnwidth]{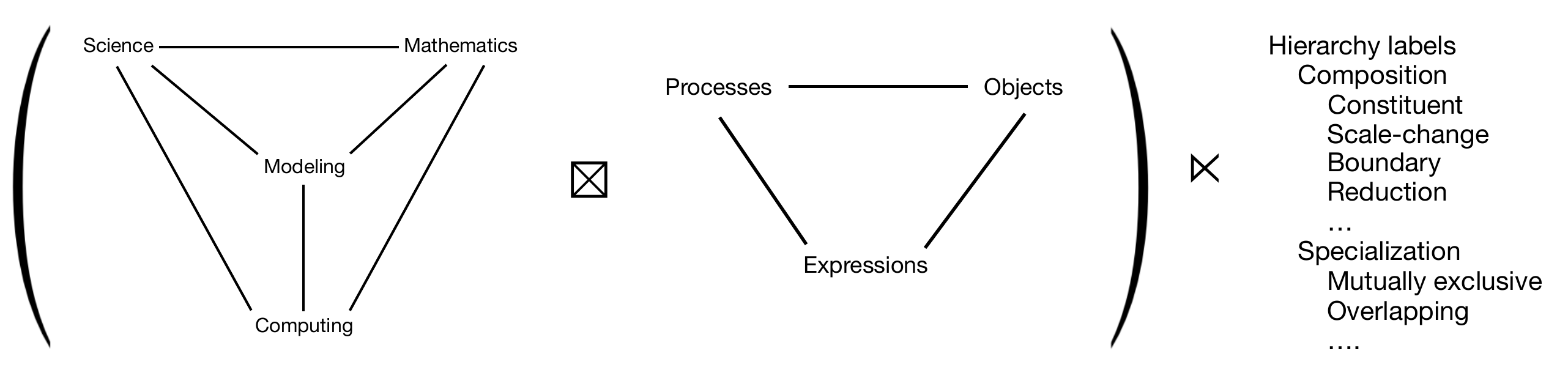}
\end{center}
\caption{ 
Outline of the Tchicoma metahierarchy for organizing formal languages for modeling and the structure-preserving maps between them.
First two labelled graphs are fully connected, and the $\boxtimes$ strong graph product between provides for plentiful potential connections.
They represent major ``knowledge domains'' and ``ontolexical categories'' respectively.
The third element is a DAG of hierarchical relationship types, as called for in the definition of a metahierarchy (see main text); 
this asymmetric construction is indicated idiosyncratically here by the ``$\ltimes$''  operator.
Actual formal languages, and structure-preserving mappings between them,
would be positioned deeply inside such a metahierarchy - not at the coarse indexing levels illustrated here.}
\label{fig:TM}       
\end{figure}

\begin{figure}
\begin{center}
  \includegraphics[width=1.0\columnwidth]{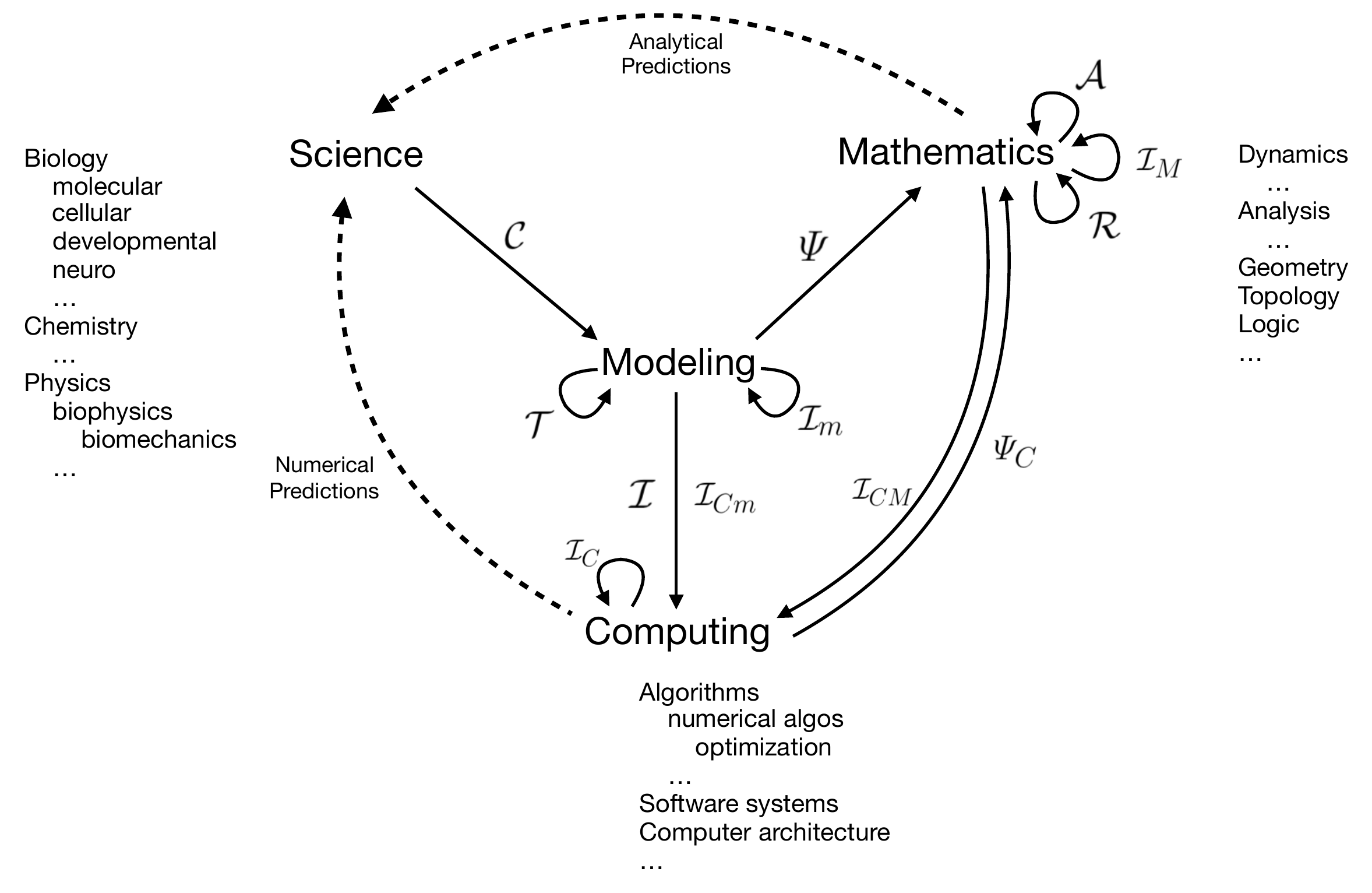}
\end{center}
\caption{
First  ``knowledge domain'' factor of the proposed Tchicoma metahierarchy for organizing formal languages
involved in complex biological modeling,
with graph edge labels showing the {\it purviews} 
of the main kinds of structure-preserving mappings between them ($\Psi, {\cal I, R, A, C, T}$),
as well as a collection of helper mapping types ($\Psi_C, \{{\cal I}_*\}$) aimed at computational implementation.
Solid arrows represent the purviews of various kinds of mappings;
the mappings themselves be would defined individually much deeper in the hierarchy. 
Mathematically founded implementation maps from 
modeling to computing should be related by         
${\cal I}_{Cm} \simeq {\cal I}_{CM} \circ \Psi$,
and/or ${\cal I}_{M} \circ \Psi \simeq  \Psi_C \circ {\cal I}_{Cm} $, where defined.
Additional mapping kinds could be defined by further such relations.
Implementation maps essential for computation can be built up by compositions such as
${\cal I} \simeq {\cal I}_{C} \circ  {\cal I}_{Cm} \circ  {\cal I}_m$, etc.,
which may vary in their computational efficiency and domain of applicability.
Dotted arrows show workflow information return to science.
Semantics maps $\Psi$ applies to modeling languages as discussed
in the text; semantics maps $\Psi_C$ apply to the semantics of
programming languages and models of computation, 
as studied in theoretical computer science.
The beginnings of a conventional specialization hierarchy for modeling-relevant topics
within several knowledge domains is also indicated.
 }
\label{fig:KD}       
\end{figure}

We propose here a particular extensible meta-hierarchy 
(as defined in Section~\ref{meta-hierarchies})
of formal languages aimed at declarative modeling
(as defined in Section~\ref{def_decl_mod})
of complex scientific domains such as developmental biology.
We 
attempt to
specify the top levels of the meta-hierarachy
(Figure~\ref{fig:TM})
but leave lower levels free to evolve with usage.
Every node in the meta-hierarchy is either a symbolic placeholder (typical for top level nodes)
or represents a formal language or sub-language to be used in a way that satisfies the definition of
declarative modeling. 
The top level nodes in the meta-hierarchy are symbolic placeholders for
a classification of deeper-level nodes along two independent labellings:
(a) ``ontolexical'', in which the labels are ``process'', ``object'', and ``expression''
as used in Section \ref{def_decl_mod}, and
(b) knowledge domain
(expanded in Figure~\ref{fig:KD}),
in which the labels are ``science'', ``mathematics'', and ``computing''.
In addition there is a ``declarative dynamical model informatics'' or simply ``models'' knowledge domain node,
aimed at mediating between the other three.
Of course many deep sub-classifications are possible especially for the knowledge domains,
beginning with specialization links to physics, chemistry, and biology, 
together with reduction links among their further specializations.
The resulting 
meta-hierarchy is named ``Tchicoma'' after a
volcanic formation in the southern Rocky Mountains.

As discussed in Section~\ref{meta-hierarchies}, 
the edges of the meta-hierarchy are labelled by relationship types
(e.g. composition vs. specialization, the latter specialized e.g.
into mutually exclusive and/or exhaustive vs. overlapping specializations;
also proven vs. machine-verifiably proven vs. unproven relationships; and so on)
that themselves stand in a specialization hierarchy.
Such relationship links could be used in the construction of maps
$\Psi$, ${\cal I}$, and ${\cal R}$ 
(for semantics, implementation and model reduction respectively)
in declarative modeling,
e.g. to retrieve similar known maps from previous work.
Specialization links can be used to insert conditions that enable theorems and algorithms to work,
and to evolve those conditions as knowledge accumulates.
Automatic curation of these link types would also provide an opportunity 
to keep usage-based statistics on the edges of each type
at each node in the meta-hierarchy
from prior successful model-building activities, 
for human visualization, for automatic heuristic search,
and for targeting the invention of new nodes in the meta-hierarchy
to regions with high previous application.
New nodes could be specializations or generalizations of single nodes
or of several nodes jointly,
resulting in a cumulative 
resculpting of the metahierarchy and its relationship type DAG under pressure of maximal utility.

The languages at the nodes of this meta-hierarchy can be specified by formal grammars, 
or they can be generated by unary and binary operators defined for all objects in some mathematical
category $C$,
including but not limited to
category-level binary operators such as universal sum, product, or function arrow
that can be defined purely in category-theoretic terms.
Such a generated language has the advantage that there is a built-in mathematical semantics
taking values in the category $C$ objects denoted by the operator expression trees.
In full generality the detection of semantic equivalence between expressions in such
languages is intractable, though it can often be specialized to a solvable problem.

The near-top nodes of the Tchicoma meta-hierarchy comprise a Cartesian product of
three ontolexical nodes and four knowledge domain nodes
(Figure~\ref{fig:TM}).
With the resulting twelve-element cross product classification 
one can identify the following potential kinds of structure-respecting inter-language mappings
(illustrated in more detail in Figure~\ref{fig:KD}),
differentiated by top-level source and target positions in the Tchicoma meta-hierarchy:

\begin{outline}[enumerate]
\1 Mappings discussed in this paper:
\2 ${\Psi}$: Semantics: Mapping {\it from} model object and/or process expressions in a modeling language $L$,
   {\it to} mathematical objects (specifically to time-evolution operators, in the case of model processes).
   This mapping is 
   an essential part of declarative modeling as defined in Section~\ref{def_decl_mod},
   and was detailed in Sections \ref{def_decl_mod} and \ref{section_extended}.
\2 ${\cal I}$: Implementation: Mapping {\it from} mathematical objects that are the semantics of model objects and/or model processes,
   {\it to} computational objects and processes respectively;
   or mapping from model expressions to computational expressions that denote these objects and processes; 
   as introduced in Section 3 and illustrated in Diagram 3.
    Computer Science seeks common target objects for many efficient implementation maps;
    current examples include Trilinos and PetSc for large-scale scientific computing.
\2 ${\cal R}$: Model reduction {\it from} model expressions {\it to} model expressions 
   with approximately the same semantics under projection to a set of observables; 
   as exemplified in Section 4 and Diagram 6.
\2 ${\cal X}$: Transformation: {\it From} expressions {\it to} expressions, 
            preserving or approximating semantics (or projections thereof); an essential part of declarative modeling as defined in Section~\ref{def_decl_mod}. 
            For example such transformations could include 
            computational implementation ${\cal I}_m$ at the model level, 
            model translation ${\cal T}$ (discussed below),
            and/or symbolic model reduction ${\cal R}_m$ = an inverse image of ${\cal R}$ under $\Psi$, if it exists.
\1 Mappings not detailed in this paper:
\2 ${\cal C}$: Model creation: {\it From} domain (object, process) expressions in a domain language
    {\it to} (object, process) expressions respectively in a modeling language.
    Domain languages may incorporate ontology formalizations such as the Web Ontology Language (``OWL'')
    or more general Description Logics; examples include 
    the BioPAX biological pathway description language and ontology (Demir et al. 2010), and
    the Gene Ontology (GO) 
    three-part hierarchy of biological objects and processes that has an object hierarchy 
    and two process hierarchies at two scales, with ``Is-A'' specialization links within each hierarchy.
    Declarative model creation packages implemented within computer algebra problem-solving environments
    include Cellerator (Shapiro et al. 2003, Shapiro et al. 2015b), Cellzilla (Shapiro et al. 2013), and Plenum (Yosiphon 2009).
\2 ${\cal T}$: Translation: {\it From} expressions {\it to} expressions, 
            preserving or approximating semantics (or projections thereof to observables).
            E.g. translation could formalize the process of translating between the inputs to different modeling software systems.
            By defining such translations first at the level of classical (and/or intuitionistic) mathematics rather than software,
            it is much easier to establish equivalences where they exist.
            Then one can explore whether and how deeply implementing or updating such translations is worth the effort.
\2 ${\cal A}$: Model analysis (phase diagrams, bifurcation diagrams, and the like): {\it From} models {\it to} mathematical analysis products
            such as reduced-parameter spaces, discontinuity or phase-change strata therein, 
            and  long-time asymptotics of observables. 
            Model analysis enables further reduction of submodels, for example by algebraic solution of fast subnetworks.
\end{outline}

For most of these  kinds of maps, 
several maps of the same kind could have the same source and target nodes
(e.g. multiple semantics maps related by refinement as discussed in 
Section~\ref{sec_refining});
in that case subscripting the mapping symbols by {\it mapping sub-kinds} could become necessary 
- although one would prefer to elaborate the meta-hierarchy nodes instead, if possible.

The curation of this meta-hierarchy may provide a fruitful application area for automatic theorem
verification software
based on theorem-proving methods,
since (a) many of these map types require the assertion of mathematical
equivalences and approximations that could be proven, possibly with computer help, in an automatically
verifiable form;  (b) the applicability of a particular map to a particular modelling problem
could be subject to logical inference on 
applicability conditions; 
and
(c)  the retrieval or synthesis of valid map {\it compositions} that achieve some formalizable
goal could be achieved by forward- and/or
reverse-chaining style theorem-proving algorithm.
Some of these applications of computer logic could be systematized 
using type theory and the Curry-Howard-Lambek
correspondence 
which in some circumstances at least maps
logical inference to type theory and to category theory
(Lambek and Scott 1986).
For ``computing'' nodes in the meta-hierarchy, and implementation maps that target them,
predictive declarative  models of computational resource use 
as a function of problem statement could also be collected and trained on past data.
Heuristic search for useful new intermediate nodes in the meta-hierarchy
could be based on the utility of constructing commutative diagrams
of inter-node mappings that ``lift'' one mapping along another,
e.g. lifting implementation maps to more 
mathematical levels of abstraction where possible,
in an internal improvement process akin to software refactoring.

\section{Conclusion}

We aim to formalize aspects of mathematical biological modeling
so that they become amenable both to computer assistance
and to cooperative human development of complex biological models.
This capability will be particularly useful in developmental biology,
where the necessity of relating genotype to phenotype 
in any fundamental ``evodevo'' research program
commits a modeler to repeated and often difficult scale-changes
within the modeling enterprise.
A conceptual framework,
based on declarative modeling,
in support of 
these goals
is presented.
The elements of the conceptual framework include an informal definition of
declarative biological modeling with formalized examples;
a nested series of declarative biological modeling languages with 
compatible mathematical semantics defined in terms of operator algebra;
a model reduction method based on machine learning
with which to tame the often necessary complexity of biological models;
and a meta-hierarchy of biological modeling sub-languages and methods,
organized and cross-linked by structure-respecting maps.

The biological modeling languages defined in this framework include
physics-derived 
operator algebra
semantics for processes expressed as reaction/rewrite rules
acting on 
discrete objects, 
parameters for such objects that can be discrete
and/or continuous variables,
extended objects with graph structure
including containment and adjacency,
and approximate spatially continuous object models.
Each of these object types receives appropriate dynamics
expressed in reaction/rewrite rules whose operator algebra semantics
is built from elementary creation and annihilation operators,
so they are compatible and can be mixed together into complex multi-rule models
by summation of time-evolution operators.
A constructive labelled-graph approach to the semantics of extended objects
including ``graded graphs'' labelled by approximation level number and/or
stratum dimension and identity
can approximate and computably implement constructive variants of classical nonconstructive
geometries such as manifolds, cell complexes, and stratified spaces.

In  many developmental biology  systems the spatial dynamics involves nontrivial changes
in geometry and/or topology of extended biological objects.
By using rewrite rules for the graph of strata defined in 
Section~\ref{sec_nonstandard_defs} 
together ODE-bearing rules for the parametric embeddings of individual strata into 3D space,
we now have in principle a way to represent such dynamics mathematically and computationally.

An essential step is to express natural graph-changing operations,
including a collection of labelled graph rewrite rules, in terms of an operator algebra
generated by the operators for the individual rules.
Each graph rewrite rule operator is expressed in terms of elementary
creation and annihilation operators, hence explicitly implementable in terms of
binary and/or integer-valued random variables.
Using this result (Propositions 1 and 2 and corollaries),
we also 
achieve summable operator algebra semantics for modeling languages with
rewrite rules at the level of graded graphs, stratified graphs, abstract cell complexes,
graded stratified graphs, and/or graded abstract cell complexes 
that could implement selected continuum limits such as mesh-approximable
stratified spaces and cell complexes.
A very expressive language of ``dynamical graph grammars'' results.
All rule operator products and commutators are explicitly calculable,
enabling the derivation of simulation and analysis methods.

Significant limitations of the approach as discussed here are of course legion and notably
include the fact that many graph structures can be defined virtually,
as the result of a function of the labels pertaining to two vertices that 
may or may not be connected, rather than in terms of explicitly represented edges
as we have generally assumed.

A potentially generic model reduction and moment closure method for such models is
based on dynamically evolving Boltzmann distributions, 
derived from fine-scale models by a form of machine learning.
Other model reduction methods may be enabled by the 
broad collection of possible mathematical model types that 
have been formalized as possible outcomes of model reduction.
In this way, both numeric (machine learning) and symbolic
(declarative model transformations) Artificial Intelligence methods
can be brought to bear on complex biological modeling problems.

Finally an  overarching meta-hierarchy of modeling sub-languages is proposed, within which
the structure-respecting maps required for declarative biological modeling could be 
defined, curated, and evolved through experience for maximal utility.
This framework 
may  provide
opportunities for mathematical biologists to 
contribute to systematic mappings for complex biological
model creation, definition, reduction, implementation and analysis
in ways that could be 
greatly amplified by
automation and artificially intelligent computational improvement.

\subsection*{Acknowledgements}

\small
The author is indebted to many discussants,
none of whom are responsible for errors herein.
This is a pre-print of an article published in
the Bulletin of Mathematical Biology.
The final authenticated BMB version is available
online at: https://link.springer.com/article/10.1007/s11538-019-00628-7.
There the paper is factored into Main text + Supplementary Material format.
The author wishes to thank the reviewers for the Bulletin, who  been particularly helpful;
also Ray Wightman of the Sainsbury Laboratory for help with sample preparation and microscopy.
The Microscopy Facility at the Sainsbury Laboratory is supported by the Gatsby Charitable Foundation.
The author wishes to acknowledge the hospitality of the Sainsbury Laboratory Cambridge University,
and the Center for Nonlinear Studies of the Los Alamos National Laboratory,
and funding from the Leverhulme Trust, 
National Institutes for Health grant R01HD073179, USAF/DARPA FA8750-14-C-0011,
National Institute of Aging grant R56AG059602, and Human Frontiers Science Program grant  HFSP - RGP0023/2018.
\par 
\normalsize

\section*{References}
\small

\hangindent=3em \noindent 
(Abelson et al. 1996)
Harold Abelson and Gerald Jay Sussman
Structure and Interpretation of Computer Programs -  Second  Edition
MIT Press.

\hangindent=3em \noindent 
(Alnaes et al. 2014)
Martin S. Alnaes, Anders Logg, Kristian B. Oelgaard, Marie E. Rognes, Garth N. Wells,
``Unified Form Language: A domain-specific language for weak formulations of partial differential equations''.
ACM Transactions on Mathematical Software.
Volume 40, Issue 2, Article No. 9, February 2014. Also arXiv:1211.4047 .

\hangindent=3em \noindent 
(Arabshahi et al. 2018)
F. Arabshahi, S. Singh, A. Anandkumar.
``Combining Symbolic Expressions and Black-box Function Evaluations for Training Neural Programs.''
Proc. International Conference on Learning Representations (ICLR). 2018

\hangindent=3em \noindent 
(Arnold et al. 2010) 
Douglas N. Arnold, Richard S. Falk, and Ragnar Winther,
``Finite Element Exterior Calculus: From Hodge Theory to Numerical Stability''.
Bulletin of the American Mathematical Society.
Volume 47, Number 2, pp. 281-354, April 2010.

\hangindent=3em \noindent 
(Backus 1978). 
J. Backus, 
``Can programming be liberated from the von Neumann style?: A functional style and its algebra of programs''. 
Communications of the ACM. 21 (8): 613, 1978. doi:10.1145/359576.359579. 

\hangindent=3em \noindent 
(Banwarth-Kuhn et al. 2018)
Mikahl Banwarth-Kuhn, Ali Nematbakhsh, Kevin W. Rodriguez, Stephen Snipes, Carolyn G. Rasmussen, G. Venugopala Reddy, Mark Alber
``Cell-Based Model of the Generation and Maintenance of the Shape and Structure of the Multilayered Shoot Apical Meristem of Arabidopsis thaliana''
Bulletin of Mathematical Biology
December 2018.

\hangindent=3em \noindent 
(Behr et al. 2016)
Nicolas Behr, Vincent Danos, Ilias Garnier,
``Stochastic mechanics of graph rewriting''. 
Proceedings of the 31st Annual ACM/IEEE Symposium on Logic in Computer Science, 
New York City, United States. pp.46 - 55, 2016.

\hangindent=3em \noindent 
(Bendich et al. 2007)
Inferring Local Homology from Sampled Stratified Spaces
Paul Bendich, David Cohen-Steiner, Herbert Edelsbrunner, John Harer, Dmitriy Morozov.
Proceedings of the 48th Annual IEEE Symposium on Foundations of Computer Science, pages 536-546, 2007.

\hangindent=3em \noindent 
(Blinov et al. 2004)
Michael L. Blinov, James R. Faeder, Byron Goldstein and William S. Hlavacek.
BioNetGen: software for rule-based modeling of signal transduction based on the interactions of molecular domains.
Bioinformatics,
Vol. 20 no. 17, pages 3289-3291, 2004.

\hangindent=3em \noindent 
(Booth and Tillotson 1980) 
P. Booth and J. Tillotson, Monoidal closed, Cartesian closed and convenient categories of topological spaces. Pacific J. Math. Volume 88, Number 1 (1980), 35-53.

\hangindent=3em \noindent 
(Brown et al. 2008)
R. Brown, I. Morris, J. Shrimpton and C.D. Wensley
Graphs of morphisms of graphs
the electronic journal of combinatorics 15 (2008), \#A1

\hangindent=3em \noindent 
(Burkardt et al. 2006) 
John Burkardt, Max Gunzburger, Hyung-Chun Lee
POD and CVT-based reduced-order modeling of Navier-Stokes flows
Computer Methods in Applied Mechanics and Engineering
Volume 196, Issues 1-3, Pages 337-355, 1 December 2006.

\hangindent=3em \noindent 
(Cardelli 2008)
Luca Cardelli
On process rate semantics 
Theoretical Computer Science 391 (2008) 190-215

\hangindent=3em \noindent 
(Chakrabortty et al., 2018)
``A Plausible Microtubule-Based Mechanism for Cell Division Orientation in Plant Embryogenesis''
Bandan Chakrabortty, Viola Willemsen, Thijs de Zeeuw, Che-Yang Liao,
Dolf Weijers, Bela Mulder, Ben Scheres.
Current Biology 28, 1-13
October 8, 2018. \\
https://doi.org/10.1016/j.cub.2018.07.025

\hangindent=3em \noindent 
(Coifman and Lafon 2006)
Coifman RR, Lafon S,  ``Diffusion maps''. Appl Comput Harm Anal 21:5-30, 2006.

\hangindent=3em \noindent 
(Danos et al. 2007)
Danos V, Feret J, Fontana W, Harmer R, Krivine J ,
``Rule-based modelling of cellular signaling''.
Lect Notes Comput Sci, Springer, 4703:17-41, 2007.

\hangindent=3em \noindent 
(Danos et al. 2012)
V. Danos, J. Feret, W. Fontana, R. Harmer, J. Hayman, J. Krivine, C. D. Thompson-Walsh, G. Winskel,
``Graphs, Rewriting and Pathway Reconstruction for Rule-Based Models''. 
FSTTCS 2012: 276-288

\hangindent=3em \noindent 
(de Goes et al. 2016)
Fernando de Goes, Mathieu Desbrun, Mark Meyer, and Tony DeRose,
``Subdivision Exterior Calculus for Geometry Processing''.
ACM Trans. Graph., 35(4), Art. 133, 2016.

\hangindent=3em \noindent 
(Demir et al. 2010)
Demir, Emek et al.
``BioPAX - A Community Standard for Pathway Data Sharing.''
Nature biotechnology 28.9 : 935-942, 2010.

\hangindent=3em \noindent 
(Desbrun et al 2005)
Mathieu Desbrun, Anil N. Hirani, Melvin Leok, Jerrold E. Marsden,
``Discrete Exterior Calculus''.
arXiv:math/0508341v2 , May 2002.

\hangindent=3em \noindent 
(Doi 1976a) M. Doi, Journal of Physics A: Mathematical and General 9, 1465
(1976).

\hangindent=3em \noindent 
(Doi 1976b) M. Doi, Journal of Physics A: Mathematical and General 9, 1479
(1976).

\hangindent=3em \noindent 
(Ehrig et al. 2006)
H. Ehrig, K. Ehrig, U. Prange, and G. Taentzer. 
Fundamentals of Algebraic Graph Transformation. 
Springer-Verlag Berlin Heidelberg, 2006. 

\hangindent=3em \noindent 
(Engwirda 2016) 
Darren Engwirda,
``Conforming restricted Delaunay mesh generation for piecewise smooth complexes''
25th International Meshing Roundtable (IMR25)
Procedia Engineering vol. 163 pp.84-96 (2016) 
doi:10.1016/j.proeng.2016.11.024

\hangindent=3em \noindent 
(Ermentrout 2004)
B. Ermentrout, Simplifying and Reducing Complex Models. 
In {\it Computational Modeling of Genetic and Biochemical Networks}, 
Bower and Bolouri, eds. MIT Press, 2004.

\hangindent=3em \noindent 
(Ernst et al. 2018)
Oliver K. Ernst, Tom Bartol, Terrence Sejnowski, and Eric Mjolsness. 
``Learning Dynamic Boltzmann Distributions as Reduced Models of Spatial Chemical Kinetics'', 
Journal of Chemical Physics 149, 034107, July 2018. 
Also arXiv 1803.01063, March 2018. 

\hangindent=3em \noindent 
(Fages and Soliman 2008)
Francois Fages and Sylvain Soliman 2008,
``Abstract Interpretation and Types for Systems Biology''.
Theoretical Computer Science 403, 52-70, 2008.

\hangindent=3em \noindent 
(Fomin 1994)
Serbey Fomin, 
Duality of Graded Graphs.
Journal of Algebraic Combinatorics 3, 357-404,  1994

\hangindent=3em \noindent 
(Frey 2003)
Brendan Frey,
``Extending Factor Graphs so as to
Unify Directed and Undirected Graphical Models''
Proceedings of the Nineteenth Conference on Uncertainty in Artificial Intelligence (UAI2003), 2003.
Also arXiv:1212.2486.

\hangindent=3em \noindent 
(Giavitto and Michel 2001)
Jean-Louis Giavitto and Olivier Michel,
``MGS: A Programming Language for the Transformations of Topological Collections''.
LaMI Technical Report 61-2001, 
Universit\'e d'Evry Val d'Essone
May 2001.

\hangindent=3em \noindent 
(Giavitto and Spicher 2008)
Jean-Louis Giavitto, Antoine Spicher
``Topological rewriting and the geometrization of programming''
Physica D 237 (2008) 1302-1314

\hangindent=3em \noindent 
(Glimm and Jaffe 1981)
Glimm and Jaffe, 
{\it Quantum Physics: A Functional Integral Point of View}. 
Springer-Verlag 1981, Secrtion 6.1, Minkowski space axiom W3.

\hangindent=3em \noindent 
(Golightly and Wilkinson 2011)
Andrew Golightly, Darren J. Wilkinson,
``Bayesian parameter inference for stochastic biochemical network models using particle Markov chain Monte Carlo''.
Interface Focus 2011 1 807-820; DOI: 10.1098/rsfs.2011.0047.  25 October 2011

\hangindent=3em \noindent 
(Hammond et al. 2011)
David K. Hammond, Pierre Vandergheynst, and R\'emi Gribonval,
``Wavelets on graphs via spectral graph theory''.
Applied and Computational Harmonic Analysis, Elsevier, 2011, 30 (2), pp.129-150.

\hangindent=3em \noindent 
(Hirsch 1976)
Morris Hirsch,
{\it Differential Topology},
Graduate Texts in Mathematics 33,
Springer-Verlag New York, 1976.

\hangindent=3em \noindent 
(Holguera et al. 2018)
 Isabel Holguera and Claude Desplan,
 ``Neuronal specification in space and time'',
Science 362, 176-180 (2018)

\hangindent=3em \noindent 
(HTT 2013) Homotopy Type Theory: Univalent Foundations of Mathematics  2013

\hangindent=3em \noindent 
(Hughes 2000)
Thomeas J. R. Hughes,
{\it Finite Element Methods}.
Dover 2000.

\hangindent=3em \noindent 
(Imrich and Klav\v{z}ar, 2000)
Wilfried Imrich and Sandi Klav\v{z}ar,
{\it Product Graphs: Structure and Recognition}.
John Wiley and Sons, 2000.

\hangindent=3em \noindent 
(Johnson 2012)
Gary Todd Johnson, 
``Dependency Diagrams and Graph- Constrained Correlation Dynamics: New Systems for Probabilistic Graphical Modeling'', 
PhD thesis, Computer Science Department, University of California, Irvine, March 2012.

\hangindent=3em \noindent 
(Johnson et al. 2015)
Todd Johnson, Thomas Bartol, Terrence Sejnowski, and Eric Mjolsness. 
``Model Reduction for Stochastic CaMKII Reaction Kinetics in Synapses by Graph-Constrained Correlation Dynamics'', 
Physical Biology 12:4, July 2015

\hangindent=3em \noindent 
(J\"{o}nsson et al. 2005)
 Henrik J\"{o}nsson, Marcus Heisler, Venugopala Reddy, Vikas Agrawal, Victoria Gor, Bruce E. Shapiro, Eric Mjolsness, Elliot M. Meyerowitz,
``Modeling the organization of the WUSCHEL expression domain in the shoot apical meristem''. 
Bioinformatics 21(supplÐ1):i232Ði240 (2005)

\hangindent=3em \noindent 
(J\"{o}nsson et al. 2006)
Henrik J\"{o}nsson, Marcus Heisler, Bruce E. Shapiro, Elliot M. Meyerowitz, Eric Mjolsness,
ÒAn auxin-driven polarized transport model for phyllotaxisÓ.
Proceedings of the National Academy of Sciences, 13 January 2006.

\hangindent=3em \noindent 
 (J\"{o}nsson et al. 2018)
 Henrik J\"{o}nsson and Sainsbury Laboratory Cambridge University research group,
 ``The Organism-Tissue Simulator'',
simulation software source code in the C++ language. URL \\
https://gitlab.com/slcu/teamhj/organism,
accessed December 2018.

\hangindent=3em \noindent 
(Jones et al. 2008)
Peter W. Jones, Mauro Maggioni, and Raanan Schul
``Manifold parametrizations by eigenfunctions of the Laplacian and heat kernels''
Proc. Nat. Acad. Science USA, vol. 105   no. 6   1803-1808, 2008 

\hangindent=3em \noindent 
(Joyner et al. 2012)
David Joyner,  Ond{\v r}ej {\v C}ertk, Aaron Meurer,  Brian E. Granger,
``Open source computer algebra systems: SymPy''.
 ACM Communications in Computer Algebra. 45 (3/4): 225-234, 2012.

\hangindent=3em \noindent 
(Julien et. al 2019)
Jean-Daniel Julien, Alain Pumir, Arezki Boudaoud
``Strain- or Stress-Sensing in Mechanochemical Patterning by the Phytohormone Auxin''
Bulletin of Mathematical Biology
March 2019.

\hangindent=3em \noindent 
(McQuarrie 1967)
Donald A. McQuarrie
``Stochastic Approach to Chemical Kinetics''.
Journal of Applied Probability, Vol. 4, No. 3 (Dec., 1967), pp. 413-478

\hangindent=3em \noindent 
(Kac 1974)
Mark Kac,
``A Stochastic Model Related to the Telegraphers Equation''.
Rocky Mountain Journal of Mathematics,
Volume 4, Number 3, Summer 1974.

\hangindent=3em \noindent 
(Kipf and Welling 2016)
Thomas N. Kipf, Max Welling,
``Semi-Supervised Classification with Graph Convolutional Networks''.
arXiv:1609.02907.

\hangindent=3em \noindent 
(Knauer 2011)
Ulrich Knauer
Algebraic Graph Theory: Morphisms, Monoids and Matrices
DeGruyter, Sept. 2011
978-3-11-025509-6
Section 4.3, Theorem 4.3.5.

\hangindent=3em \noindent 
(Lambek and Scott 1986)
J. Lambek and P. J. Scott,
``Introduction to Higher Order Categorical Logic'',
Cambridge University Press 1986.

\hangindent=3em \noindent 
(Lane 2015) Brendan Lane,
``Cell Complexes: The Structure of Space and the Mathematics of Modularity''',
PhD thesis, University of Calgary, September 2015.

\hangindent=3em \noindent 
(Lauritzen 1995)
Steffen L. Lauritzen,
Graphical Models,
Oxford Science Publications 1995.

\hangindent=3em \noindent 
(Logg et al. 2012)
Anders Logg and Kent-Andre Mardal and Garth N. Wells and others,
{\it Automated Solution of Differential Equations by the Finite Element Method}.
Springer, 2012.

\hangindent=3em \noindent 
(Maignan et al. 2015)
``Global Graph Transformations''
Luidnel Maignan, Antoine Spicher
Proceedings of the 6th International Workshop on Graph Computation Models.
L'Aquila, Italy.
Edited by Detlef Plump.
CUER Workshop Proceedings, Vol 1403.pp. 34-49. 
July 20, 2015.
http://ceur-ws.org/Vol-1403/

\hangindent=3em \noindent 
(Mandl and Shaw 2010)
Mandl and Shaw,
Quantum Field Theory, 2nd Edition.
J. Wiley. 
Section 13.2, requiring a Wick rotation $t \rightarrow \pm i t$ 
to relate quantum and statistical field theories.

\hangindent=3em \noindent 
(Martinelli et al 1982)
A. Martelli and U. Montanari,
``An Efficient Unification Algorithm''.
ACM Transactions on Programming Languages and Systems,Vol.4, No. 2, April 1982,Pages 258-282.
April 1982.

\hangindent=3em \noindent 
(Milnor 1980)
Robin Milner, {\it A Calculus of Communicating Systems}, Springer Verlag,1980.

\hangindent=3em \noindent 
(Mattis and Glasser 1998) D. C. Mattis and M. L. Glasser, Rev. Mod. Phys. 70, 979 (1998).

\hangindent=3em \noindent 
(McGrew et al. 2018)
McGrew, W. F., Zhang, X.,Fasano, R. J., SchŠffer, S. A., Beloy, K., Nicolodi, D. , Brown, R. C., Hinkley, N., Milani, G., Schioppo, M. and Yoon, T. H., and Ludlow, A. D.,
``Atomic clock performance enabling geodesy below the centimetre level''.
Nature 564(7734), pp. 87-90, 2018.

\hangindent=3em \noindent 
(Mironova et al. 2012)
Mironova VV, Omelyanchuk NA, Novoselova ES, Doroshkov AV, 
Kazantsev FV, Kochetov AV, Kolchanov NA, Mjolsness E., 
Likhoshvai VA. 
``Combined in silico/in vivo analysis of mechanisms providing for root apical meristem self-organization and maintenance.''
\\
Annals of Botany 110:2 pp 349-360 (DOI: 10.1093/aob/mcs069), July 2012.

\hangindent=3em \noindent 
(Mjolsness et al. 1991)
E. Mjolsness, D. H. Sharp, and J. Reinitz, 
``A Connectionist Model of Development''.
Journal of Theoretical Biology, vol 152 no 4, pp 429-454, 1991. 

\hangindent=3em \noindent 
(Mjolsness 2005)
Eric Mjolsness
``Stochastic Process Semantics for Dynamical Grammar Syntax: An Overview''
https://arxiv.org/abs/cs/0511073 ,
20 November 2005

\hangindent=3em \noindent 
(Mjolsness and Yosiphon 2006)
Eric Mjolsness and Guy Yosiphon,
``Stochastic Process Semantics for Dynamical Grammars'', 
Annals of Mathematics and Artificial Intelligence, 47(3-4) August 2006.

\hangindent=3em \noindent 
(Mjolsness et al. 2009)
E. Mjolsness, D. Orendorff, P. Chatelain, P. Koumoutsakos,
``An Exact Accelerated Stochastic Simulation Algorithm''. 
Journal of Chemical Physics 130 144110, 2009.

\hangindent=3em \noindent 
(Mjolsness  2010)
``Towards Measurable Types for Dynamical Process Modeling Languages'', Eric Mjolsness. 
Proceedings of the 26th Conference on Mathematical Foundations of Programming Semantics (MFPS 2010).
Electronic Notes in Theoretical Computer Science (ENTCS), vol. 265, pp. 123-144, 6 Sept. 2010, Elsevier. 

\hangindent=3em \noindent 
(Mjolsness and  Cunha 2012)
Eric Mjolsness and Alexandre Cunha,
``Topological object types for morphodynamic modeling languages''.
PMA 2012 : IEEE Fourth International Symposium on Plant Growth Modeling, Visualization and Applications. 
Shanghai China, October 2012. IEEE Press.

\hangindent=3em \noindent 
(Mjolsness and  Prasad 2013)
Eric Mjolsness and Upendra Prasad,
``Mathematics of Small Stochastic Reaction Networks: A Boundary Layer Theory for Eigenstate Analysis''.
Journal of Chemical Physics 138, 104111 (DOI: 10.1063/1.4794128), March 2013.

\hangindent=3em \noindent 
(Mjolsness 2013)
Eric Mjolsness, ``Time-ordered product expansions for computational stochastic systems biology''. Physical Biology, v 10, 035009, June 2013.

\hangindent=3em \noindent 
(Moore and Purugganan 2003)
Richard C. Moore and Michael D. Purugganan,
``The early stages of duplicate gene evolution''.
Proceedings of the National Academy of the United States of America,
 100 (26) 15682-15687; December 23, 2003.
https://doi.org/10.1073/pnas.2535513100

\hangindent=3em \noindent 
(Morrison and Kinney 2016)
Muir J. Morrison and Justin B. Kinney,
``Modeling multi-particle complexes in stochastic chemical systems''.
arXiv1603.07369v1.
March 2016.

\hangindent=3em \noindent 
(Murphy et al. 2001)
Michael Murphy, David M. Mount, and Carl W. Gable,
``A point-placement strategy for conforming Delaunay tetrahedralization''.
International Journal of Computational Geometry and Applications Vol. 11, No. 6  pp. 669-682, 2001.?

\hangindent=3em \noindent 
(Nedelec et al. 2007) 
Francois Nedelec and Dietrich Foethke
``Collective Langevin dynamics of flexible cytoskeletal fibers''
New J. Phys. 9 427,  2007.

\hangindent=3em \noindent 
(Orendorff and Mjolsness 2012)
David Orendorff and Eric Mjolsness. 
``A Hierarchical Exact Accelerated Stochastic Simulation Algorithm'', 
Journal of Chemical Physics 137, 214104 
(DOI: 10.1063/1.4766353 ; 
arXiv:1212.4080), December 2012.

\hangindent=3em \noindent 
(Peliti 1985) Peliti, L., J. Phys. France 46, 1469 (1985).

\hangindent=3em \noindent 
(Perlis 1982) 
Alan Perlis, epigram no.
102, in
``Epigrams in Programming'', ACM SIGPLAN September, 1982.
https://cpsc.yale.edu/epigrams-programming ,
accessed March 2018.

\hangindent=3em \noindent 
(Pierce 2002) Benjamin C. Pierce, {\it Types and Programming Languages}. MIT Press 2002.

\hangindent=3em \noindent 
(Plotkin 2004)
Gordon D. Plotkin. The Origins of Structural Operational Semantics. 
 The Journal of Logic and Algebraic Programming. 60-61:3-15, 2004.

\hangindent=3em \noindent 
(Poggio and Girosi 1990)
T. Poggio; F. Girosi
``Regularization Algorithms for Learning that are Equivalent to Multilayer Networks''
Science, New Series, Vol. 247, No. 4945. , pp. 978-982, Feb. 23, 1990.

\hangindent=3em \noindent 
(Prusinkiewicz and  Lindenmeyer 1990)
Przemyslaw Prusinkiewicz and Aristed Lindenmeyer 
{\it Algorithmic Beauty of Plants}
Springer-Verlag 1990 .

\hangindent=3em \noindent 
(Prusinkiewicz et al 1993)
P. Prusinkiewicz, M. S. Hammel, and E. Mjolsness, 
 ``Animation of Plant Development''.
SIGGRAPH '93 Conference Proceedings, ACM 1993. 

\hangindent=3em \noindent 
(Rand and Walkington 2009) Alexander Rand and Noel Walkington
``Collars and Intestines: Practical Conforming Delaunay Refinement''
In: Clark B.W. (eds) Proceedings of the 18th International Meshing Roundtable. Springer, Berlin, Heidelberg 2009.
https://doi.org/10.1007/978-3-642-04319-2\_28   

\hangindent=3em \noindent 
(Schulze and Tarkhanov 2003)
B.-W. Schulze N. Tarkhanov
``Differential analysis on stratified spaces''
In: {\it Hyperbolic Differential Operators and Related Problems}
(Eds. Ancona and J. Vaillant), 
Lecture Notes in Pure and Appl. Math., vol. 233, 
Marcel Dekker, New York, pp. 157-178, 2003.

\hangindent=3em \noindent 
(Shapiro et al. 2003) Bruce E. Shapiro, Andre Levchenko, Elliot M. Meyerowitz, Barbara  J. Wold and Eric D. Mjolsness.
``Cellerator: Extending a computer algebra system to include biochemical arrows for signal transduction simulations.''  
Bioinformatics, 19(5):677-678, 2003.

\hangindent=3em \noindent 
(Shapiro et al. 2012)
Bruce E Shapiro, Henrik J\"{o}nsson, Patrick Sahlin, Marcus Heisler, Adrienne Roeder, Michael Burl, Elliot M Meyerowitz, Eric D Mjolsness,
``Tessellations and Pattern Formation in Plant Growth and Development''.
arXiv:1209.2937 , 13 Sept. 2012.

\hangindent=3em \noindent 
(Shapiro et al. 2013)
Bruce E. Shapiro, Elliot Meyerowitz, Eric Mjolsness,
``Using Cellzilla for Plant Growth Simulations at the Cellular Level''.
Frontiers in Plant Biophysics and Modeling, 4:00408, 2013.

\hangindent=3em \noindent 
(Shapiro et al. 2015a)
Bruce E. Shapiro, Cory Tobin, Eric Mjolsness, and Elliot M. Meyerowitz,
``Analysis of Cell Divisions Patterns in the Arabidopsis Shoot Apical Meristem'', 
Proceedings of the National Academy of Sciences 112:15 pp 4815-4820, 2015.

\hangindent=3em \noindent 
(Shapiro et al. 2015b) Bruce E. Shapiro and Eric Mjolsness,
``Pycellerator: An arrow-based reaction-like modelling language for biological simulations''.
Bioinformatics, Oct 26. 2015. 
DOI: 10.1093/bioinformatics/btv596.

\hangindent=3em \noindent 
(Shaw et al. 2003)
S. L. Shaw, R. Kamyar, and D. W. Ehrhardt. Sustained Microtubule Treadmilling 
in Arabidopsis Cortical Arrays. Science, DOI: 10.1126/science.1083529, 2003.
Table 1 shows measured transition probabilities.

\hangindent=3em \noindent 
(Shellard et al 2018)
A. Shellard, A. Szabo, X. Trepat, R. Mayor
``Supracellular contraction at the rear of neural crest cell groups
drives collective chemotaxis'', 
Science vol 362 no 6412, 19 October 2018.

\hangindent=3em \noindent 
(Si and Gartner 2005) 
Si H. and Gartner K., 
Meshing Piecewise Linear Complexes by Constrained Delaunay Tetrahedralizations, 
Proceeding of the 14th International Meshing Roundtable, September 2005.

\hangindent=3em \noindent 
(Smith et al. 2017)
J. S. Smith  O. Isayev  and  A. E. Roitberg
``ANI-1: an extensible neural network potential with DFT accuracy at force field computational cost''
Chemical Science Chem. Sci.,8, 3192-3203, 2017.

\hangindent=3em \noindent 
(Spicher et al. 2007 )
Antoine Spicher and Olivier Michel,
``Declarative modeling of a neurulation-like process''.
Bio Systems, vol 87 2-3, pp. 281-8, 2007.

\hangindent=3em \noindent 
(Steenrod 1967) N. E. Steenrod, ``A convenient category of topological spaces''. Michigan Mathematical Journal, Volume 14, Issue 2, 133-152, 1967.

\hangindent=3em \noindent 
(Thompson et al. 2016)
Ammon Thompson, Harold H. Zakon, and Mark Kirkpatrick
``Compensatory Drift and the Evolutionary Dynamics of Dosage-Sensitive Duplicate Genes''
Genetics, vol. 202 no. 2 765-774; February 1, 2016 .
https://doi.org/10.1534/genetics.115.178137

\hangindent=3em \noindent 
(Tindemans et al. 2014)
Simon H. Tindemans, Eva E. Deinum, Jelmer J. Lindeboom and Bela M. Mulder,
``Efficient event-driven simulations shed new light on microtubule organization in the plant cortical array''
Frontiers in Physics, v. 2  Article 19, April 1014.
doi: 10.3389/fphy.2014.00019

\hangindent=3em \noindent 
(Van Kampen 1981) van Kampen, N. G.: \textit{Stochastic Processes
in Physics and Chemistry.} North-Holland (1981)\label{Van Kampen}

\hangindent=3em \noindent 
(Vemu et al. 2018)
``Severing enzymes amplify microtubule arrays through lattice GTP-tubulin incorporation''
Annapurna Vemu, Ewa Szczesna, Elena A. Zehr, Jeffrey O. Spector, Nikolaus Grigorieff, Alexandra M. Deaconescu, Antonina Roll-Mecak
Science Vol. 361, Issue 6404, eaau1504, 24 Aug 2018
DOI: 10.1126/science.aau1504

\hangindent=3em \noindent 
(Wang et al. 2010)
Yuanfeng Wang, Scott Christley, Eric Mjolsness, and Xiaohui Xie,
``Parameter inference for discretely observed stochastic kinetic models using stochastic gradient descent''.
BMC Systems Biology 4:99, 2010.

\hangindent=3em \noindent 
(Weinberger 1994)
S. Weinberger,
{\it The Topological Classification of Stratified Spaces},
University of Chicago Press, 1994.

\hangindent=3em \noindent 
(Wightman and Turner 2007)
R. Wightman, S.R. Turner. 
``Severing at sites of microtubule crossover contributes to microtubule alignment in cortical arrays''.
 The Plant Journal
 52:742-751, 2007.

\hangindent=3em \noindent 
(Winograd 1975)
Terry Winograd,  ``Frame Representations and the Procedural - Declarative Controversy,''
in D. Bobrow and A. Collins, eds., Representation and Understanding: Studies in Cognitive Science, 
Academic Press, pp. 185-210, 1975. 
Available as \\
http://hci.stanford.edu/winograd/ papers/FrameRep.pdf , accessed October 2018.

\hangindent=3em \noindent 
(Wolff et al. 2019)
Henri B. Wolff, Lance A. Davidson, Roeland M. H. Merks
``Adapting a Plant Tissue Model to Animal Development: Introducing Cell Sliding into VirtualLeaf''
Bulletin of Mathematical Biology
March 2019.

\hangindent=3em \noindent 
(Wolpert 1969)
L. Wolpert, ``Positional information and the spatial pattern of cellular differentiation''. 
J. Theoretical Biology 25 (1): 1Ð47, 1969.

\hangindent=3em \noindent 
(Wolfram Research 2017)
Wolfram Research, Inc.
``Mathematica Version 11''
Publisher: Wolfram Research, Inc.,
Champaign, Illinois 2018

\hangindent=3em \noindent 
(Yang et al. 2005)
Chin-Rang Yang, Bruce E. Shapiro, Eric D. Mjolsness and G. Wesley Hatfield
``An enzyme mechanism language for the mathematical modeling of metabolic pathways''
Bioinformatics, vol. 21 no. 6, pages 774-0780, 2005.
\\
doi:10.1093/bioinformatics/bti068

\hangindent=3em \noindent 
(Yosiphon 2009) 
Guy Yosiphon, 
``Stochastic Parameterized Grammars: Formalization, Inference, and Modeling Applications'', 
PhD thesis, Computer Science Department,
University of California, Irvine,
June 2009.
Thesis and code available at
\\
http://computableplant.ics.uci.edu/theses/ guy/downloads/DGPublications.html ,
accessed October 2018.

\normalsize

\end{document}